\documentclass[longauth]{aa}

\usepackage{graphicx}
\usepackage{natbib}
\usepackage{scalerel}
\usepackage{adjustbox}
\usepackage{ulem}

\usepackage[table]{xcolor}

\usepackage[nolist,nohyperlinks]{acronym}

\usepackage{txfonts}
\usepackage[pdfencoding=auto,psdextra]{hyperref}
\hypersetup{
    colorlinks=true,
    linkcolor=blue,
    filecolor=magenta,      
    urlcolor=blue,
    citecolor=blue
}
\usepackage{orcidlink} 
\newcommand{\orcid}[1]{\orcidlink{#1}}

\usepackage[nameinlink,capitalise]{cleveref}
\crefname{section}{Sect.}{Sects.}
\Crefname{section}{Section}{Sections}
\crefname{figure}{Fig.}{Figs.}
\Crefname{figure}{Figure}{Figures}
\crefname{equation}{Eq.}{Eqs.}
\Crefname{equation}{Equation}{Equations}

\makeatletter
\renewcommand*\aa@pageof{, page \thepage{} of \pageref*{LastPage}}
\makeatother
\usepackage{lastpage}

\usepackage[utf8]{inputenc}

\usepackage[switch, modulo]{lineno}

\usepackage{euclid}

\begin{document}

\acrodef{ASIC}{application specific integrated circuit}
\acrodef{BFE}{brighter-fatter effect}
\acrodef{CaLA}{camera-lens assembly}
\acrodef{CCD}{charge-coupled device}
\acrodef{CoLA}{corrector-lens assembly}
\acrodef{CDS}{Correlated Double Sampling}
\acrodef{CFC}{cryo-flex cable}
\acrodef{CFHT}{Canada-France-Hawaii Telescope}
\acrodef{CGH}{computer-generated hologram}
\acrodef{CNES}{Centre National d'Etude Spacial}
\acrodef{CPPM}{Centre de Physique des Particules de Marseille}
\acrodef{CPU}{central processing unit}
\acrodef{CR}{cosmic ray}
\acrodef{CTE}{coefficient of thermal expansion}
\acrodef{DCU}{Detector Control Unit}
\acrodef{DES}{Dark Energy Survey}
\acrodef{DPU}{Data Processing Unit}
\acrodef{DS}{Detector System}
\acrodef{EDS}{\Euclid Deep Survey}
\acrodef{EE}{encircled energy}
\acrodef{ERO}{Early Release Observations}
\acrodef{ESA}{European Space Agency}
\acrodef{EWS}{\Euclid Wide Survey}
\acrodef{FDIR}{Fault Detection, Isolation and Recovery}
\acrodef{FGS}{fine guidance sensor}
\acrodef{FOM}[FoM]{figure of merit}
\acrodef{FOV}[FoV]{field of view}
\acrodef{FPA}{focal-plane array}
\acrodef{FWA}{filter-wheel assembly}
\acrodef{FWC}{full-well capacity}
\acrodef{FWHM}{full width at half maximum}
\acrodef{GWA}{grism-wheel assembly}
\acrodef{H2RG}{HAWAII-2RG}
\acrodef{IP2I}{Institut de Physique des 2 Infinis de Lyon}
\acrodef{JWST}{{\em James Webb} Space Telescope}
\acrodef{IAD}{ion-assisted deposition}
\acrodef{ICU}{Instrument Control Unit}
\acrodef{IPC}{inter-pixel capacitance}
\acrodef{LAM}{Laboratoire d'Astrophysique de Marseille}
\acrodef{LED}{light-emitting diode}
\acrodef{LSB}{low-surface brightness}
\acrodef{MACC}{Multiple Accumulated}
\acrodef{MLI}{multi-layer insulation}
\acrodef{MMU}{Mass Memory Unit}
\acrodef{MPE}{Max-Planck-Institut für extraterrestrische Physik}
\acrodef{MPIA}{Max-Planck-Institut für Astronomie}
\acrodef{NA}{numerical aperture}
\acrodef{NASA}{National Aeronautics and Space Administration}
\acrodef{JPL}{NASA Jet Propulsion Laboratory}
\acrodef{MZ-CGH}{multi-zonal computer-generated hologram}
\acrodef{NI-CU}{NISP calibration unit}
\acrodef{NI-OA}{near-infrared optical assembly}
\acrodef{NI-GWA}{NISP Grism Wheel Assembly}
\acrodef{NIR}{near-infrared}
\acrodef{NISP}{Near-Infrared Spectrometer and Photometer}
\acrodef{PARMS}{plasma-assisted reactive magnetron sputtering}
\acrodef{PLM}{payload module}
\acrodef{PTFE}{polytetrafluoroethylene}
\acrodef{PV}{performance-verification}
\acrodef{PWM}{pulse-width modulation}
\acrodef{PSF}{point spread function}
\acrodef{QE}{quantum efficiency}
\acrodef{ROI}[RoI]{region of interest}
\acrodef{ROIC}{readout-integrated chip}
\acrodef{ROS}{reference observing sequence}
\acrodef{SCA}{sensor chip array}
\acrodef{SCE}{sensor chip electronics}
\acrodef{SCS}{sensor chip system}
\acrodef{SED}{spectral energy distribution}
\acrodef{SGS}{science ground segment}
\acrodef{SHS}{Shack-Hartmann sensor}
\acrodef{SNR}[S/N]{signal-to-noise ratio}
\acrodef{SED}{spectral energy distribution}
\acrodef{SiC}{silicon carbide}
\acrodef{SVM}{service module}
\acrodef{VIS}{visible imager}
\acrodef{WD}{white dwarf}
\acrodef{WCS}{world coordinate system}
\acrodef{WFE}{wavefront error}
\acrodef{ZP}{zero point}

%
%

\title{\Euclid: Early Release Observations -- Programme overview and pipeline for compact- and diffuse-emission photometry \thanks{This paper is published on behalf of the Euclid Consortium}}

\maxdeadcycles=500 
\author{J.-C.~Cuillandre\orcid{0000-0002-3263-8645}\thanks{\email{jc.cuillandre@cea.fr}}\inst{\ref{aff1}}
\and E.~Bertin\orcid{0000-0002-3602-3664}\inst{\ref{aff1}}
\and M.~Bolzonella\orcid{0000-0003-3278-4607}\inst{\ref{aff2}}
\and H.~Bouy\orcid{0000-0002-7084-487X}\inst{\ref{aff3},\ref{aff4}}
\and S.~Gwyn\orcid{0000-0001-8221-8406}\inst{\ref{aff5}}
\and S.~Isani\inst{\ref{aff6}}
\and M.~Kluge\orcid{0000-0002-9618-2552}\inst{\ref{aff7}}
\and O.~Lai\orcid{0000-0001-5656-7346}\inst{\ref{aff8}}
\and A.~Lan\c{c}on\orcid{0000-0002-7214-8296}\inst{\ref{aff9}}
\and D.~A.~Lang\orcid{0000-0002-1172-0754}\inst{\ref{aff10}}
\and R.~Laureijs\inst{\ref{aff11}}
\and T.~Saifollahi\orcid{0000-0002-9554-7660}\inst{\ref{aff9},\ref{aff12}}
\and M.~Schirmer\orcid{0000-0003-2568-9994}\inst{\ref{aff13}}
\and C.~Stone\orcid{0000-0002-9086-6398}\inst{\ref{aff14}}
\and Abdurro'uf\orcid{0000-0002-5258-8761}\inst{\ref{aff15}}
\and N.~Aghanim\orcid{0000-0002-6688-8992}\inst{\ref{aff16}}
\and B.~Altieri\orcid{0000-0003-3936-0284}\inst{\ref{aff17}}
\and F.~Annibali\inst{\ref{aff2}}
\and H.~Atek\orcid{0000-0002-7570-0824}\inst{\ref{aff18}}
\and P.~Awad\orcid{0000-0002-0428-849X}\inst{\ref{aff12}}
\and M.~Baes\orcid{0000-0002-3930-2757}\inst{\ref{aff19}}
\and E.~Ba\~nados\orcid{0000-0002-2931-7824}\inst{\ref{aff13}}
\and D.~Barrado\orcid{0000-0002-5971-9242}\inst{\ref{aff20}}
\and S.~Belladitta\orcid{0000-0003-4747-4484}\inst{\ref{aff13},\ref{aff2}}
\and V.~Belokurov\orcid{0000-0002-0038-9584}\inst{\ref{aff21}}
\and A.~Boselli\orcid{0000-0002-9795-6433}\inst{\ref{aff22},\ref{aff23}}
\and F.~Bournaud\inst{\ref{aff1}}
\and J.~Bovy\orcid{0000-0001-6855-442X}\inst{\ref{aff24}}
\and R.~A.~A.~Bowler\orcid{0000-0003-3917-1678}\inst{\ref{aff25}}
\and G.~Buenadicha\inst{\ref{aff17}}
\and F.~Buitrago\orcid{0000-0002-2861-9812}\inst{\ref{aff26},\ref{aff27}}
\and M.~Cantiello\orcid{0000-0003-2072-384X}\inst{\ref{aff28}}
\and D.~Carollo\orcid{0000-0002-0005-5787}\inst{\ref{aff29}}
\and S.~Codis\inst{\ref{aff1}}
\and M.~L.~M.~Collins\orcid{0000-0002-1693-3265}\inst{\ref{aff30}}
\and G.~Congedo\orcid{0000-0003-2508-0046}\inst{\ref{aff31}}
\and E.~Dalessandro\orcid{0000-0003-4237-4601}\inst{\ref{aff2}}
\and V.~de~Lapparent\orcid{0009-0007-1622-1974}\inst{\ref{aff18}}
\and F.~De~Paolis\orcid{0000-0001-6460-7563}\inst{\ref{aff32},\ref{aff33},\ref{aff34}}
\and J.~M.~Diego\orcid{0000-0001-9065-3926}\inst{\ref{aff35}}
\and P.~Dimauro\orcid{0000-0001-7399-2854}\inst{\ref{aff36},\ref{aff37}}
\and J.~Dinis\orcid{0000-0001-5075-1601}\inst{\ref{aff38},\ref{aff39}}
\and H.~Dole\orcid{0000-0002-9767-3839}\inst{\ref{aff16}}
\and P.-A.~Duc\orcid{0000-0003-3343-6284}\inst{\ref{aff40}}
\and D.~Erkal\orcid{0000-0002-8448-5505}\inst{\ref{aff30}}
\and M.~Ezziati\orcid{0009-0003-6065-1585}\inst{\ref{aff22}}
\and A.~M.~N.~Ferguson\inst{\ref{aff31}}
\and A.~Ferr\'e-Mateu\orcid{0000-0002-6411-220X}\inst{\ref{aff41},\ref{aff42}}
\and A.~Franco\orcid{0000-0002-4761-366X}\inst{\ref{aff33},\ref{aff32},\ref{aff34}}
\and R.~Gavazzi\orcid{0000-0002-5540-6935}\inst{\ref{aff22},\ref{aff18}}
\and K.~George\orcid{0000-0002-1734-8455}\inst{\ref{aff43}}
\and W.~Gillard\orcid{0000-0003-4744-9748}\inst{\ref{aff44}}
\and J.~B.~Golden-Marx\orcid{0000-0002-6394-045X}\inst{\ref{aff45}}
\and B.~Goldman\orcid{0000-0002-2729-7276}\inst{\ref{aff46},\ref{aff9}}
\and A.~H.~Gonzalez\orcid{0000-0002-0933-8601}\inst{\ref{aff47}}
\and R.~Habas\orcid{0000-0002-4033-3841}\inst{\ref{aff28}}
\and W.~G.~Hartley\inst{\ref{aff48}}
\and N.~A.~Hatch\orcid{0000-0001-5600-0534}\inst{\ref{aff45}}
\and R.~Kohley\inst{\ref{aff17}}
\and J.~Hoar\inst{\ref{aff17}}
\and J.~M.~Howell\orcid{0009-0002-2242-6515}\inst{\ref{aff31}}
\and L.~K.~Hunt\orcid{0000-0001-9162-2371}\inst{\ref{aff49}}
\and P.~Jablonka\orcid{0000-0002-9655-1063}\inst{\ref{aff50}}
\and M.~Jauzac\orcid{0000-0003-1974-8732}\inst{\ref{aff51},\ref{aff52},\ref{aff53},\ref{aff54}}
\and Y.~Kang\orcid{0009-0000-8588-7250}\inst{\ref{aff48}}
\and J.~H.~Knapen\orcid{0000-0003-1643-0024}\inst{\ref{aff42},\ref{aff41}}
\and J.-P.~Kneib\orcid{0000-0002-4616-4989}\inst{\ref{aff50}}
\and R.~Kohley\inst{\ref{aff17}}
\and P.~B.~Kuzma\orcid{0000-0003-1980-8838}\inst{\ref{aff31},\ref{aff55}}
\and S.~S.~Larsen\orcid{0000-0003-0069-1203}\inst{\ref{aff56}}
\and O.~Marchal\inst{\ref{aff9}}
\and J.~Mart\'{i}n-Fleitas\orcid{0000-0002-8594-569X}\inst{\ref{aff57}}
\and P.~Marcos-Arenal\orcid{0000-0003-1549-9396}\inst{\ref{aff58}}
\and F.~R.~Marleau\orcid{0000-0002-1442-2947}\inst{\ref{aff59}}
\and E.~L.~Mart\'in\orcid{0000-0002-1208-4833}\inst{\ref{aff42},\ref{aff41}}
\and D.~Massari\orcid{0000-0001-8892-4301}\inst{\ref{aff2}}
\and A.~W.~McConnachie\orcid{0000-0003-4666-6564}\inst{\ref{aff5}}
\and M.~Meneghetti\orcid{0000-0003-1225-7084}\inst{\ref{aff2},\ref{aff60}}
\and M.~Miluzio\inst{\ref{aff17},\ref{aff58}}
\and J.~Miro~Carretero\orcid{0000-0003-1808-1753}\inst{\ref{aff61},\ref{aff62}}
\and H.~Miyatake\orcid{0000-0001-7964-9766}\inst{\ref{aff63},\ref{aff64},\ref{aff65}}
\and M.~Mondelin\orcid{0009-0004-5954-0930}\inst{\ref{aff1}}
\and M.~Montes\orcid{0000-0001-7847-0393}\inst{\ref{aff42},\ref{aff41}}
\and A.~Mora\orcid{0000-0002-1922-8529}\inst{\ref{aff57}}
\and O.~M\"uller\orcid{0000-0003-4552-9808}\inst{\ref{aff50}}
\and C.~Nally\orcid{0000-0002-7512-1662}\inst{\ref{aff31}}
\and K.~Noeske\inst{\ref{aff66}}
\and A.~A.~Nucita\inst{\ref{aff32},\ref{aff33},\ref{aff34}}
\and P.~A.~Oesch\orcid{0000-0001-5851-6649}\inst{\ref{aff48},\ref{aff67},\ref{aff68}}
\and M.~Oguri\orcid{0000-0003-3484-399X}\inst{\ref{aff69},\ref{aff70}}
\and R.~F.~Peletier\orcid{0000-0001-7621-947X}\inst{\ref{aff12}}
\and M.~Poulain\orcid{0000-0002-7664-4510}\inst{\ref{aff71}}
\and L.~Quilley\orcid{0009-0008-8375-8605}\inst{\ref{aff72}}
\and G.~D.~Racca\inst{\ref{aff11}}
\and M.~Rejkuba\orcid{0000-0002-6577-2787}\inst{\ref{aff73}}
\and J.~Rhodes\orcid{0000-0002-4485-8549}\inst{\ref{aff74}}
\and P.-F.~Rocci\inst{\ref{aff16}}
\and J.~Rom\'an\orcid{0000-0002-3849-3467}\inst{\ref{aff41},\ref{aff42}}
\and S.~Sacquegna\orcid{0000-0002-8433-6630}\inst{\ref{aff32},\ref{aff33},\ref{aff34}}
\and E.~Saremi\orcid{0000-0002-5075-1764}\inst{\ref{aff75}}
\and R.~Scaramella\orcid{0000-0003-2229-193X}\inst{\ref{aff36}}
\and E.~Schinnerer\orcid{0000-0002-3933-7677}\inst{\ref{aff13}}
\and S.~Serjeant\orcid{0000-0002-0517-7943}\inst{\ref{aff76}}
\and E.~Sola\orcid{0000-0002-2814-3578}\inst{\ref{aff21}}
\and J.~G.~Sorce\orcid{0000-0002-2307-2432}\inst{\ref{aff77},\ref{aff16},\ref{aff78}}
\and F.~Tarsitano\orcid{0000-0002-5919-0238}\inst{\ref{aff48}}
\and I.~Tereno\inst{\ref{aff38},\ref{aff27}}
\and S.~Toft\orcid{0000-0003-3631-7176}\inst{\ref{aff68},\ref{aff67}}
\and C.~Tortora\orcid{0000-0001-7958-6531}\inst{\ref{aff79}}
\and M.~Urbano\orcid{0000-0001-5640-0650}\inst{\ref{aff9}}
\and A.~Venhola\orcid{0000-0001-6071-4564}\inst{\ref{aff71}}
\and K.~Voggel\orcid{0000-0001-6215-0950}\inst{\ref{aff40}}
\and J.~R.~Weaver\orcid{0000-0003-1614-196X}\inst{\ref{aff80}}
\and X.~Xu\orcid{0000-0001-7980-8202}\inst{\ref{aff12}}
\and M.~\v{Z}erjal\orcid{0000-0001-6023-4974}\inst{\ref{aff42},\ref{aff41}}
\and R.~Z\"oller\orcid{0000-0002-0938-5686}\inst{\ref{aff43},\ref{aff7}}
\and S.~Andreon\orcid{0000-0002-2041-8784}\inst{\ref{aff81}}
\and N.~Auricchio\orcid{0000-0003-4444-8651}\inst{\ref{aff2}}
\and M.~Baldi\orcid{0000-0003-4145-1943}\inst{\ref{aff82},\ref{aff2},\ref{aff60}}
\and A.~Balestra\orcid{0000-0002-6967-261X}\inst{\ref{aff83}}
\and S.~Bardelli\orcid{0000-0002-8900-0298}\inst{\ref{aff2}}
\and A.~Basset\inst{\ref{aff84}}
\and R.~Bender\orcid{0000-0001-7179-0626}\inst{\ref{aff7},\ref{aff43}}
\and C.~Bodendorf\inst{\ref{aff7}}
\and E.~Branchini\orcid{0000-0002-0808-6908}\inst{\ref{aff85},\ref{aff86},\ref{aff81}}
\and S.~Brau-Nogue\inst{\ref{aff87}}
\and M.~Brescia\orcid{0000-0001-9506-5680}\inst{\ref{aff88},\ref{aff79},\ref{aff89}}
\and J.~Brinchmann\orcid{0000-0003-4359-8797}\inst{\ref{aff90},\ref{aff91}}
\and S.~Camera\orcid{0000-0003-3399-3574}\inst{\ref{aff92},\ref{aff93},\ref{aff94}}
\and V.~Capobianco\orcid{0000-0002-3309-7692}\inst{\ref{aff94}}
\and C.~Carbone\orcid{0000-0003-0125-3563}\inst{\ref{aff95}}
\and J.~Carretero\orcid{0000-0002-3130-0204}\inst{\ref{aff96},\ref{aff97}}
\and S.~Casas\orcid{0000-0002-4751-5138}\inst{\ref{aff98}}
\and F.~J.~Castander\orcid{0000-0001-7316-4573}\inst{\ref{aff99},\ref{aff100}}
\and M.~Castellano\orcid{0000-0001-9875-8263}\inst{\ref{aff36}}
\and S.~Cavuoti\orcid{0000-0002-3787-4196}\inst{\ref{aff79},\ref{aff89}}
\and A.~Cimatti\inst{\ref{aff101}}
\and C.~J.~Conselice\orcid{0000-0003-1949-7638}\inst{\ref{aff25}}
\and L.~Conversi\orcid{0000-0002-6710-8476}\inst{\ref{aff102},\ref{aff17}}
\and Y.~Copin\orcid{0000-0002-5317-7518}\inst{\ref{aff103}}
\and F.~Courbin\orcid{0000-0003-0758-6510}\inst{\ref{aff50}}
\and H.~M.~Courtois\orcid{0000-0003-0509-1776}\inst{\ref{aff104}}
\and M.~Cropper\orcid{0000-0003-4571-9468}\inst{\ref{aff105}}
\and J.-G.~Cuby\orcid{0000-0002-8767-1442}\inst{\ref{aff6},\ref{aff22}}
\and A.~Da~Silva\orcid{0000-0002-6385-1609}\inst{\ref{aff38},\ref{aff39}}
\and H.~Degaudenzi\orcid{0000-0002-5887-6799}\inst{\ref{aff48}}
\and A.~M.~Di~Giorgio\orcid{0000-0002-4767-2360}\inst{\ref{aff106}}
\and M.~Douspis\orcid{0000-0003-4203-3954}\inst{\ref{aff16}}
\and C.~A.~J.~Duncan\inst{\ref{aff25}}
\and X.~Dupac\inst{\ref{aff17}}
\and S.~Dusini\orcid{0000-0002-1128-0664}\inst{\ref{aff107}}
\and M.~Fabricius\orcid{0000-0002-7025-6058}\inst{\ref{aff7},\ref{aff43}}
\and M.~Farina\orcid{0000-0002-3089-7846}\inst{\ref{aff106}}
\and S.~Farrens\orcid{0000-0002-9594-9387}\inst{\ref{aff1}}
\and S.~Ferriol\inst{\ref{aff103}}
\and S.~Fotopoulou\orcid{0000-0002-9686-254X}\inst{\ref{aff108}}
\and M.~Frailis\orcid{0000-0002-7400-2135}\inst{\ref{aff29}}
\and E.~Franceschi\orcid{0000-0002-0585-6591}\inst{\ref{aff2}}
\and S.~Galeotta\orcid{0000-0002-3748-5115}\inst{\ref{aff29}}
\and B.~Garilli\orcid{0000-0001-7455-8750}\inst{\ref{aff95}}
\and B.~Gillis\orcid{0000-0002-4478-1270}\inst{\ref{aff31}}
\and C.~Giocoli\orcid{0000-0002-9590-7961}\inst{\ref{aff2},\ref{aff109}}
\and P.~G\'omez-Alvarez\orcid{0000-0002-8594-5358}\inst{\ref{aff110},\ref{aff17}}
\and A.~Grazian\orcid{0000-0002-5688-0663}\inst{\ref{aff83}}
\and F.~Grupp\inst{\ref{aff7},\ref{aff43}}
\and L.~Guzzo\orcid{0000-0001-8264-5192}\inst{\ref{aff111},\ref{aff81}}
\and S.~V.~H.~Haugan\orcid{0000-0001-9648-7260}\inst{\ref{aff112}}
\and J.~Hoar\inst{\ref{aff17}}
\and H.~Hoekstra\orcid{0000-0002-0641-3231}\inst{\ref{aff61}}
\and W.~Holmes\inst{\ref{aff74}}
\and I.~Hook\orcid{0000-0002-2960-978X}\inst{\ref{aff113}}
\and F.~Hormuth\inst{\ref{aff114}}
\and A.~Hornstrup\orcid{0000-0002-3363-0936}\inst{\ref{aff115},\ref{aff116}}
\and P.~Hudelot\inst{\ref{aff18}}
\and K.~Jahnke\orcid{0000-0003-3804-2137}\inst{\ref{aff13}}
\and M.~Jhabvala\inst{\ref{aff117}}
\and E.~Keih\"anen\orcid{0000-0003-1804-7715}\inst{\ref{aff118}}
\and S.~Kermiche\orcid{0000-0002-0302-5735}\inst{\ref{aff44}}
\and A.~Kiessling\orcid{0000-0002-2590-1273}\inst{\ref{aff74}}
\and M.~Kilbinger\orcid{0000-0001-9513-7138}\inst{\ref{aff1}}
\and T.~Kitching\orcid{0000-0002-4061-4598}\inst{\ref{aff105}}
\and B.~Kubik\orcid{0009-0006-5823-4880}\inst{\ref{aff103}}
\and K.~Kuijken\orcid{0000-0002-3827-0175}\inst{\ref{aff61}}
\and M.~K\"ummel\orcid{0000-0003-2791-2117}\inst{\ref{aff43}}
\and M.~Kunz\orcid{0000-0002-3052-7394}\inst{\ref{aff119}}
\and H.~Kurki-Suonio\orcid{0000-0002-4618-3063}\inst{\ref{aff120},\ref{aff121}}
\and O.~Lahav\orcid{0000-0002-1134-9035}\inst{\ref{aff122}}
\and S.~Ligori\orcid{0000-0003-4172-4606}\inst{\ref{aff94}}
\and P.~B.~Lilje\orcid{0000-0003-4324-7794}\inst{\ref{aff112}}
\and V.~Lindholm\orcid{0000-0003-2317-5471}\inst{\ref{aff120},\ref{aff121}}
\and I.~Lloro\inst{\ref{aff123}}
\and D.~Maino\inst{\ref{aff111},\ref{aff95},\ref{aff124}}
\and E.~Maiorano\orcid{0000-0003-2593-4355}\inst{\ref{aff2}}
\and O.~Mansutti\orcid{0000-0001-5758-4658}\inst{\ref{aff29}}
\and O.~Marggraf\orcid{0000-0001-7242-3852}\inst{\ref{aff125}}
\and K.~Markovic\orcid{0000-0001-6764-073X}\inst{\ref{aff74}}
\and N.~Martinet\orcid{0000-0003-2786-7790}\inst{\ref{aff22}}
\and F.~Marulli\orcid{0000-0002-8850-0303}\inst{\ref{aff126},\ref{aff2},\ref{aff60}}
\and R.~Massey\orcid{0000-0002-6085-3780}\inst{\ref{aff52}}
\and S.~Maurogordato\inst{\ref{aff8}}
\and H.~J.~McCracken\orcid{0000-0002-9489-7765}\inst{\ref{aff18}}
\and E.~Medinaceli\orcid{0000-0002-4040-7783}\inst{\ref{aff2}}
\and Y.~Mellier\inst{\ref{aff127},\ref{aff18}}
\and G.~Meylan\inst{\ref{aff50}}
\and J.~J.~Mohr\orcid{0000-0002-6875-2087}\inst{\ref{aff43},\ref{aff7}}
\and M.~Moresco\orcid{0000-0002-7616-7136}\inst{\ref{aff126},\ref{aff2}}
\and L.~Moscardini\orcid{0000-0002-3473-6716}\inst{\ref{aff126},\ref{aff2},\ref{aff60}}
\and E.~Munari\orcid{0000-0002-1751-5946}\inst{\ref{aff29},\ref{aff128}}
\and R.~Nakajima\inst{\ref{aff125}}
\and R.~C.~Nichol\orcid{0000-0003-0939-6518}\inst{\ref{aff30}}
\and S.-M.~Niemi\inst{\ref{aff11}}
\and C.~Padilla\orcid{0000-0001-7951-0166}\inst{\ref{aff129}}
\and S.~Paltani\orcid{0000-0002-8108-9179}\inst{\ref{aff48}}
\and F.~Pasian\orcid{0000-0002-4869-3227}\inst{\ref{aff29}}
\and J.~A.~Peacock\orcid{0000-0002-1168-8299}\inst{\ref{aff31}}
\and K.~Pedersen\inst{\ref{aff130}}
\and W.~J.~Percival\orcid{0000-0002-0644-5727}\inst{\ref{aff131},\ref{aff132},\ref{aff10}}
\and V.~Pettorino\inst{\ref{aff11}}
\and S.~Pires\orcid{0000-0002-0249-2104}\inst{\ref{aff1}}
\and G.~Polenta\orcid{0000-0003-4067-9196}\inst{\ref{aff133}}
\and M.~Poncet\inst{\ref{aff84}}
\and L.~A.~Popa\inst{\ref{aff134}}
\and L.~Pozzetti\orcid{0000-0001-7085-0412}\inst{\ref{aff2}}
\and F.~Raison\orcid{0000-0002-7819-6918}\inst{\ref{aff7}}
\and R.~Rebolo\inst{\ref{aff42},\ref{aff41}}
\and A.~Refregier\inst{\ref{aff135}}
\and A.~Renzi\orcid{0000-0001-9856-1970}\inst{\ref{aff136},\ref{aff107}}
\and G.~Riccio\inst{\ref{aff79}}
\and Hans-Walter~Rix\orcid{0000-0003-4996-9069}\inst{\ref{aff13}}
\and E.~Romelli\orcid{0000-0003-3069-9222}\inst{\ref{aff29}}
\and M.~Roncarelli\orcid{0000-0001-9587-7822}\inst{\ref{aff2}}
\and E.~Rossetti\orcid{0000-0003-0238-4047}\inst{\ref{aff82}}
\and R.~Saglia\orcid{0000-0003-0378-7032}\inst{\ref{aff43},\ref{aff7}}
\and D.~Sapone\orcid{0000-0001-7089-4503}\inst{\ref{aff137}}
\and P.~Schneider\orcid{0000-0001-8561-2679}\inst{\ref{aff125}}
\and T.~Schrabback\orcid{0000-0002-6987-7834}\inst{\ref{aff59}}
\and A.~Secroun\orcid{0000-0003-0505-3710}\inst{\ref{aff44}}
\and G.~Seidel\orcid{0000-0003-2907-353X}\inst{\ref{aff13}}
\and S.~Serrano\orcid{0000-0002-0211-2861}\inst{\ref{aff100},\ref{aff138},\ref{aff99}}
\and C.~Sirignano\orcid{0000-0002-0995-7146}\inst{\ref{aff136},\ref{aff107}}
\and G.~Sirri\orcid{0000-0003-2626-2853}\inst{\ref{aff60}}
\and J.~Skottfelt\orcid{0000-0003-1310-8283}\inst{\ref{aff139}}
\and L.~Stanco\orcid{0000-0002-9706-5104}\inst{\ref{aff107}}
\and P.~Tallada-Cresp\'{i}\orcid{0000-0002-1336-8328}\inst{\ref{aff96},\ref{aff97}}
\and A.~N.~Taylor\inst{\ref{aff31}}
\and H.~I.~Teplitz\orcid{0000-0002-7064-5424}\inst{\ref{aff140}}
\and R.~Toledo-Moreo\orcid{0000-0002-2997-4859}\inst{\ref{aff141}}
\and A.~Tsyganov\inst{\ref{aff142}}
\and I.~Tutusaus\orcid{0000-0002-3199-0399}\inst{\ref{aff87}}
\and E.~A.~Valentijn\inst{\ref{aff12}}
\and L.~Valenziano\orcid{0000-0002-1170-0104}\inst{\ref{aff2},\ref{aff143}}
\and T.~Vassallo\orcid{0000-0001-6512-6358}\inst{\ref{aff43},\ref{aff29}}
\and G.~Verdoes~Kleijn\orcid{0000-0001-5803-2580}\inst{\ref{aff12}}
\and Y.~Wang\orcid{0000-0002-4749-2984}\inst{\ref{aff140}}
\and J.~Weller\orcid{0000-0002-8282-2010}\inst{\ref{aff43},\ref{aff7}}
\and O.~R.~Williams\orcid{0000-0003-0274-1526}\inst{\ref{aff142}}
\and G.~Zamorani\orcid{0000-0002-2318-301X}\inst{\ref{aff2}}
\and E.~Zucca\orcid{0000-0002-5845-8132}\inst{\ref{aff2}}
\and C.~Baccigalupi\orcid{0000-0002-8211-1630}\inst{\ref{aff128},\ref{aff29},\ref{aff144},\ref{aff145}}
\and C.~Burigana\orcid{0000-0002-3005-5796}\inst{\ref{aff146},\ref{aff143}}
\and P.~Casenove\inst{\ref{aff84}}
\and P.~Liebing\inst{\ref{aff105}}
\and V.~Scottez\inst{\ref{aff127},\ref{aff147}}
\and P.~Simon\inst{\ref{aff125}}
\and D.~Scott\orcid{0000-0002-6878-9840}\inst{\ref{aff148}}\\ \vspace{1cm} }
										   
\institute{Universit\'e Paris-Saclay, Universit\'e Paris Cit\'e, CEA, CNRS, AIM, 91191, Gif-sur-Yvette, France\label{aff1}
\and
INAF-Osservatorio di Astrofisica e Scienza dello Spazio di Bologna, Via Piero Gobetti 93/3, 40129 Bologna, Italy\label{aff2}
\and
Laboratoire d'Astrophysique de Bordeaux, CNRS and Universit\'e de Bordeaux, All\'ee Geoffroy St. Hilaire, 33165 Pessac, France\label{aff3}
\and
Institut universitaire de France (IUF), 1 rue Descartes, 75231 PARIS CEDEX 05, France\label{aff4}
\and
NRC Herzberg, 5071 West Saanich Rd, Victoria, BC V9E 2E7, Canada\label{aff5}
\and
Canada-France-Hawaii Telescope, 65-1238 Mamalahoa Hwy, Kamuela, HI 96743, USA\label{aff6}
\and
Max Planck Institute for Extraterrestrial Physics, Giessenbachstr. 1, 85748 Garching, Germany\label{aff7}
\and
Universit\'e C\^{o}te d'Azur, Observatoire de la C\^{o}te d'Azur, CNRS, Laboratoire Lagrange, Bd de l'Observatoire, CS 34229, 06304 Nice cedex 4, France\label{aff8}
\and
Observatoire Astronomique de Strasbourg (ObAS), Universit\'e de Strasbourg - CNRS, UMR 7550, Strasbourg, France\label{aff9}
\and
Perimeter Institute for Theoretical Physics, Waterloo, Ontario N2L 2Y5, Canada\label{aff10}
\and
European Space Agency/ESTEC, Keplerlaan 1, 2201 AZ Noordwijk, The Netherlands\label{aff11}
\and
Kapteyn Astronomical Institute, University of Groningen, PO Box 800, 9700 AV Groningen, The Netherlands\label{aff12}
\and
Max-Planck-Institut f\"ur Astronomie, K\"onigstuhl 17, 69117 Heidelberg, Germany\label{aff13}
\and
Department of Physics, Universit\'{e} de Montr\'{e}al, 2900 Edouard Montpetit Blvd, Montr\'{e}al, Qu\'{e}bec H3T 1J4, Canada\label{aff14}
\and
Johns Hopkins University 3400 North Charles Street Baltimore, MD 21218, USA\label{aff15}
\and
Universit\'e Paris-Saclay, CNRS, Institut d'astrophysique spatiale, 91405, Orsay, France\label{aff16}
\and
ESAC/ESA, Camino Bajo del Castillo, s/n., Urb. Villafranca del Castillo, 28692 Villanueva de la Ca\~nada, Madrid, Spain\label{aff17}
\and
Institut d'Astrophysique de Paris, UMR 7095, CNRS, and Sorbonne Universit\'e, 98 bis boulevard Arago, 75014 Paris, France\label{aff18}
\and
Sterrenkundig Observatorium, Universiteit Gent, Krijgslaan 281 S9, 9000 Gent, Belgium\label{aff19}
\and
Centro de Astrobiolog\'ia (CAB), CSIC-INTA, ESAC Campus, Camino Bajo del Castillo s/n, 28692 Villanueva de la Ca\~nada, Madrid, Spain\label{aff20}
\and
Institute of Astronomy, University of Cambridge, Madingley Road, Cambridge CB3 0HA, UK\label{aff21}
\and
Aix-Marseille Universit\'e, CNRS, CNES, LAM, Marseille, France\label{aff22}
\and
INAF - Osservatorio Astronomico di Cagliari, Via della Scienza 5, 09047 Selargius (CA), Italy\label{aff23}
\and
David A. Dunlap Department of Astronomy \& Astrophysics, University of Toronto, 50 St George Street, Toronto, Ontario M5S 3H4, Canada\label{aff24}
\and
Jodrell Bank Centre for Astrophysics, Department of Physics and Astronomy, University of Manchester, Oxford Road, Manchester M13 9PL, UK\label{aff25}
\and
Departamento de F\'{i}sica Te\'{o}rica, At\'{o}mica y \'{O}ptica, Universidad de Valladolid, 47011 Valladolid, Spain\label{aff26}
\and
Instituto de Astrof\'isica e Ci\^encias do Espa\c{c}o, Faculdade de Ci\^encias, Universidade de Lisboa, Tapada da Ajuda, 1349-018 Lisboa, Portugal\label{aff27}
\and
INAF - Osservatorio Astronomico d'Abruzzo, Via Maggini, 64100, Teramo, Italy\label{aff28}
\and
INAF-Osservatorio Astronomico di Trieste, Via G. B. Tiepolo 11, 34143 Trieste, Italy\label{aff29}
\and
School of Mathematics and Physics, University of Surrey, Guildford, Surrey, GU2 7XH, UK\label{aff30}
\and
Institute for Astronomy, University of Edinburgh, Royal Observatory, Blackford Hill, Edinburgh EH9 3HJ, UK\label{aff31}
\and
Department of Mathematics and Physics E. De Giorgi, University of Salento, Via per Arnesano, CP-I93, 73100, Lecce, Italy\label{aff32}
\and
INFN, Sezione di Lecce, Via per Arnesano, CP-193, 73100, Lecce, Italy\label{aff33}
\and
INAF-Sezione di Lecce, c/o Dipartimento Matematica e Fisica, Via per Arnesano, 73100, Lecce, Italy\label{aff34}
\and
Instituto de F\'isica de Cantabria, Edificio Juan Jord\'a, Avenida de los Castros, 39005 Santander, Spain\label{aff35}
\and
INAF-Osservatorio Astronomico di Roma, Via Frascati 33, 00078 Monteporzio Catone, Italy\label{aff36}
\and
Observatorio Nacional, Rua General Jose Cristino, 77-Bairro Imperial de Sao Cristovao, Rio de Janeiro, 20921-400, Brazil\label{aff37}
\and
Departamento de F\'isica, Faculdade de Ci\^encias, Universidade de Lisboa, Edif\'icio C8, Campo Grande, PT1749-016 Lisboa, Portugal\label{aff38}
\and
Instituto de Astrof\'isica e Ci\^encias do Espa\c{c}o, Faculdade de Ci\^encias, Universidade de Lisboa, Campo Grande, 1749-016 Lisboa, Portugal\label{aff39}
\and
Universit\'e de Strasbourg, CNRS, Observatoire astronomique de Strasbourg, UMR 7550, 67000 Strasbourg, France\label{aff40}
\and
Departamento de Astrof\'isica, Universidad de La Laguna, 38206, La Laguna, Tenerife, Spain\label{aff41}
\and
Instituto de Astrof\'isica de Canarias, Calle V\'ia L\'actea s/n, 38204, San Crist\'obal de La Laguna, Tenerife, Spain\label{aff42}
\and
Universit\"ats-Sternwarte M\"unchen, Fakult\"at f\"ur Physik, Ludwig-Maximilians-Universit\"at M\"unchen, Scheinerstrasse 1, 81679 M\"unchen, Germany\label{aff43}
\and
Aix-Marseille Universit\'e, CNRS/IN2P3, CPPM, Marseille, France\label{aff44}
\and
School of Physics and Astronomy, University of Nottingham, University Park, Nottingham NG7 2RD, UK\label{aff45}
\and
International Space University, 1 rue Jean-Dominique Cassini, 67400 Illkirch-Graffenstaden, France\label{aff46}
\and
Department of Astronomy, University of Florida, Bryant Space Science Center, Gainesville, FL 32611, USA\label{aff47}
\and
Department of Astronomy, University of Geneva, ch. d'Ecogia 16, 1290 Versoix, Switzerland\label{aff48}
\and
INAF-Osservatorio Astrofisico di Arcetri, Largo E. Fermi 5, 50125, Firenze, Italy\label{aff49}
\and
Institute of Physics, Laboratory of Astrophysics, Ecole Polytechnique F\'ed\'erale de Lausanne (EPFL), Observatoire de Sauverny, 1290 Versoix, Switzerland\label{aff50}
\and
Department of Physics, Centre for Extragalactic Astronomy, Durham University, South Road, DH1 3LE, UK\label{aff51}
\and
Department of Physics, Institute for Computational Cosmology, Durham University, South Road, DH1 3LE, UK\label{aff52}
\and
Astrophysics Research Centre, University of KwaZulu-Natal, Westville Campus, Durban 4041, South Africa\label{aff53}
\and
School of Mathematics, Statistics \& Computer Science, University of KwaZulu-Natal, Westville Campus, Durban 4041, South Africa\label{aff54}
\and
National Astronomical Observatory of Japan, 2-21-1 Osawa, Mitaka, Tokyo 181-8588, Japan\label{aff55}
\and
Department of Astrophysics/IMAPP, Radboud University, PO Box 9010, 6500 GL Nijmegen, The Netherlands\label{aff56}
\and
Aurora Technology for European Space Agency (ESA), Camino bajo del Castillo, s/n, Urbanizacion Villafranca del Castillo, Villanueva de la Ca\~nada, 28692 Madrid, Spain\label{aff57}
\and
HE Space for European Space Agency (ESA), Camino bajo del Castillo, s/n, Urbanizacion Villafranca del Castillo, Villanueva de la Ca\~nada, 28692 Madrid, Spain\label{aff58}
\and
Universit\"at Innsbruck, Institut f\"ur Astro- und Teilchenphysik, Technikerstr. 25/8, 6020 Innsbruck, Austria\label{aff59}
\and
INFN-Sezione di Bologna, Viale Berti Pichat 6/2, 40127 Bologna, Italy\label{aff60}
\and
Leiden Observatory, Leiden University, Einsteinweg 55, 2333 CC Leiden, The Netherlands\label{aff61}
\and
Departamento de F{\'\i}sica de la Tierra y Astrof{\'\i}sica, Universidad Complutense de Madrid, Plaza de las Ciencias 2, E-28040 Madrid, Spain\label{aff62}
\and
Kobayashi-Maskawa Institute for the Origin of Particles and the Universe, Nagoya University, Chikusa-ku, Nagoya, 464-8602, Japan\label{aff63}
\and
Institute for Advanced Research, Nagoya University, Chikusa-ku, Nagoya, 464-8601, Japan\label{aff64}
\and
Kavli Institute for the Physics and Mathematics of the Universe (WPI), University of Tokyo, Kashiwa, Chiba 277-8583, Japan\label{aff65}
\and
European Space Agency, 8-10 rue Mario Nikis, 75738 Paris Cedex 15, France\label{aff66}
\and
Niels Bohr Institute, University of Copenhagen, Jagtvej 128, 2200 Copenhagen, Denmark\label{aff67}
\and
Cosmic Dawn Center (DAWN)\label{aff68}
\and
Center for Frontier Science, Chiba University, 1-33 Yayoi-cho, Inage-ku, Chiba 263-8522, Japan\label{aff69}
\and
Department of Physics, Graduate School of Science, Chiba University, 1-33 Yayoi-Cho, Inage-Ku, Chiba 263-8522, Japan\label{aff70}
\and
Space physics and astronomy research unit, University of Oulu, Pentti Kaiteran katu 1, FI-90014 Oulu, Finland\label{aff71}
\and
Centre de Recherche Astrophysique de Lyon, UMR5574, CNRS, Universit\'e Claude Bernard Lyon 1, ENS de Lyon, 69230, Saint-Genis-Laval, France\label{aff72}
\and
European Southern Observatory, Karl-Schwarzschild Str. 2, 85748 Garching, Germany\label{aff73}
\and
Jet Propulsion Laboratory, California Institute of Technology, 4800 Oak Grove Drive, Pasadena, CA, 91109, USA\label{aff74}
\and
School of Physics \& Astronomy, University of Southampton, Highfield Campus, Southampton SO17 1BJ, UK\label{aff75}
\and
School of Physical Sciences, The Open University, Milton Keynes, MK7 6AA, UK\label{aff76}
\and
Univ. Lille, CNRS, Centrale Lille, UMR 9189 CRIStAL, 59000 Lille, France\label{aff77}
\and
Leibniz-Institut f\"{u}r Astrophysik (AIP), An der Sternwarte 16, 14482 Potsdam, Germany\label{aff78}
\and
INAF-Osservatorio Astronomico di Capodimonte, Via Moiariello 16, 80131 Napoli, Italy\label{aff79}
\and
Department of Astronomy, University of Massachusetts, Amherst, MA 01003, USA\label{aff80}
\and
INAF-Osservatorio Astronomico di Brera, Via Brera 28, 20122 Milano, Italy\label{aff81}
\and
Dipartimento di Fisica e Astronomia, Universit\`a di Bologna, Via Gobetti 93/2, 40129 Bologna, Italy\label{aff82}
\and
INAF-Osservatorio Astronomico di Padova, Via dell'Osservatorio 5, 35122 Padova, Italy\label{aff83}
\and
Centre National d'Etudes Spatiales -- Centre spatial de Toulouse, 18 avenue Edouard Belin, 31401 Toulouse Cedex 9, France\label{aff84}
\and
Dipartimento di Fisica, Universit\`a di Genova, Via Dodecaneso 33, 16146, Genova, Italy\label{aff85}
\and
INFN-Sezione di Genova, Via Dodecaneso 33, 16146, Genova, Italy\label{aff86}
\and
Institut de Recherche en Astrophysique et Plan\'etologie (IRAP), Universit\'e de Toulouse, CNRS, UPS, CNES, 14 Av. Edouard Belin, 31400 Toulouse, France\label{aff87}
\and
Department of Physics "E. Pancini", University Federico II, Via Cinthia 6, 80126, Napoli, Italy\label{aff88}
\and
INFN section of Naples, Via Cinthia 6, 80126, Napoli, Italy\label{aff89}
\and
Instituto de Astrof\'isica e Ci\^encias do Espa\c{c}o, Universidade do Porto, CAUP, Rua das Estrelas, PT4150-762 Porto, Portugal\label{aff90}
\and
Faculdade de Ci\^encias da Universidade do Porto, Rua do Campo de Alegre, 4150-007 Porto, Portugal\label{aff91}
\and
Dipartimento di Fisica, Universit\`a degli Studi di Torino, Via P. Giuria 1, 10125 Torino, Italy\label{aff92}
\and
INFN-Sezione di Torino, Via P. Giuria 1, 10125 Torino, Italy\label{aff93}
\and
INAF-Osservatorio Astrofisico di Torino, Via Osservatorio 20, 10025 Pino Torinese (TO), Italy\label{aff94}
\and
INAF-IASF Milano, Via Alfonso Corti 12, 20133 Milano, Italy\label{aff95}
\and
Centro de Investigaciones Energ\'eticas, Medioambientales y Tecnol\'ogicas (CIEMAT), Avenida Complutense 40, 28040 Madrid, Spain\label{aff96}
\and
Port d'Informaci\'{o} Cient\'{i}fica, Campus UAB, C. Albareda s/n, 08193 Bellaterra (Barcelona), Spain\label{aff97}
\and
Institute for Theoretical Particle Physics and Cosmology (TTK), RWTH Aachen University, 52056 Aachen, Germany\label{aff98}
\and
Institute of Space Sciences (ICE, CSIC), Campus UAB, Carrer de Can Magrans, s/n, 08193 Barcelona, Spain\label{aff99}
\and
Institut d'Estudis Espacials de Catalunya (IEEC),  Edifici RDIT, Campus UPC, 08860 Castelldefels, Barcelona, Spain\label{aff100}
\and
Dipartimento di Fisica e Astronomia "Augusto Righi" - Alma Mater Studiorum Universit\`a di Bologna, Viale Berti Pichat 6/2, 40127 Bologna, Italy\label{aff101}
\and
European Space Agency/ESRIN, Largo Galileo Galilei 1, 00044 Frascati, Roma, Italy\label{aff102}
\and
Universit\'e Claude Bernard Lyon 1, CNRS/IN2P3, IP2I Lyon, UMR 5822, Villeurbanne, F-69100, France\label{aff103}
\and
UCB Lyon 1, CNRS/IN2P3, IUF, IP2I Lyon, 4 rue Enrico Fermi, 69622 Villeurbanne, France\label{aff104}
\and
Mullard Space Science Laboratory, University College London, Holmbury St Mary, Dorking, Surrey RH5 6NT, UK\label{aff105}
\and
INAF-Istituto di Astrofisica e Planetologia Spaziali, via del Fosso del Cavaliere, 100, 00100 Roma, Italy\label{aff106}
\and
INFN-Padova, Via Marzolo 8, 35131 Padova, Italy\label{aff107}
\and
School of Physics, HH Wills Physics Laboratory, University of Bristol, Tyndall Avenue, Bristol, BS8 1TL, UK\label{aff108}
\and
Istituto Nazionale di Fisica Nucleare, Sezione di Bologna, Via Irnerio 46, 40126 Bologna, Italy\label{aff109}
\and
FRACTAL S.L.N.E., calle Tulip\'an 2, Portal 13 1A, 28231, Las Rozas de Madrid, Spain\label{aff110}
\and
Dipartimento di Fisica "Aldo Pontremoli", Universit\`a degli Studi di Milano, Via Celoria 16, 20133 Milano, Italy\label{aff111}
\and
Institute of Theoretical Astrophysics, University of Oslo, P.O. Box 1029 Blindern, 0315 Oslo, Norway\label{aff112}
\and
Department of Physics, Lancaster University, Lancaster, LA1 4YB, UK\label{aff113}
\and
Felix Hormuth Engineering, Goethestr. 17, 69181 Leimen, Germany\label{aff114}
\and
Technical University of Denmark, Elektrovej 327, 2800 Kgs. Lyngby, Denmark\label{aff115}
\and
Cosmic Dawn Center (DAWN), Denmark\label{aff116}
\and
NASA Goddard Space Flight Center, Greenbelt, MD 20771, USA\label{aff117}
\and
Department of Physics and Helsinki Institute of Physics, Gustaf H\"allstr\"omin katu 2, 00014 University of Helsinki, Finland\label{aff118}
\and
Universit\'e de Gen\`eve, D\'epartement de Physique Th\'eorique and Centre for Astroparticle Physics, 24 quai Ernest-Ansermet, CH-1211 Gen\`eve 4, Switzerland\label{aff119}
\and
Department of Physics, P.O. Box 64, 00014 University of Helsinki, Finland\label{aff120}
\and
Helsinki Institute of Physics, Gustaf H{\"a}llstr{\"o}min katu 2, University of Helsinki, Helsinki, Finland\label{aff121}
\and
Department of Physics and Astronomy, University College London, Gower Street, London WC1E 6BT, UK\label{aff122}
\and
NOVA optical infrared instrumentation group at ASTRON, Oude Hoogeveensedijk 4, 7991PD, Dwingeloo, The Netherlands\label{aff123}
\and
INFN-Sezione di Milano, Via Celoria 16, 20133 Milano, Italy\label{aff124}
\and
Universit\"at Bonn, Argelander-Institut f\"ur Astronomie, Auf dem H\"ugel 71, 53121 Bonn, Germany\label{aff125}
\and
Dipartimento di Fisica e Astronomia "Augusto Righi" - Alma Mater Studiorum Universit\`a di Bologna, via Piero Gobetti 93/2, 40129 Bologna, Italy\label{aff126}
\and
Institut d'Astrophysique de Paris, 98bis Boulevard Arago, 75014, Paris, France\label{aff127}
\and
IFPU, Institute for Fundamental Physics of the Universe, via Beirut 2, 34151 Trieste, Italy\label{aff128}
\and
Institut de F\'{i}sica d'Altes Energies (IFAE), The Barcelona Institute of Science and Technology, Campus UAB, 08193 Bellaterra (Barcelona), Spain\label{aff129}
\and
Department of Physics and Astronomy, University of Aarhus, Ny Munkegade 120, DK-8000 Aarhus C, Denmark\label{aff130}
\and
Waterloo Centre for Astrophysics, University of Waterloo, Waterloo, Ontario N2L 3G1, Canada\label{aff131}
\and
Department of Physics and Astronomy, University of Waterloo, Waterloo, Ontario N2L 3G1, Canada\label{aff132}
\and
Space Science Data Center, Italian Space Agency, via del Politecnico snc, 00133 Roma, Italy\label{aff133}
\and
Institute of Space Science, Str. Atomistilor, nr. 409 M\u{a}gurele, Ilfov, 077125, Romania\label{aff134}
\and
Institute for Particle Physics and Astrophysics, Dept. of Physics, ETH Zurich, Wolfgang-Pauli-Strasse 27, 8093 Zurich, Switzerland\label{aff135}
\and
Dipartimento di Fisica e Astronomia "G. Galilei", Universit\`a di Padova, Via Marzolo 8, 35131 Padova, Italy\label{aff136}
\and
Departamento de F\'isica, FCFM, Universidad de Chile, Blanco Encalada 2008, Santiago, Chile\label{aff137}
\and
Satlantis, University Science Park, Sede Bld 48940, Leioa-Bilbao, Spain\label{aff138}
\and
Centre for Electronic Imaging, Open University, Walton Hall, Milton Keynes, MK7~6AA, UK\label{aff139}
\and
Infrared Processing and Analysis Center, California Institute of Technology, Pasadena, CA 91125, USA\label{aff140}
\and
Universidad Polit\'ecnica de Cartagena, Departamento de Electr\'onica y Tecnolog\'ia de Computadoras,  Plaza del Hospital 1, 30202 Cartagena, Spain\label{aff141}
\and
Centre for Information Technology, University of Groningen, P.O. Box 11044, 9700 CA Groningen, The Netherlands\label{aff142}
\and
INFN-Bologna, Via Irnerio 46, 40126 Bologna, Italy\label{aff143}
\and
INFN, Sezione di Trieste, Via Valerio 2, 34127 Trieste TS, Italy\label{aff144}
\and
SISSA, International School for Advanced Studies, Via Bonomea 265, 34136 Trieste TS, Italy\label{aff145}
\and
INAF, Istituto di Radioastronomia, Via Piero Gobetti 101, 40129 Bologna, Italy\label{aff146}
\and
Junia, EPA department, 41 Bd Vauban, 59800 Lille, France\label{aff147}
\and
Department of Physics and Astronomy, University of British Columbia, Vancouver, BC V6T 1Z1, Canada\label{aff148}}  

%
%
\abstract{ 
The \Euclid \ac{ERO} showcase \Euclid's capabilities in advance of its main mission, targeting 17 astronomical objects, from galaxy clusters, nearby galaxies, globular clusters, to star-forming regions.
A total of 24 hours observing time was allocated in the early months of operation, engaging the scientific community through an early public data release.
We describe the development of the \ac{ERO} pipeline to create visually compelling images while simultaneously meeting the scientific demands within months of launch, leveraging a pragmatic, data-driven development strategy. 
The pipeline's key requirements are to preserve the image quality and to provide flux calibration and photometry for compact and extended sources.
The pipeline's five pillars are: removal of instrumental signatures; astrometric calibration; photometric calibration; image stacking; and the production of science-ready catalogues for both the VIS and NISP instruments.
We report a \ac{PSF} with a full width at half maximum of \ang{;;0.16} in the optical \IE-band, and \ang{;;0.49} in the \ac{NIR} bands \YE, \JE, and \HE. 
Our VIS mean absolute flux calibration is accurate to about 1\%, and 10\% for NISP due to a limited calibration set; both instruments have considerable colour terms for individual sources.
The median depth is 25.3 and 23.2\,AB\,mag with a \ac{SNR} of 10 for galaxies, and 27.1 and 24.5\,AB\,mag at an \ac{SNR} of 5 for point sources for VIS and NISP, respectively. 
\Euclid's ability to observe diffuse emission is exceptional due to its extended \ac{PSF} nearly matching a pure diffraction halo, the best ever achieved by a wide-field, high-resolution imaging telescope.
\Euclid offers unparalleled capabilities for exploring the \ac{LSB} Universe across all scales, providing high precision within a wide \ac{FOV}, and opening a new observational window in the \ac{NIR}.
Median surface-brightness levels of 29.9 and 28.3, \,AB\,mag\,arcsec$^{-2}$ are achieved for VIS and NISP, respectively, 
for detecting a $\ang{;;10}\times\ang{;;10}$ extended feature at the 1\,$\sigma$ level.
}
 
%
%
%
%
   \titlerunning{\Euclid: ERO --  Programme overview, pipeline, and performance}
   \authorrunning{Cuillandre, J.-C., et al.}
   
   \maketitle
%
%
%
%

\section{\label{sec:intro}Introduction}

\Euclid is an on-going space mission, part of the European Space Agency (ESA) Cosmic Vision programme, originating from a 2007 call for medium-sized mission. \Euclid spawned from proposals focused on dark energy and is now conducting an extragalactic survey using optical imaging, and \ac{NIR} imaging and spectroscopy. The six-year survey is designed to study galaxy clustering and weak gravitational lensing, essential probes of the Universe's large-scale structure and the processes that govern its expansion. The mission's primary scientific objectives are outlined in a key publication that led to its official selection and adoption by ESA in 2011 and 2012 \citep{Laureijs11}, respectively. \Euclid was successfully launched on 1 July 2023. \cite{EuclidSkyOverview} describes the spacecraft, discusses the mission's early phase in orbit, its survey strategy, the data it collects, and the scientific research it enables. The two scientific instruments, VIS and the Near-Infrared Spectrometer and Photometer (NISP), are described in great depth in \cite{EuclidSkyVIS} and \cite{EuclidSkyNISP}, respectively. These three defining articles from the Euclid Consortium act as a cornucopia of \Euclid knowledge.

The Early Release Observations (EROs) programme is a special project by the ESA \Euclid science team, aimed at gathering and sharing scientific observations for public engagement and communication purposes before the main mission activities start \citep{EROcite}. The goal was to highlight \Euclid's capabilities through visually engaging astronomical objects that are not central to the mission's core cosmological goals. This entailed observations of extended objects that fill most of \Euclid's large \ac{FOV}, and naturally led to the selection of proposals from the \Euclid scientific community that showcased nearby objects. The \ac{ERO} programme considered inclusion of objects at increasing distances to cover the rich variety of science topics that can be addressed. The sequence starts with Galactic nebulae in the Orion star-forming region at a distance of 500\,pc \citep{EROOrion}, followed by globular clusters in the Milky Way \citep{EROGalGCs}, nearby galaxies \citep{ERONearbyGals}, and more distant galaxy clusters with the nearest Dorado and Fornax clusters \citep{EROFornaxGCs} at 15-20\,Mpc, subsequently the Perseus cluster at 72 Mpc \citep{EROPerseusOverview, EROPerseusICL, EROPerseusDGs}, and finally two Abell clusters with the most distant one at $z=0.228$ \citep{EROLensData}. Since these observations are not part of the main mission, there was a push to publicly release the collected data to the scientific community as quickly as possible. Due to the focus on unique sky regions, containing extended emission and not covered by the main survey, together with the quick turnaround needed for public communication, the \ac{ERO} data set was processed differently than the nominal \Euclid survey data.

In this paper, we describe in \cref{sec:ero} the \ac{ERO}'s objectives and methods. In \cref{sec:pipeline} we delve into the \ac{ERO} pipeline strategy and implementation, focusing on the pipeline's origins and requirements, followed by our implementation strategy. In \cref{sec:detrending} we focus on image detrending, an essential step for preparing the images for scientific analysis. This part is comprehensive, starting with the ingestion and initial evaluation of the \ac{ERO} images. For the detrending of the optical VIS data we explain procedures for correcting bad pixels, overscan, bias structure, dark current, and stray light, as well as applying flat-field corrections, detector-to-detector image scaling, and identifying and removing cosmic rays. Similarly, for detrending \ac{NIR} data from the NISP instrument, we cover charge-persistence correction, bad-pixel masks, electronic pedestal correction, dark current correction, flat-field correction, row correlated-noise correction, and cosmic rays. The astrometric calibration follows in \cref{sec:astrometry}, where we outline the process to accurately anchor the data to the \textit{Gaia} data release 3 (DR3) astrometric reference \citep{Prusti2016,GaiaDR3}. We describe in \cref{sec:stacking} our resampling and stacking methods, and cover the photometric calibration for both instruments in \cref{sec:photometry}. In  \cref{sec:cats} on compact-source catalogues and general performance of the \ac{ERO} data, we examine \ac{PSF} modelling, the production of the science-validation catalogues, and provide a performance summary of the \ac{ERO} data to assess their scientific utility. In \cref{sec:lsb} we assess the performance for \ac{LSB} science in support of the early science conducted with the \ac{ERO} images. We derive the extended \ac{PSF} in all four \Euclid bands. This involves studying a simple model of the optical design, modelling the encompassed energy of the  \ac{PSF}, and evaluating the consequences for the \ac{ERO} \ac{LSB} science cases.

The paper concludes in \cref{sec:summary} with an executive summary that encapsulates the main points and findings showcased. We first present in \cref{sec:obssummary} a concise summary of the \ac{ERO} data. \Cref{sec:datasummary} then presents two complete tables summarising details on the \ac{ERO} data set (depth, etc.). In \cref{apdx:DoradoFluxcal} we discuss the selection of relevant stars along a given line of sight for identifying and cataloguing stars for follow-up studies. Lastly, in \cref{apdx:OpticalModel}, we describe the optical model of the telescope used to compare with results from \cref{sec:lsb}.

\section{\label{sec:ero}\Euclid \texorpdfstring{\ac{ERO}}{ERO} overview}

\subsection{Programme description}

The \ac{ERO} programme was an initiative by the \Euclid science team. At inception, we aimed to acquire scientific observations for communication and early scientific results purposes before the nominal mission begins. Through the \ac{ERO} programme, we aimed to showcase the unique instrumental capabilities of \Euclid by selecting large and nearby astronomical targets that are completely separate from the cosmological objectives of \Euclid. We seized the opportunity to schedule specific fields in the sky during the early operations phase, ensuring our activities did not interfere with the planning process of the nominal survey and allowing us a greater degree of operational flexibility. Since this programme fell outside the scope of the nominal mission, we committed to making the resulting scientific data products publicly available as promptly as possible.

In March 2023, we issued a call for proposals to the Euclid Collaboration. The total time allocated was limited to 24 accumulated hours. After evaluating the visibility of the fields during the \ac{PV} phase and ensuring the absence of nearby overly-bright stars, we ranked the proposals based on their societal impact merit, scientific merit, and uniqueness. The selected proposals and their approved targets are listed in \cref{table:obsspecs}.

Due to the focused attention on non-standard \Euclid fields outside the main mission survey area (\cref{fig:AllSky}) and the relatively short timescales for preparing communication products, we handled the processing and release of the data products separately from the development of the \Euclid science ground segment. The \ac{ERO} image processing drew on common knowledge and extensive experience in astronomical imaging with charge-coupled devices (CCDs) and \ac{H2RG} HgCdTe sensors, akin to those used by VIS and NISP, respectively. The \ac{ERO} programme required the use of the \ac{EWS} \ac{ROS} \citep[see][]{Scaramella-EP1,EuclidSkyOverview} that also collects slitless spectra. Given the tight schedule and reliance on common knowledge, we could not process the spectroscopic data in time for inclusion in the first public data release.

\subsection{Observing strategy}

\begin{figure*}
	\includegraphics[angle=0,width=1.0\hsize]{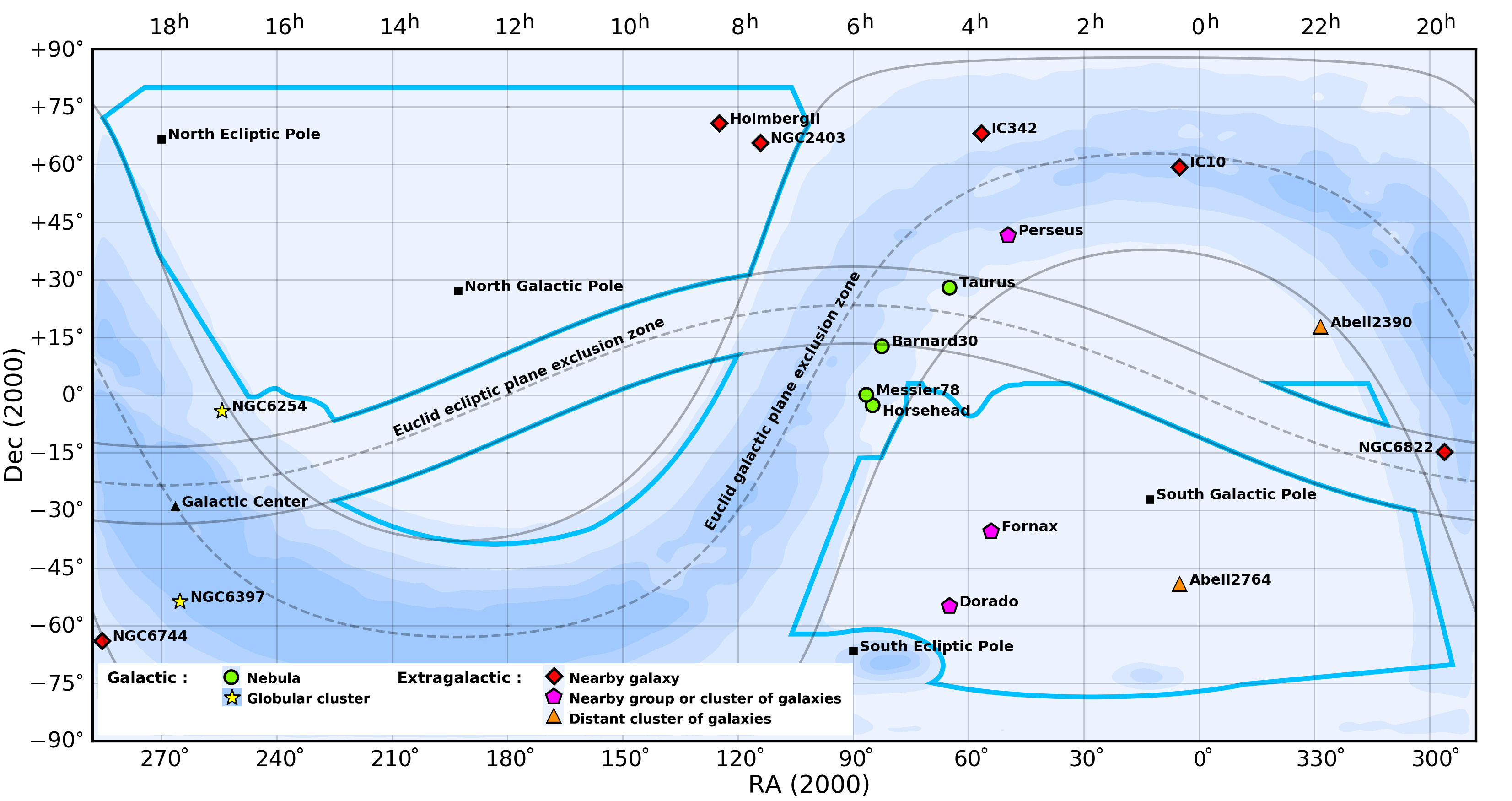}
	\caption{
 Location, name, and nature of the 17 \ac{ERO} fields on an all-sky map, with the general \ac{ROI} of the \ac{EWS} highlighted by the four blue contours. These outlines signify that some of the \ac{ERO} targets will be revisited in the coming years. The distinctive nature of the \ac{ERO} programme facilitated explorations spanning from the Galactic plane to the southern Galactic cap, areas that were accessible during the observation period. This range of coverage showcases the \ac{ERO} programme's goal to venture across a wide variety of astronomical phenomena and regions. The blue background depicts the stellar density across the sky.
\label{fig:AllSky}}
\end{figure*}

The \ac{ROS} provides a single standard \ac{EWS} field together with a range of inline calibration data. It has been highly optimised to provide a maximum amount of scientific data in a minimum amount of time, in a consistent way during the entire six-year survey; it guarantees sufficient a \ac{SNR} and depth for \Euclid's core science. The \ac{ERO} programme permitted multiple \ac{ROS} observations of certain fields to enhance the depth. Additionally, the programme allowed for observations outside the \ac{ROI}, which delineates the useful extragalactic sky area for the wide and deep surveys (\cref{fig:AllSky}).

An \ac{ROS} field is composed of four dither pointings designed to fill the detector gaps. For each dither, the same measurement sequence is executed. First comes a VIS $\IE$-band nominal-science exposure of 566\,s, with a concurrent NISP spectroscopic exposure of 574\,s in one of the four red-grism orientations. These are followed by a sequence of three NISP images in the \JE, \HE, and \YE bands, each lasting 112\,s. For the \ac{ERO} programme, each dither also included an $\IE$ short-science exposure of 95\,s simultaneously to the $\YE$ exposure, yielding four such images per \ac{ROS}; for the EWS, two of the VIS short-science exposures are replaced by VIS calibration images \citep[see][]{EuclidSkyOverview}.

The \ac{ERO} fields were scheduled during the available time slots in the \ac{PV} phase dedicated to calibration observations. The \ac{PV} phase commenced on 6 August 2023 and concluded on 3 December 2023. The pre-launch allocation for the \ac{PV} phase was two months, however, it was soon interrupted due to failures of the spacecraft's \ac{FGS} in fields with a low density of suitable guide stars. \ac{PV} observations resumed on 28 September 2023 after concluding the development and validation of improved \ac{FGS} software. Prior to this date, we proceeded with observing \ac{ERO} fields that were about to lose their visibility, accepting the risk of poor guiding for some or all exposures in these fields. The programme allowed observation of suitable backup sources in case fields were lost due to closure of their visibility window or an operational anomaly.

\subsection{Summary of the observations}

A summary of the observed \ac{ERO} fields is provided in \cref{table:obsspecs}. Observations conducted before 28 September 2023 -- which covered high stellar density fields such as the globular cluster NGC\,6254, the irregular galaxy IC\,10, and the Perseus galaxy cluster -- were unaffected by the \ac{FGS} anomaly. However, for the reflection nebula NGC\,1333, the Fornax galaxy cluster, and one \ac{ROS} on the Taurus molecular cloud, only a limited number of exposures per \ac{ROS} achieved the required guiding performance. Despite this, the Fornax galaxy cluster, observed during three epochs, yielded a sufficient number of good exposures in all four bands to enable scientific analysis.

All observations conducted before 16 September 2023 were executed with a dither pattern anomalously rotated by \ang{90;;} relative to the nominal direction defined in the \ac{ROS}. This misalignment resulted in zero-coverage gaps in the VIS and NISP stacked images, attributable to the lack of sensor coverage from either VIS or NISP. These gaps account for a few percent of the total field area, yet the exposures remain viable for scientific investigation.

\section{\label{sec:pipeline}\texorpdfstring{\ac{ERO}}{ERO} pipeline strategy and implementation}

\subsection{Origins and requirements}

The \ac{ERO} pipeline was initially developed to create aesthetically striking images of astronomical sources within three months of the telescope's launch, celebrating the advent of a new telescope. 
The objective was to occupy a significant portion of the \Euclid \ac{FOV} with large, colourful objects. Such objects are categorised as extended emission, whether due to their combined stellar density (as in a globular cluster), their nebulous nature (star-forming regions), or their diffuse aspect (unresolved stars), encompassing both high and lowsurface-brightness sources. Throughout this reduction process, it was imperative that the images highlight \Euclid's unparalleled sharpness from the optical to the \ac{NIR} across such a large \ac{FOV}.

Prior to launch, the project rapidly evolved to address the need for the early public release of associated scientific data and related science results. Given that many of the proposed science projects depended on the precise measurement of physical properties derived from extended emission, an alternate approach to the official scientific processing of \Euclid data was required. Consequently, the \ac{ERO} pipeline was tasked with delivering both outreach images and scientific products for all six \ac{ERO} science teams (see \cref{table:obsspecs}). The minimum requirements for the quality of processing and calibration at the onset of the effort have been met and some exceeded as detailed in this paper. This achievement has facilitated a rich early showcase of \Euclid science, highlighting its unique observing capabilities.

Creating images for early release to the world necessitated starting from raw data to produce contiguous images of the sky that are free of visual flaws, such as detector mosaic gaps, cosmic rays, detector persistence, and variations in detector sensitivity, among others. These requirements also enhanced the production of science-ready products. The \ac{ERO} pipeline adeptly managed both domains, with the outreach effort diverging partway through the process. This divergence occurs post-detrending, a step that is detailed below. The subsequent steps involved in producing visually engaging images are not the focus of this paper, which is dedicated to the production of science products.

The selected science projects drove the ultimate requirements for the development of the \ac{ERO} pipeline:
{\begin{itemize}
    \item preservation of the intrinsic delivered image quality;
    \item correction of optical distortions to anchor on the \ac{WCS};
    \item uniformity of the flux calibration across the \ac{FOV};
    \item photometry of compact sources and extended emission;
    \item matched processing for the two instruments (VIS and NISP).
\end{itemize}}

These fundamental requirements translated into the five main pillars of the \ac{ERO} pipeline, based on the adoption of a uniform set of image processing tools:

{\begin{itemize}
    \item optimal removal of instrumental signatures (detrending);
    \item astrometric calibration (internal and absolute);
    \item photometric calibration (internal and external);
    \item stacking of images into a contiguous region of the sky;
    \item production of science-ready catalogues based on the stacks.
\end{itemize}}

\subsection{Implementation strategy}

Due to the tight schedule (3 months to deliver images and the first science data release, 6 months for final products for the first science publications), we adopted for the detrending part the existing C code pipeline developed for similar optical and \ac{NIR} wide-field imaging instruments operated over the past couple of decades at the \ac{CFHT}: MegaCam \citep{Magnier2004} and WIRCam \citep{Pipien2018}. For astrometric calibration, stacking, \ac{PSF} modelling, and source extraction, we chose the {\tt AstrOmatic} suite \citep{Bertin1996,Bertin2002,Bertin2006}, widely adopted across the scientific community. Many of its developments have been driven by these two \ac{CFHT} instruments as well, making them particularly well-suited for \Euclid's wide-field imaging data. Additional key community-based resources adopted in the \ac{ERO} pipeline include {\tt Astrometry.net} \citep{Lang2010}, and various Python packages such as {\tt deepCR} \citep{Zhang2020}.

The calendar necessitated a pragmatic approach to the development of the \ac{ERO} pipeline, including enhancements of some {\tt AstrOmatic} tools to fully leverage the data set's quality. Development of the \ac{ERO} pipeline commenced within weeks of the first space data availability post-launch. The tight schedule mandated the formulation and optimisation of processing recipes based on an empirical, data-driven approach, without prior knowledge of the specifics of the \Euclid instruments and detectors. Consequently, the resulting \ac{ERO} pipeline, with relaxed requirements for photometry of compact sources and more stringent ones for extended sources, was bound to be inherently distinct and entirely separate from the main mission pipeline.

The following sections address all aspects that led to the \ac{ERO} science products for both VIS and NISP: data detrending, astrometric calibration, photometric calibration, stacking, \ac{PSF} extraction, and catalogue production.

\section{\label{sec:detrending}Image detrending}

\subsection{Ingestion and initial evaluation of the ERO images}

The Flexible Image Transport System (FITS) is a universally adopted file format designed for the efficient transfer of both metadata (contained within a FITS header) and pixel-based data. In the context of the \Euclid mission, our adopted software has been enhanced to include a comprehensive set of keywords, a development that became common in the astronomical community following the widespread adoption of detector mosaics in the late 1990s. This evolution was spurred by innovations related to the Image Reduction and Analysis Facility \citep[{\tt IRAF};][]{IRAF1986}, leading to the integration of these standards across various image processing applications. These specific keywords help delineate the physical coordinates of particular pixel regions, such as prescan and overscan areas, and the active imaging pixels within each detector (e.g. {\tt PRESCAN}, {\tt OVERSCAN}, {\tt BIASSEC}, {\tt DATASEC}), and provide details about their physical layout within the detector mosaic (e.g. {\tt DETSIZE}, {\tt DETSEC}).

The initial stage in the \ac{ERO} pipeline involved enhancing Level 1 \Euclid (LE1) Multi-Extension FITS (MEF) images for both VIS and NISP by incorporating the aforementioned FITS keywords. Utilising the {\tt libfh} library\footnote{\url{https://software.cfht.hawaii.edu/fits_guide.html}} from the \ac{CFHT}, this step generates a new MEF file that preserves the original structure of 144 extensions for VIS and 16 for NISP. Simultaneously, it compiles detailed statistics about the images (such as bias levels and overall image quality), creates JPEG previews, and fills a text-based database with a comprehensive summary of key attributes for each \Euclid image. This database created within days of the observations became a crucial resource for all later stages of the pipeline.

Upon the availability of various previews, the pipeline began a preliminary visual validation process aimed at identifying and excluding images affected by sub-optimal guiding. This step involved enhancing the database with validation flags. Notably, to support the development of the \ac{ERO} pipeline, data from both the commissioning phase and \ac{PV} activities conducted alongside the \ac{ERO} observation period were integrated into the \ac{ERO} database.

\subsection{Detrending optical data from the VIS instrument}

\subsubsection{Bad pixel masks}

The cosmetic quality of the Teledyne e2v CCD273-84 CCDs in the 6$\times$6 mosaic is excellent (each CCD measures $4096\times4132$ pixels), necessitating minimal masking for single bad pixels, clusters of bad pixels, and blocked columns. Such pixels that do need to be masked were identified through their nonlinear response by comparing an internal calibration \ac{LED} illumination image of \num{30000} analog-to-digital units (ADUs) with one at 1/10th that intensity; deviations above or below a 1.2\% threshold led to masking. Furthermore, as a conservative measure for photometry accuracy, we masked five lines at the bottom and four lines at the top of each of the four imaging quadrants per CCD (with four parallel readouts, each quadrant measures $2048\times2066$ pixels). This action was necessary due to slight nonlinearity issues related to the geometry of the pixels at the top of the quadrant, influenced by the presence of an injection charge channel in the middle of the CCD, and the slight instability in the electronic chains at the start of readout for the bottom lines. Overall, the \ac{ERO} pipeline mask (0/1) affects merely 0.53\% of all imaging pixels, with a significant portion (0.44\%) originating from the nine lines masked per quadrant.

\subsubsection{Overscan correction}

\begin{figure}
	\includegraphics[width=\linewidth]{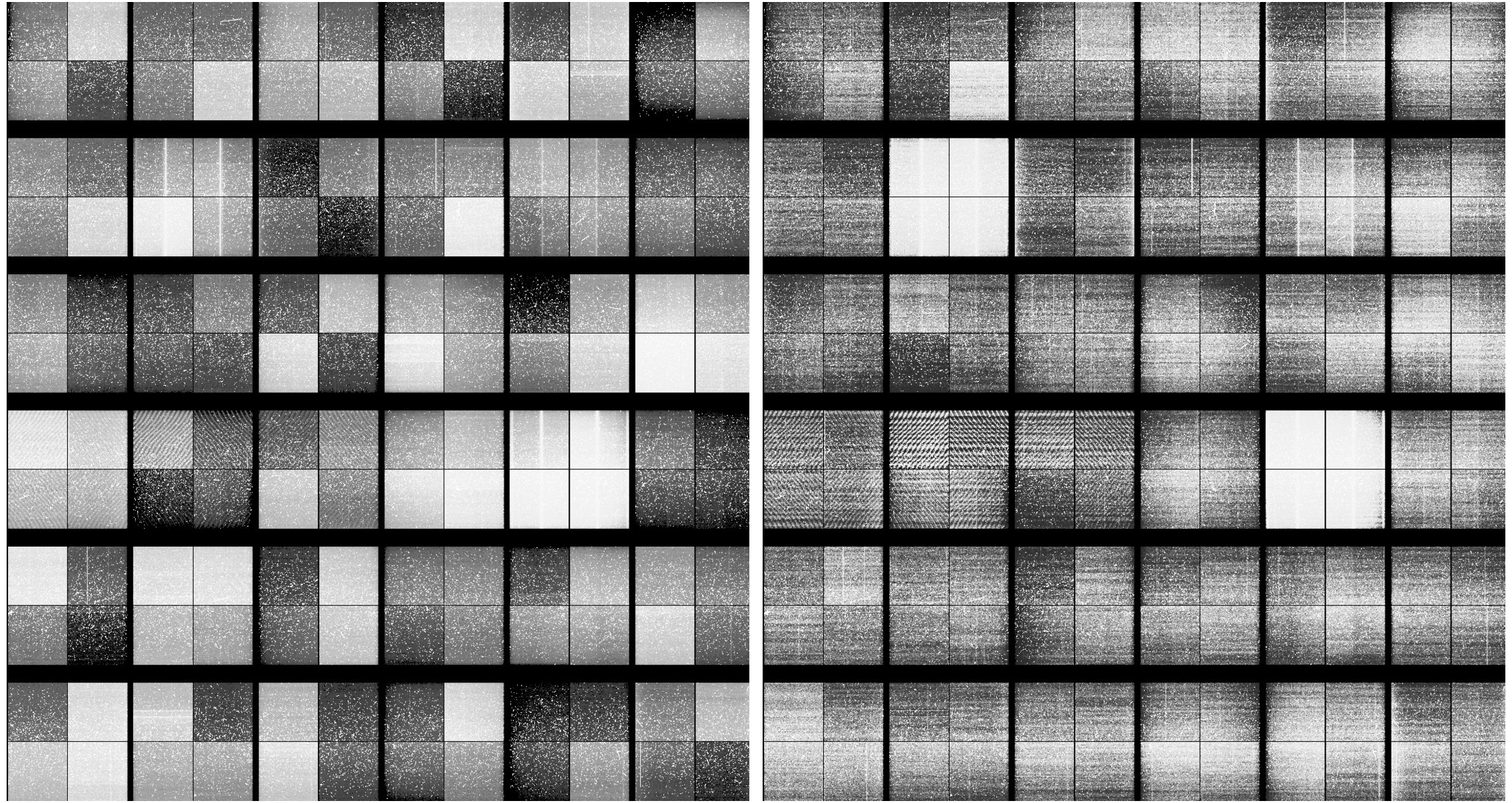}
	\caption{\textit{Left}: single raw VIS bias frame captured at L2, showing numerous cosmic rays, corrected with a fixed pedestal level per quadrant based on the median value in the overscan area.  \textit{Right}: correction using a smoothed vertical vector of the overscan area. The contrast here is maximised to highlight effects at the sub-ADU level. The solution adopted on the right still exhibits an instrumental signature (some quadrants, or entire detectors, have a non-zero signal), necessitating an additional two-dimensional bias correction. 
\label{fig:Overscan}}
\end{figure}

Close examination of the overscan region per quadrant (28 columns wide) revealed a subtle modulation at the roughly 1\,ADU level across scales of 50 pixels along the vertical axis. The median electronic gain of VIS is 3.5 electrons per ADU, indicating slow fluctuations at the 3--4 electrons level of the readout pedestal drift. This phenomenon is linked to a temporal instability in the readout electronic chains on a timescale of seconds throughout the 72-s-long readout. This random effect is believed to be caused by the power supply, which generates faint ripples during the readout. These ripples occur at a wide range of frequencies, affecting anything from individual lines to several hundred lines, and can have an amplitude of up to 1\,ADU in that second regime. Sudden transitions to high signals -- for example saturated stars -- can also cause jumps on the order of 1\,ADU during the readout. A typical 566-s VIS integration is dominated by the background from zodiacal light, at a median level of 40\,ADUs in the \ac{ERO} raw data (this translates to 22.2\,mag\,arcsec$^{-2}$). This modulation of the readout pedestal, constituting a third of the photon-noise level, was evident in all raw VIS images (\cref{fig:Bias}, left) and required correction. It is effectively mitigated in the \ac{ERO} pipeline by subtracting a vector from each column of the imaging area. This median vector was constructed for each quadrant per image across the overscan and smoothed by a 50-pixel tall median filter, matching the typical scale of modulation. Subtracting this vector from each column in the imaging area corrected the intrinsic additive pedestal introduced by the electronic chain in a single step. Since this slow modulation varies from exposure to exposure, it cannot be accounted for in a median bias, necessitating per-exposure execution. This correction of the electronic pedestal does not impact the noise properties of the images on small (pixel) scales, while significantly enhancing the overall background flatness of the VIS images (see \cref{fig:Overscan}).

\subsubsection{Bias structure correction}

\begin{figure}
	\includegraphics[width=\linewidth]{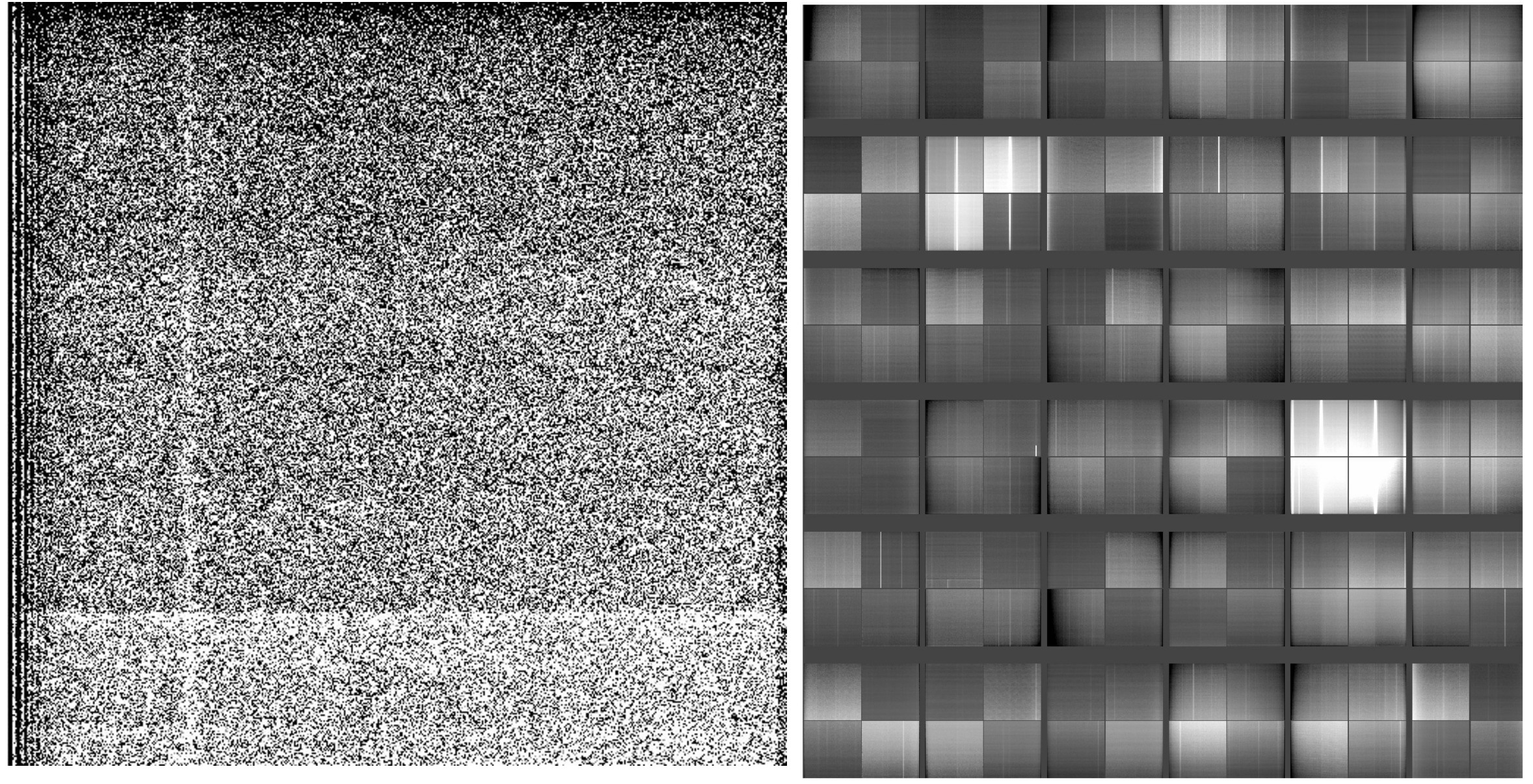}
	\caption{\textit{Left}: single quadrant from a raw bias image showing: (i) a noticeable jump in the readout pedestal during readout (with more subtle variations observable throughout the readout); (ii) a top-down intensity gradient indicative of light injection during readout; and (iii) an electronic ringing effect on the left at the start of each line readout. \textit{Right}: high-\ac{SNR} master bias frame (median of tens of raw images) used to process all \ac{ERO} data. The brightness variation here does not exceed 0.6\,ADU, while the typical raw \ac{ERO} signal -- zodiacal light background -- is around 40\,ADUs, necessitating this correction.
\label{fig:Bias}}
\end{figure}

The overscan correction eliminates the varying electronic pedestal, but structures can still be observed on an overscan-corrected bias frame (right panel of \cref{fig:Overscan}), indicating the need to remove two-dimensional structures by subtracting a master bias image. Two CCDs in the mosaic exhibit a particularly high pedestal level due to a glow from the serial register protection circuity when clocking the serial register, adding a signal of up to 0.6\,ADU to the pixels of those detectors. A similar effect is seen at a lower level (0.1 to 0.3\,ADU maximum) across the other 34 CCDs. This systematic additive effect is effectively mitigated by subtracting a full bias (2-dimensional array) constructed from a median of over 100 bias frames, initially corrected for their overscan level (see \cref{fig:Bias}). Oscillations in amplitude over the first 100 pixels at the start of each line readout (ringing) are attributed to the electronic chains stabilising after each end-of-line reset operation. This purely additive effect is corrected through this process. Following overscan and bias correction of science images, the noise properties remain unaffected, with the contribution of the native readout noise of 3.2 electrons (0.93\,ADU, with a dispersion of 0.06 across all 144 outputs) unchanged.

\subsubsection{Dark current and stray light}

At the operational temperature in space, approximately 150\,K, dark-current generation is negligible compared to all other sources, especially the dominant zodiacal light. The signature of dark current is undetectable in the space data at the sub-ADU level, and consequently, no correction is applied.

Stray light, an additive contaminant that occurs even with the VIS shutter closed, severely affected early commissioning data \citep{EuclidSkyOverview}. Positioning the spacecraft safely with respect to the Sun during the \ac{ERO} programme reduced this contamination to a negligible level, except for the very first and two very last fields, Fornax, Dorado and Holmberg\,II, which were captured under borderline conditions. The current version of the \ac{ERO} pipeline, used for our first scientific publications, does not yet incorporate this correction, and those two fields should be treated cautiously when exploring their \ac{LSB} features at the 28--30\,mag\,arcsec$^{-2}$ level; we note that all magnitudes in this paper are in the AB system \citep{oke1983}. The stray-light correction for these three fields will be implemented in the next \ac{ERO} data release.

\subsubsection{Flat-field correction for large and small scales}

\begin{figure*}
	\includegraphics[angle=0,width=1.0\hsize]{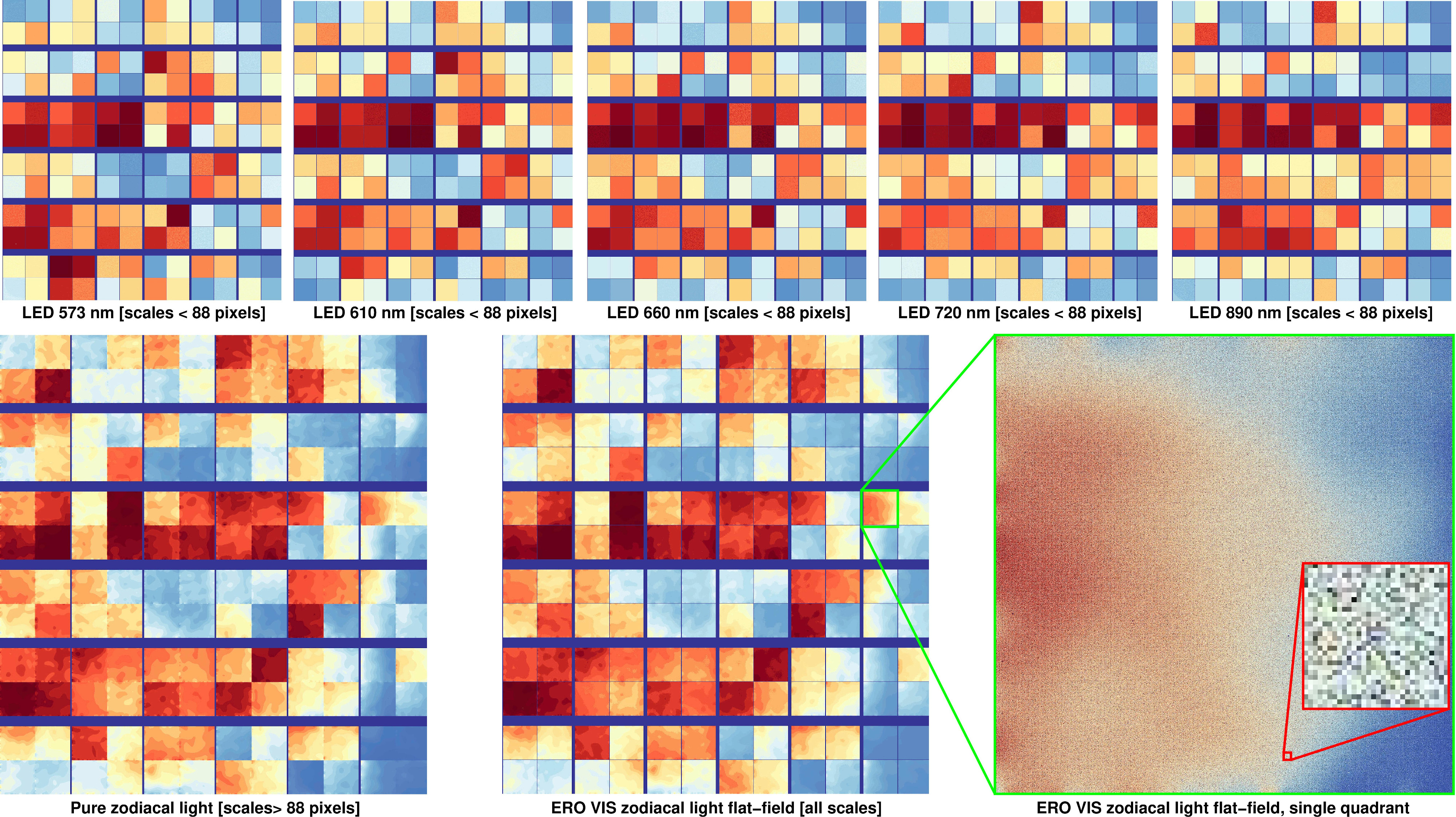}
	\caption{\ac{ERO} zodiacal light flat-field.  This is a combination of large scales ($>\ang{;;10}$) derived solely from the zodiacal light observed in any long VIS exposures, and small scales ($<\ang{;;10}$) coming from a weighted average of five (LED 573\,nm,  610\,nm, 660\,nm, 720\,nm, 890\,nm,) high-\ac{SNR} \ac{LED} internal calibration images, each adjusted according to the VIS throughput. The colour scale used here is arbitrary, set to explore the full range of intensity within each image. The top five images display the relative evolution of \{gain $\times$ quantum efficiency\} with wavelength across all 36 VIS detectors, a large-scale signature that is, however, removed in the flat-field in favour of the one emanating from the pure zodiacal light image (bottom left, the general top-left to bottom-right gradient being related to an internal illumination gradient of VIS). The final \ac{ERO} VIS zodiacal flat-field (bottom centre) incorporates all scales present in the \Euclid signal, from pixel-to-pixel sensitivity variations (bottom right, highlighting a quadrant covering $\ang{;;205}\times\ang{;;206}$, with a zoom on a 30 by 30 pixels area showing the pixel-to-pixel sensitivity differences) to the scale of the \ac{FOV}.
\label{fig:ZodiFlat}}
\end{figure*}

A fundamental aspect of the \ac{ERO} pipeline strategy is the zodiacal light flat-field, based on the principle that the zodiacal light at L2 is the flattest light source observable across the entire sky. The dust in the ecliptic plane scatters sunlight, producing a glowing haze with a well-modelled behaviour \citep[see figure 12 in ][]{Scaramella-EP1}. This distribution peaks at the ecliptic plane and diminishes rapidly, averaging 22.1\,mag\,arcsec$^{-2}$ across the 17 \ac{ERO} fields. Beyond an ecliptic latitude of \ang{15;;}, the distribution's steepest part shows a brightness slope of 0.007\,MJy\,sr$^{-1}$ per latitude degree. In a worst-case scenario (15$^\circ$ ecliptic latitude, at an absolute level of 0.31\,MJy\,sr$^{-1}$), this translates to a gradient of $3\times10^{-3}$\,e$^{-}$\,pixel$^{-1}$\,s$^{-1}$ across the \Euclid \ac{FOV}. This results in a variation of 0.8\,ADU -- compared to a total background level of 60\,ADU at that location -- from one corner to the opposite one in a standard 566\,s raw exposure, rendering the gradient virtually undetectable in this worst case scenario and validating this background as our reference for flattening. Consequently, a VIS image that is properly processed to preserve the zodiacal light as an integral part of the \Euclid signal should display a perfectly flat background across the entire \ac{FOV}, in the absence of other faint sources of emission such as Galactic cirrus.

To construct a zodiacal light reference, it is essential to base this on a median of fields characterised by a relatively low density of extended astronomical sources, rendering most \ac{ERO} data inappropriate for this. Consequently, we selected a 6-hour observation window of a calibration field at the south ecliptic pole where the zodiacal light is perfectly uniform (no gradient by nature, hence no risk of contamination of the flat-field), conducted during the commissioning phase under pristine conditions (good guiding and absence of stray light), and observed with extremely large dithers. This approach allowed for the effective exclusion of all sources in the median stack. Despite the limited number of input frames (20), this field, devoid of any extended sources, produced a high-quality image of the zodiacal light once the stack was filtered to retain scales above \ang{;;10}.

\begin{figure*}
	\includegraphics[width=\linewidth]{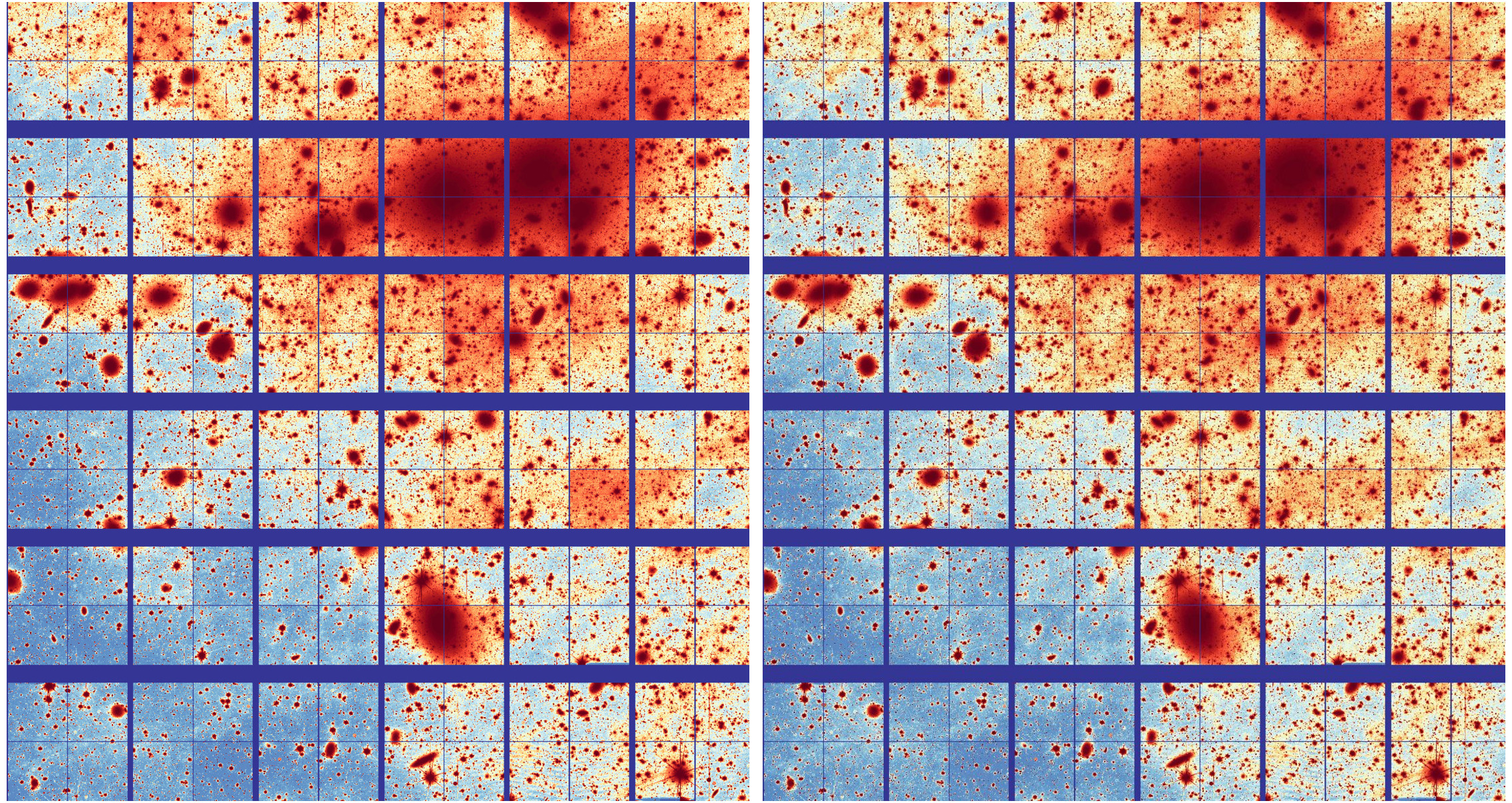}
	\caption{\textit{Left}: VIS image of the Perseus cluster following the \ac{ERO} mask, overscan, bias, flat-field, and {\tt deepCR} corrections reveals a checkerboard pattern in a small number of quadrants. The effect is 2\% of the background for the two most affected quadrants (if left uncorrected this would leave residuals at the 27 to 28th \,mag\,arcsec$^{-2}$ level). \textit{Right}: final image, after applying a low-flux nonlinearity correction to approximately 30 quadrants.  This now displays uniform flatness of the background across all borders---both between detectors and within quadrants---indicating reliable photometry for extended emission such as galaxy stellar halos, intra-cluster light, and Galactic cirrus, as showcased here. The limiting surface brightness of these single frames exceeds 29\,mag\,arcsec$^{-2}$ throughout the entire \ac{FOV} (direct detection of faint contrasts at the \ang{;;10} scale).
\label{fig:NSCALE}}
\end{figure*}

While the zodiacal light served as the optimal light source for correcting medium-to-large scale structures in \Euclid images, the total flux collected per image was relatively low, amounting to just a few tens of ADUs in a 566\,s exposure. This low level did not provide robust statistics for smaller scales, such as pixel-to-pixel variations \citep{Borlaff-EP16}. To address this, we utilised internal light calibration images generated by sequentially activating a series of narrowband \acp{LED} mounted within \Euclid. We employed five of these sources, with central wavelengths of 573\,nm, 610\,nm, 660\,nm, 720\,nm, and 890\,nm, each providing an illumination level of approximately 10\,000\,ADU. Up to 30 images were stacked for each \ac{LED}, then combined into a single VIS broadband \ac{LED} flat-field based on their relative throughput across the \IE-band (530\,nm to 920\,nm). To isolate the small-scale variations from this stack, a model of the medium- to large-scale variations per quadrant was subtracted. This subtraction was done at a scale low enough to also include the vignetted corners of the \ac{LED} exposures, ensuring that science images, which are not vignetted, remained unaffected. The \ac{SNR} on these \ac{LED} exposures is exceptionally high, particularly with the stacking of tens of exposures, thereby not compromising the signal quality in those corners. The top five panels of \cref{fig:ZodiFlat} illustrate the final outcome of this process: each quadrant appears perfectly flat.
 
This flat-field strategy meticulously selected the cut-on/cut-off filtering scales to ensure no physical scales from the \ac{LED} flat-field are present in the pure zodiacal light flat-field, and vice versa. To flatten the \ac{LED} flat-field, we subtracted a map created with a median boxcar filter 16 pixels on the side and an additional $3\times3$ Gaussian convolution kernel. For the pure zodiacal light, the filtering process used a median boxcar filter 88 pixels on the side and an extra $3\times3$ Gaussian convolution kernel. The CCDs in the VIS mosaic are pristine, with few structures at these intermediate scales, further ensuring that no crucial component of the flat-field was suppressed. The final step involved multiplying the pure zodiacal light flat-field by the \ac{LED} flat-field on a per-quadrant basis, after normalising the \ac{LED} flux to match the flux of the pure zodiacal light, respecting the relative scaling factors of those corrections in the final flat-field. \Cref{fig:ZodiFlat} demonstrates the process: the end result (bottom centre) is a normalised flat-field where medium to large scales (beyond \ang{;;10}) originate solely from the VIS response to the zodiacal light, and small scales (below \ang{;;10}) are derived solely from the VIS response to the \acp{LED}.

The final \ac{ERO} VIS flat-field was normalised to the average of the mode of all 144 quadrants. This approach ensured that the detrended images retained properties similar to those of the raw data.

The overarching goal of the \ac{ERO} zodiacal flat-field was to produce images that appear background-flat. However, this approach is likely to introduce a bias in the photometry of astronomical sources other than the zodiacal light, due to their differing \acp{SED} across the very large width of the \IE-band. The zodiacal light \acp{SED} peaks in the optical around 500\,nm, decreasing steadily until 3\,$\mu$m \citep{Leinert1998}. Any astronomical source with a \acp{SED} departing from this simple slope will be inevitably biased since the relative contribution to its total flux across the \IE-band will be normalised to that of the zodiacal light.
For this initial release, there was not enough time to investigate a form of illumination correction \citep{Regnault+2009} that could eventually standardise the photometry for a specific class of stars and provide colour-term corrections across the \ac{FOV}. The photometric accuracy of the \ac{ERO} programme is further discussed below. As for other potential multiplicative corrections, both VIS nonlinearity at high flux levels and the so-called `brighter-fatter' effect \citep{antilogus2014} are considered second-order effects with negligible impact on \ac{ERO} science.

\subsubsection{Detector-to-detector image scaling}

A VIS image, once corrected for additive instrumental components through overscan and bias corrections, and for multiplicative instrumental components with the main flat-field, results in an image that is uniformly flat. We anticipated a precise continuity of the background level across all borders, both detector-to-detector and quadrant-to-quadrant. However, during the processing of the 17 \ac{ERO} fields, we observed that an additional step was necessary to adjust a few quadrants for an unexplained weak effect resembling low-level flux nonlinearity. The left panel of \cref{fig:NSCALE} illustrates that some quadrants across the VIS mosaic appear either brighter or fainter than expected, leading to a residual checkerboard pattern that required correction.

As previously discussed, the zodiacal-light component in the \ac{ERO} flat-field stems from observations of the south Galactic cap, where the zodiacal-light intensity is near its minimum, registering 29.7\,ADU in our standard 566\,s exposures. Conversely, the zodiacal background in most \ac{ERO} fields is around 40\,ADU. An investigation into this effect, utilising images with shorter integration times (and thus a lower total background in ADU), revealed that this discrepancy is purely multiplicative, scaling with the background flux. On average, only 30 to 40 quadrants, out of a total of 144 quadrants, required adjustment on a per-\ac{ERO} project basis. This adjustment is necessary because the absolute zodiacal-light level varies from field to field. However, the correction factor for specific quadrants remained consistent across all fields. The average flux scaling needed was about 1\%, with a maximum of 2\% for the two most affected quadrants (if left uncorrected this would leave residuals at the 27 to 28th \,mag\,arcsec$^{-2}$ level). This correction was implemented through a single multiplicative factor per quadrant derived visually. This is a first order correction at the percent level enabling the first ERO science effort (some slight residuals can still be perceived after correction). The next ERO data release will include an automated recipe based on optimisation of gradients across all quadrant borders and mosaic gaps.

This ultimate multiplicative adjustment applied to the VIS images ensured that the continuity of extended emission was perfectly preserved (\cref{fig:NSCALE}, right).  However, this highlights the limitation of the \ac{ERO} photometry to an accuracy within a few percent.

\subsubsection{Quantisation noise}

\begin{figure}
	\includegraphics[width=\linewidth]{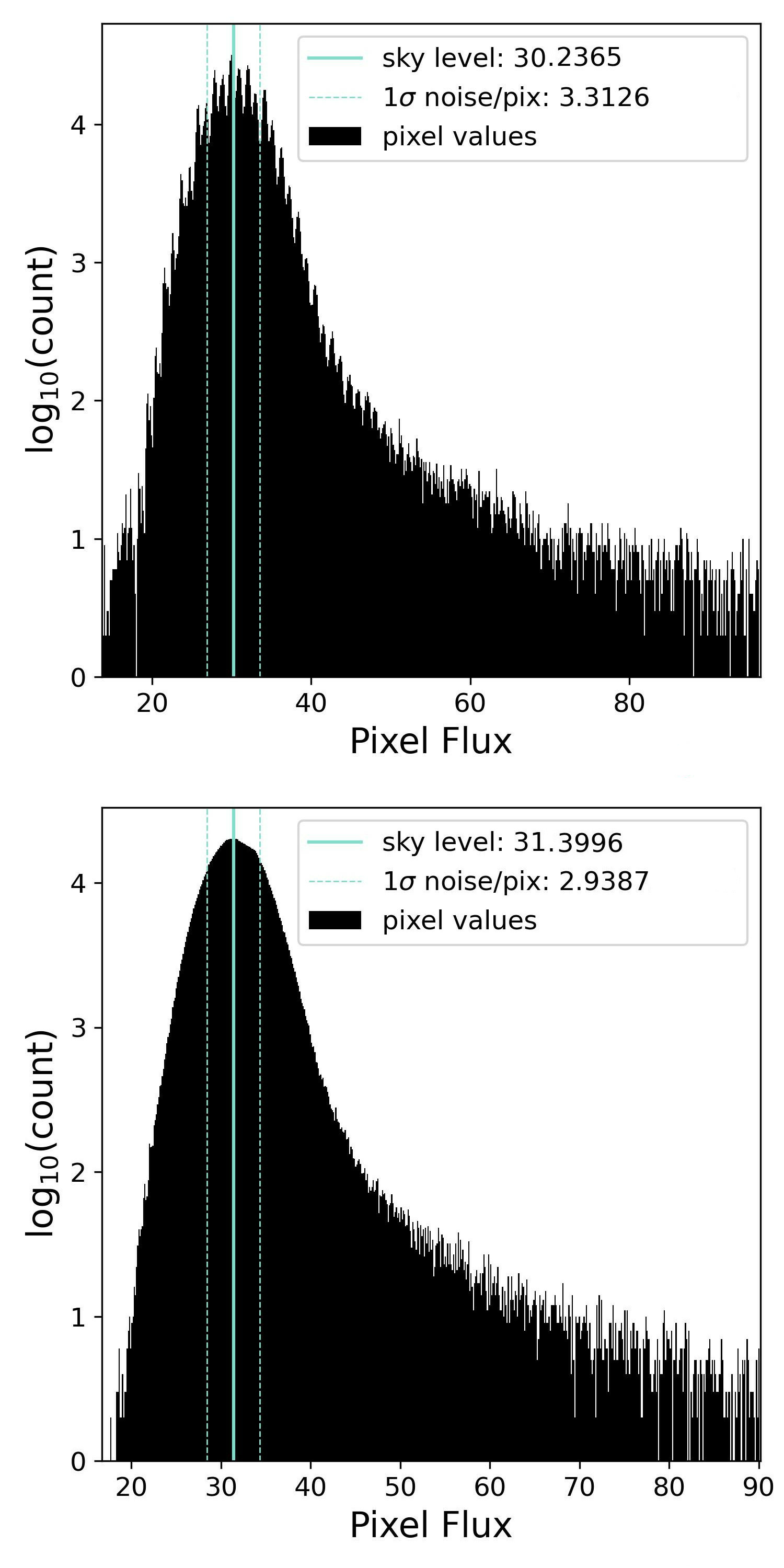}
	\caption{\textit{Left}: histogram of all pixels across a 200\,arcsec$^2$ region from a VIS detrended image.  This shows regular spikes at the 1\,ADU frequency, indicative of quantisation noise (this effect is also observed in the raw data). \textit{Right}: same region after resampling the detrended image with {\tt SWarp} using a Lanczos3 function, showing that the pixels have been correlated and the quantisation error has been smoothed out.
\label{fig:Quantization}}
\end{figure}


Upon analysing the noise characteristics of VIS images processed via the \ac{ERO} pipeline, it was evident that the signature of quantisation noise from the analogue-to-digital converter (ADC) was present across all quadrants. This phenomenon, already noticeable in the raw data, persisted through detrending processes. Quantisation noise emerges from the rounding differences between the analogue input voltage to the ADC and its resulting digital output, an effect distinctly visible in a histogram of pixel intensity (\cref{fig:Quantization}, top). This noise is nonlinear and varies depending on the signal being converted.

It is important to recognise that the majority of \Euclid's scientific endeavours in the forthcoming years will focus on faint objects with relatively low \ac{SNR}, meaning these objects will be faint sources superimposed on the image background. Such sources typically lie just to the right of the peak in the histogram of \cref{fig:Quantization} (top), where quantisation noise is especially pronounced. This contributes to the photometric error budget, which remains primarily influenced by photon noise from the zodiacal light background, as well as readout noise. After resampling the image for stacking (\cref{fig:Quantization}, bottom), where neighbouring pixels are correlated, the effect of quantisation noise naturally diminishes.

\subsubsection{Identification and removal of cosmic rays}

\begin{figure}
	\includegraphics[width=\linewidth]{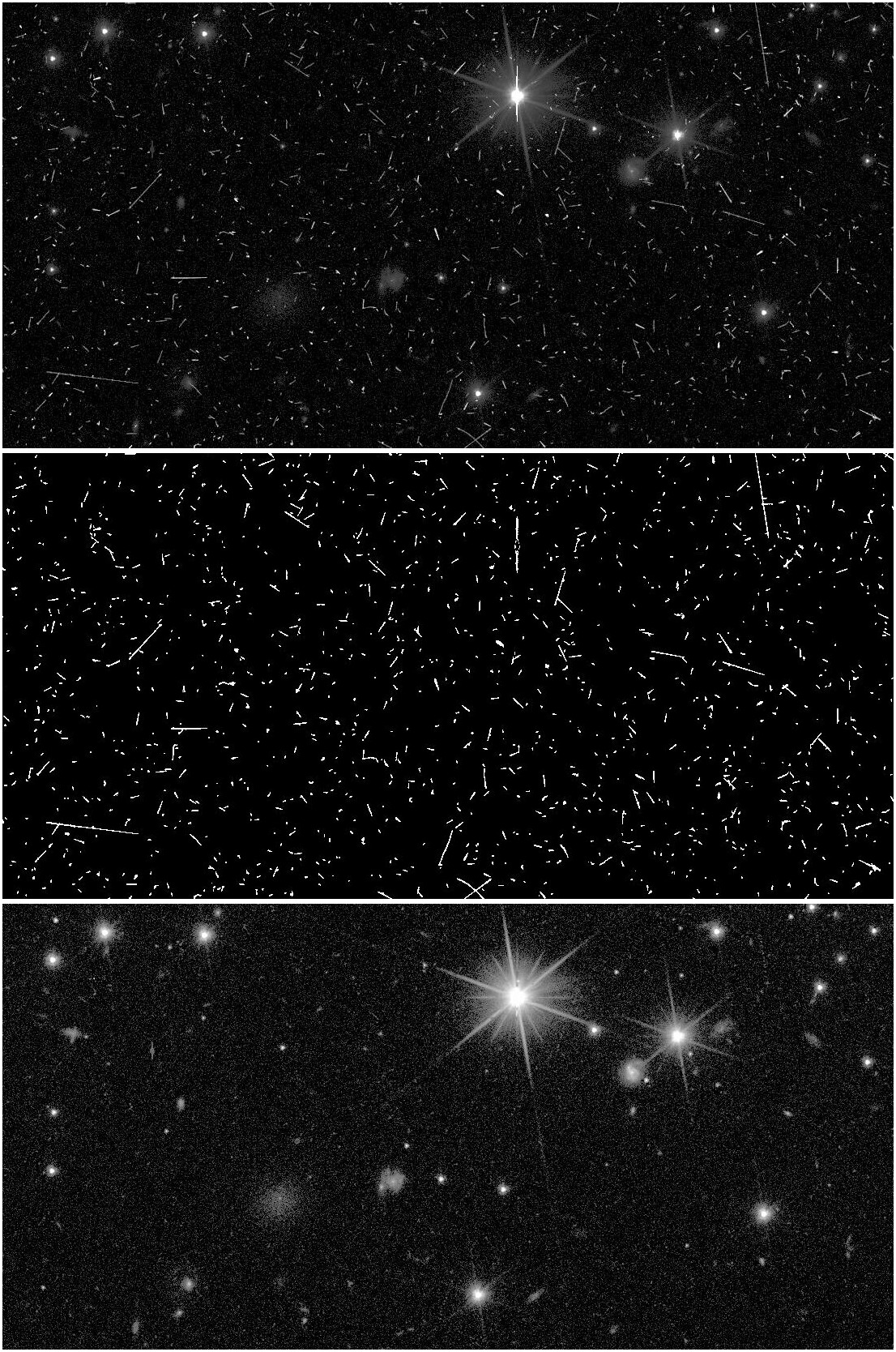}
	\caption{The Python-based, machine learning-driven tool {\tt deepCR} effectively identifies (centre) and repairs (bottom) pixels affected by cosmic rays (top). Each image segment (single rectangle) measures $\ang{;2;}\times\ang{;1;}$, demonstrating {\tt deepCR}'s effectiveness in enhancing image quality.
\label{fig:DeepCR}}
\end{figure}

Situated at the Eath-Sun Lagrangian point 2, \Euclid is continually bombarded by \acp{CR} of extragalactic, Galactic, and Solar origin. The CCDs' deep-depleted silicon layer, 40\,$\micron$ thick and designed for enhanced red sensitivity, is particularly susceptible to interactions with \acp{CR} (\cref{fig:DeepCR}). Combined with the long VIS integration times, this results in raw images that are heavily contaminated: each 566-second-long image over the 0.52\,deg$^2$ \ac{FOV} has approximately 1.4$\times10^6$ affected pixels. The prevalence of these transients impacts nearly all astronomical sources in the image and initially hindered our efforts to achieve a precise astrometric solution. Consequently, a method for repairing the affected pixels was explored and evaluated for widespread application in the \ac{ERO} pipeline.

Cosmic ray hits in individual images were identified and corrected using {\tt deepCR}, a deep-learning-based \ac{CR} removal tool specifically designed for astronomical images \citep{deepCR}. {\tt deepCR} was developed and trained with images from the \HST (HST) Advanced Camera for Surveys (ACS) using the F814W filter. Given the similarities in environment and detector characteristics between HST/ACS and \Euclid/VIS, the {\tt deepCR} model demonstrated remarkable efficiency in processing VIS images. Compared to other well-known methods like {\tt LAcosmic} \citep{LACosmic}, {\tt deepCR} not only offers superior performance in both detecting and replacing affected pixels (by utilising in-painting rather than interpolation), but also operates relatively quickly on hardware equipped with a graphics processing unit. For instance, processing an entire VIS frame, including its 144 quadrants, takes about 50\,s on an NVIDIA RTX 6000 Ada graphics card.

\begin{figure}
	\includegraphics[width=\linewidth]{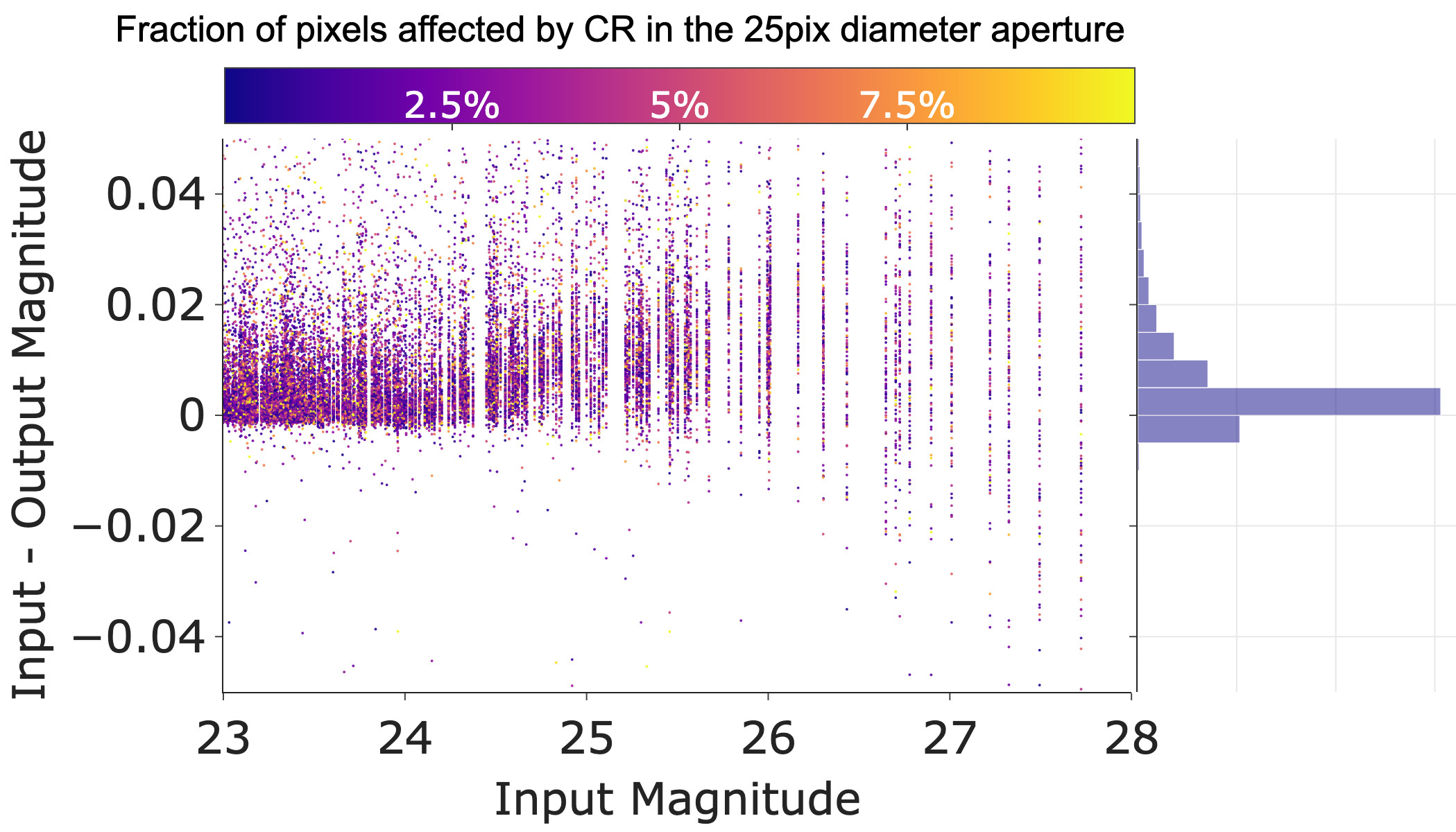}
	\caption{Magnitude difference between the input synthetic stars and their measured photometry, after applying {\tt deepCR} masking and in-painting, is plotted against the input magnitude (using an arbitrary zero point). The colour scale on the graph represents the number of pixels within a 25-pixel aperture diameter that were affected by cosmic rays. On the right side of the graph, a histogram displays the distribution of the magnitude differences, providing a visual representation of the photometric accuracy and the impact of cosmic ray corrections across various levels of cosmic ray contamination.
\label{fig:DeepCRphotometry}}
\end{figure}

To evaluate the impact of the {\tt deepCR} correction on final photometry, we conducted a set of tests. We added synthetic Gaussian profiles with a \ac{FWHM} corresponding to that of VIS images and fluxes ranging from 5000 to 500\,000\,ADU (as an integrated flux, meaning the sum of all pixels within 25-pixel apertures) to real VIS exposures impacted by cosmic ray hits. These modified exposures were then processed with {\tt deepCR}. Following this, we performed source detection and photometric measurements on the corrected images using {\tt SourceExtractor} \citep{Bertin1996}, employing a conservative 25-pixel aperture diameter for analysis (noting that smaller aperture photometry, like \ac{PSF} photometry, would be even less affected). We find that about 20\% of the artificial stars are unaffected by CR hits within the 25 pixels aperture, and that the largest fraction of pixels affected within these apertures reaches approximately 10\%. The comparison of output to input instrumental magnitudes for the sources affected by CR is depicted in \cref{fig:DeepCRphotometry}, with the colour scale indicating the fraction of pixels within the 25-pixel diameter aperture affected by cosmic rays, as identified by {\tt deepCR}. On average, the effect on magnitude is minimal, showcasing a skewed distribution with a mode around 1\,mmag and 25\%,75\% quartiles at 0.2 and 6\,mmag. This negligible impact occurs almost independently of the number of pixels affected by cosmic rays in the aperture, demonstrating the in-painting's robustness and efficiency. 

This approach not only maximises \Euclid's capabilities for astrometry but also enhances the cleanliness of the images for subsequent steps in the pipeline. Specifically, {\tt deepCR} ensures that even with the standard \ac{ROS} consisting of four dithered exposures, the resulting image stack is completely free of blacked-out pixels (out of approximately 606 million pixels) across the \ac{FOV} due to \ac{CR} contamination. This is particularly noteworthy because small areas of the \ac{FOV} are exposed only once throughout the 4-exposure dither, highlighting the role of {\tt deepCR} in maintaining image integrity.

\subsection{Detrending NIR data from the NISP instrument}

NISP uses 16 Teledyne HAWAII-2RG detectors of $2048\times2048$ pixels, arranged in a 4$\times$4 mosaic. Similar to the procedures outlined for VIS, this section details the processing steps for NISP as they occur sequentially within the \ac{ERO} pipeline, beginning with the correction of purely additive components of the instrumental signature.

\subsubsection{Charge-persistence correction\label{sec:persistence}}

Charge persistence is the process of trapping charge carriers within lattice defects of pixels and their slow release with a rate $\dot{R}$ during subsequent exposures \citep{Tulloch2018}. This effect manifests as a faint version of a previous image contaminating the following images. Strongly saturated pixels remain bright for many hours. Masking is a suboptimal solution. The absence of dithering between the NISP spectra (captured in parallel to the long VIS exposures) and images taken in different filter bands creates persistence always at the same sky coordinates. Masking would result in large holes in the stacked images. A better approach is modelling and subtracting the persistence from each single exposure.

Ideally, $\dot{R}$ is measured using a bright \ac{LED} flat image and a subsequent long series of dark images \citep{Serra2015}. Such data were taken during the \ac{PV} phase in September 2023. The series was repeated three times with almost identical results. Therefore, we took the median of all three series to remove cosmic rays. The persistence model was then derived from these median images.

\begin{figure}
	\includegraphics[width=\linewidth]{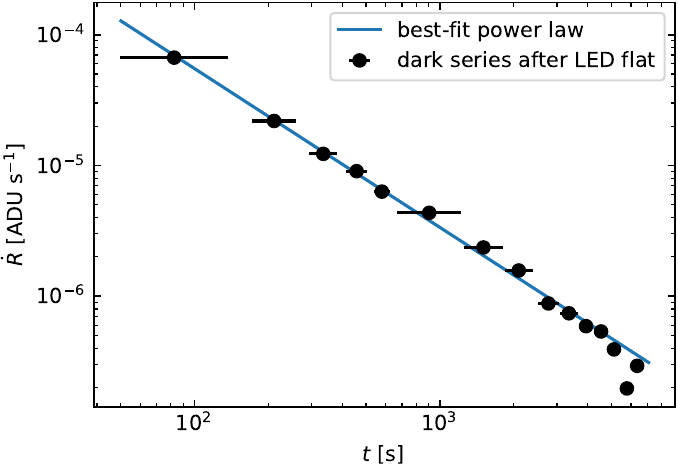}
	\caption{Release rate $\dot{R}$ for an example pixel derived from an \ac{LED} flat with a subsequent series of dark exposures. The data points show the persistence signal $P$ in the dark images with the bias pedestal (1024\,ADU) subtracted and normalised by the exposure time and initial signal $S$. The horizontal error bars mark the beginning and end of each dark exposure. The best-fit power law ($A=-1.215, B=-1.829$) is shown in blue. \label{fig:powerlaw_example}}
\end{figure}

\begin{figure*}
	\includegraphics[width=\linewidth]{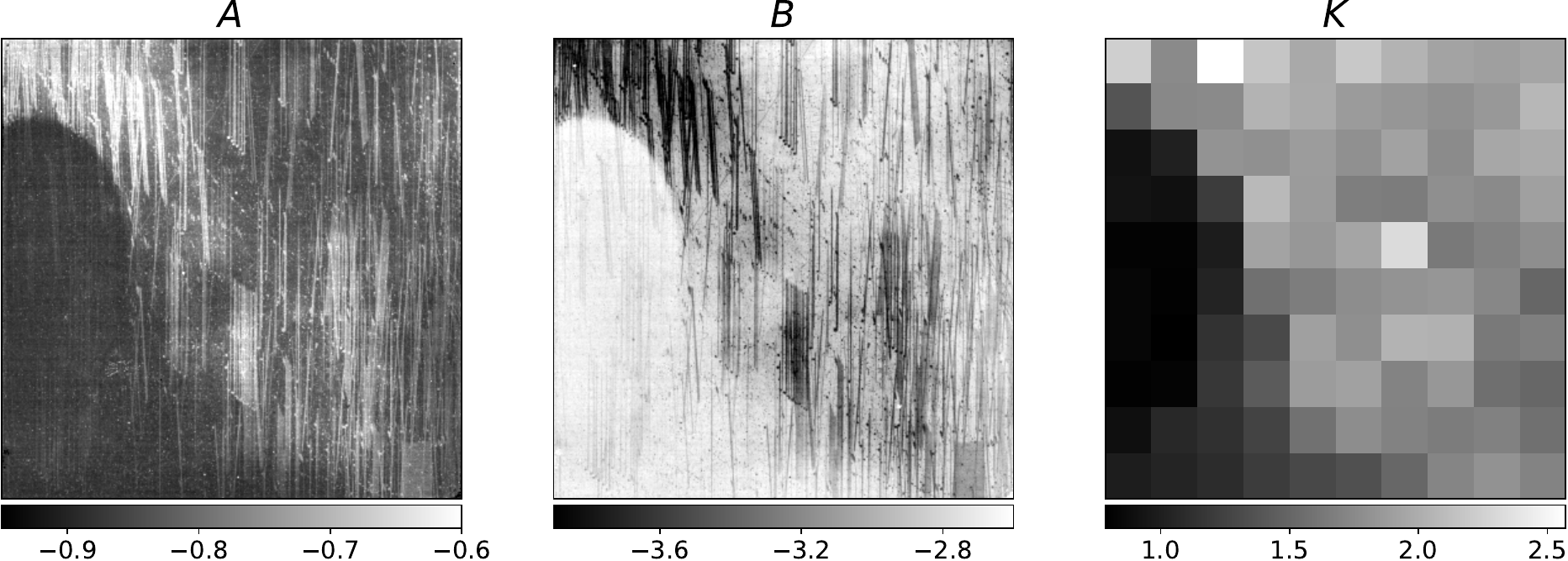}
	\caption{Persistence model parameters for NISP detector 1. The parameters $A$ and $B$ are defined in Eq.~ (\ref{eq:persistence_model}). The correction factor $K$ is shown for the date 2023-09-16. It is defined in Eq.~(\ref{eq:persistence_corr}). \label{fig:persistence_modelparams}}
\end{figure*}

We calculated $\dot{R}(x,y)$ from the signal in each dark image at pixel $(x,y)$, minus the bias pedestal of 1024\,ADU, and normalised by the integration time and initial signal in the flat image ($S\approx \num{40000}$\,ADU). 
The measurements are shown in \cref{fig:powerlaw_example}. To reduce the noise, we took the median value around each pixel in the interval $[x-2:x+2,y-2:y+2]$.
Similar to the situation in CCD detectors \citep[e.g.][]{Kluge2020}, we find that $\dot{R}(t)$ follows a power law,
\begin{equation}
	\logten(\dot R(x,y,t)[{\rm ADU\,s^{-1}}]) = A(x,y) \logten(t[{\rm s}]) + B(x,y)\;. \label{eq:persistence_model}
\end{equation}
The slope $A$ and offset $B$ are fit for each pixel. An example for detector 1 is shown in the left two panels of \cref{fig:persistence_modelparams}. 
We notice strong spatial variations and an anti-correlation between $A$ and $B$ that cannot be explained purely by fitting uncertainties. The anti-correlation implies that the persistence signal is more stable than $A$ and $B$ individually.

The persistence signal $P(x,y,t_{\rm start},t_{\rm end})$ was then estimated by integrating $\dot{R}$ from the start $t_{\rm start}$ to the end $t_{\rm end}$ of the subsequent exposures. We did not find any dependence of $\dot{R}$ on the initial signal $S$. Therefore, we simply scale the predicted persistence by $S$:

\begin{equation}
P = S \int_{t_{\rm start}}^{t_{\rm end}} \dot{R}(t)\,{\rm d}t = S \frac{10^B}{A+1} \Big(t_{\rm end}^{A+1} - t_{\rm start}^{A+1}\Big)\; .
\end{equation}

For a typical overhead of 60\,s and integration time of $t_{\rm end} - t_{\rm start}=87$\,s (effective exposure time from the entire duration of 112\,s), we get $P\approx0.0011\, S$ to $0.0033\, S$, that is, a few permille of the previous signal remain in the next image.
This model is limited to unsaturated pixels. When saturation occurs, the true $S$ value is unknown and cannot be estimated. However, masking those pixels with $S>\num{50000}$\,ADU for 24 hours would result in too many mildly affected pixels being discarded, inevitably leading to some areas across the stack of four images with no signal. In the \ac{ERO} pipeline implementation we left these pixels untouched and relied on sigma-clipping iterative algorithms to reject pixels affected by this specific persistence regime. 

Complications arise because $\dot{R}$ varies over long periods of time. We suspect that the cause is related to the small-scale pattern in the dark images that is also visible in \cref{fig:persistence_modelparams}. The stripe pattern resembles spectra. Probably, saturation in a pixel affects the parameters $A$ and $B$ for longer time scales, an effect that has been observed in NISP ground tests and also in other \acp{H2RG}  \citep[see e.g.][]{mcleod2016}. 
The true $\dot{R}$ can deviate by a factor of around 2 or more from the model prediction. We mitigate this effect by empirically rescaling $P$ as $P'=P\times K$ on a $10\times10$ grid ($X,Y$) for each detector (see \cref{fig:persistence_modelparams}, right panel). We refer to the elements of this grid as `blocks'. 
For each sufficiently large contaminated region $i$, we calculated the clipped median flux before subtraction $F_{\rm in,org}$, and after subtraction $F_{\rm in,sub}$, as well as outside of it $F_{\rm out}$. The local correction is then

\begin{equation}
	K_i = 1 - \frac{F_{\rm in,sub} - F_{\rm out}}{F_{\rm in,org} - F_{\rm out}}\; . \label{eq:persistence_corr}
\end{equation}

This matches the flux inside the region to the surrounding flux.
Corrections for single regions can be strongly affected by outliers. To increase the robustness, we take the median correction $K(X,Y) = {\rm med}[K_i(X,Y)]$ within each block. Because corrections are calculated for each single exposure and we do not observe strong short-term variations in $K$ between exposures, we then combined all $K$ for each day and \ac{ERO} project by taking the median. The result is shown in the right panel of \cref{fig:persistence_modelparams}.
The matrix $K(X,Y)$ is then linearly interpolated on the finer grid $K(x,y)$ to obtain a correction for each NISP pixel. The mean correction is ${\rm mean}(K)=1.32$ with a standard deviation of ${\rm std}(K)=0.26$.

For all spectra and images taken up to 1\,hour prior to the current exposure, we modelled and subtracted the predicted persistence. First, we subtracted the clipped median signal from each preceding exposure before modelling its persistence. Although this is not correct in principle (because the background also creates persistence), this step is important to not deform the background signal due to imperfect estimations of $\dot{R}(x,y)$. We have visually verified that the effect on the predicted persistence is negligible.
We only modelled persistence $P<70$\,ADU. For brighter persistence, we relied on outlier rejection using sigma clipping during the stacking procedure because modelling uncertainties would leave visible residuals. Masking these pixels could be beneficial but we decided against it to avoid empty pixels in the stacks. Consequently, the inner regions of bright objects can still contain persistence from previous spectra.

\begin{figure*}
	\includegraphics[width=\linewidth]{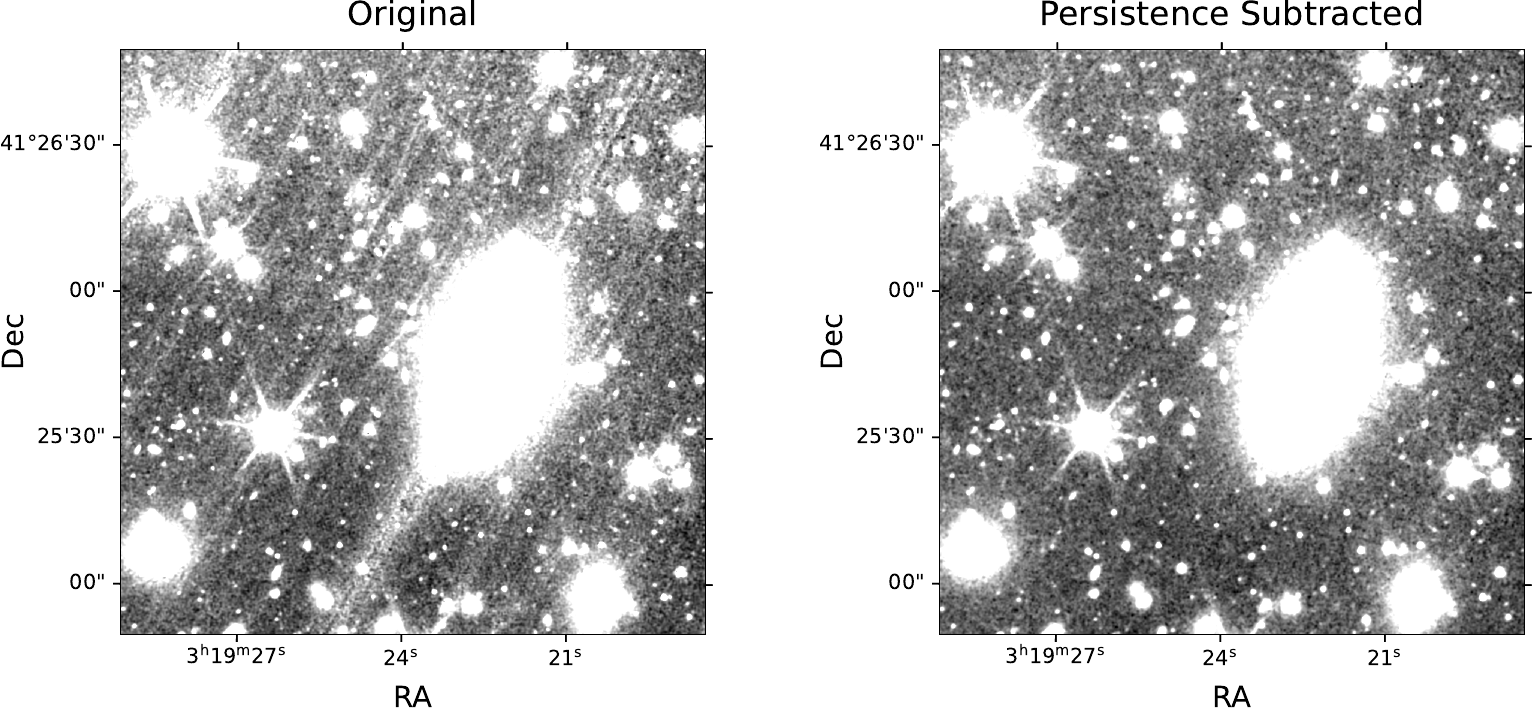}
    \caption{Example region in the Perseus \ac{ERO} \JE-band stack before (left) and after (right) subtracting the predicted persistence from the single exposures. White corresponds to a surface brightness of approximately 25\,mag\,arcsec$^{-2}$. \label{fig:persistence_before_after}}
\end{figure*}

\Cref{fig:persistence_before_after}, left panel, shows a region in the stacked Perseus \ac{ERO} project that is strongly affected by persistence. The image in the right panel shows the result after successfully subtracting the predicted persistence from the single exposures. The diagonal stripe pattern that arose from the spectra taken just before the \JE-band images is mostly gone. The effect on point-source photometry is $P'/S\approx0.15$--$0.45\%$. More affected are the colour profiles of extended galaxies. The persistence from the spectra on the subsequently taken \JE-band images makes the surface-brightness profile locally brighter by approximately 0.05\,mag. 
Overall, we estimate from our models that 2\% (8\%, 30\%) of the area in the Perseus ERO \JE-band stack was affected by persistence brighter than 25 (26, 27) mag\,arcsec$^{-2}$ before our correction was applied.
This highlights the importance of correcting for persistence in the NISP data.

\subsubsection{Bad pixel mask}

The 16 \acp{H2RG} in NISP are of exceptional quality, featuring technology distinct from and not directly comparable to that of CCDs. Such detectors invariably have a higher fraction of pixels with marginal response. In applying the same approach as used for the VIS instrument -- flagging nonlinear pixels through the analysis of the ratio of two internal \ac{LED} illumination images -- a significantly higher threshold of 10\% (versus 1.2\% for VIS) is required to avoid excessively flagging pixels. Below this threshold, pixels are corrected at first order by flat-fielding. Compared to VIS, the NISP mosaic exhibits significantly larger gaps between detectors, and with most \ac{ERO} projects only involving four exposures per dither, overly aggressive flagging would result in numerous gaps in the final image stack, especially in the areas close to the mosaic gaps. The threshold was therefore adjusted upwards until all single \ac{ROS} observations for \ac{ERO} projects (comprising four exposures per NISP band) resulted in a stack with no sky pixels left unexposed.

Each detector in the NISP instrument consists of a $2040\times2040$ pixel array sensitive to light, with a total of 0.4\% of pixels being masked. This proportion is comparable to that of the VIS instrument, despite a much higher threshold for identifying nonlinear pixels in NISP and a smaller total number of pixels (67 million for NISP). The distribution of masked pixels is consistent throughout the mosaic, with the notable exception of the top-right corner detector (DET16) that exhibits a 40\% excess of masked pixels. The outermost four pixels around the periphery of the NISP \acp{H2RG} are insensitive to light. They are used for detector monitoring, do not significantly improve the \ac{ERO} pipeline processing, and are thus simply masked.

\subsubsection{Electronic pedestal correction}

Due to the onboard multi-frame sampling and subtraction performed before transmitting NISP images to Earth \citep[for details see][]{EuclidSkyNISP}, the pedestal of all raw NISP data is artificially set to 1024\,ADU. This value is hardcoded into the \ac{ERO} pipeline. We observed a low-level time-dependent variation in the relative background level between detectors on a per-exposure basis. This leads to occasional mid- to large-scale background inhomogeneities in the final image stacks at the sub-percent level of the main background.

\subsubsection{Dark current correction}

\begin{figure}
	\includegraphics[width=\linewidth]{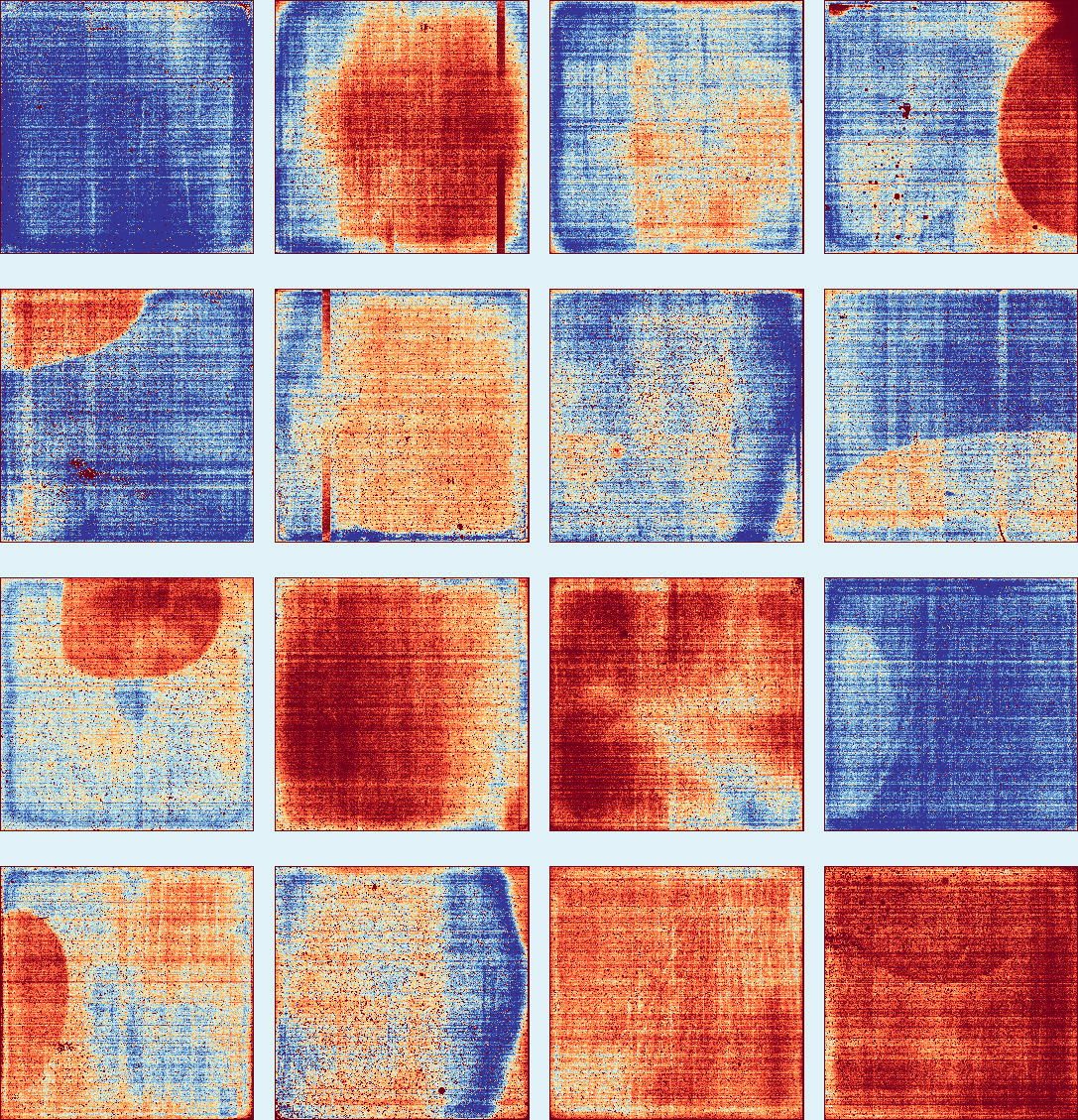}
	\caption{Dark current map for the 16 NISP detectors, derived from a stack of 110-s integration dark frames, which matches the duration of the science exposures. The amplitude of the dark current varies, reaching up to 1.6\,ADU at most across a single detector, with an average value of 0.8\,ADU across the entire mosaic (minimum in dark blue is 0.0\,ADU, maximum in deep red is 2.6\,ADU).
\label{fig:NISPdark}}
\end{figure}

Operating at a temperature of 95\,K, the NISP detectors exhibit a low dark current, averaging 0.8\,ADU per pixel over the duration of science exposures (112\,s, leading to an effective integration time of 87.2\,s). The dark current distribution across the detectors' surfaces is highly structured (\cref{fig:NISPdark}), necessitating the subtraction of a dark frame from the science images. The master dark frame for the \ac{ERO} data was generated using a median stack of 100 dark frames, each with an integration time matching that of the science exposures. These dark frames were captured following a prolonged period without any exposure to illumination, from astronomical sources or \acp{LED}, to prevent any residual signal contamination due to image persistence.

\subsubsection{Flat-field correction for large and small scales}

\begin{figure*}
	\includegraphics[angle=0,width=1.0\hsize]{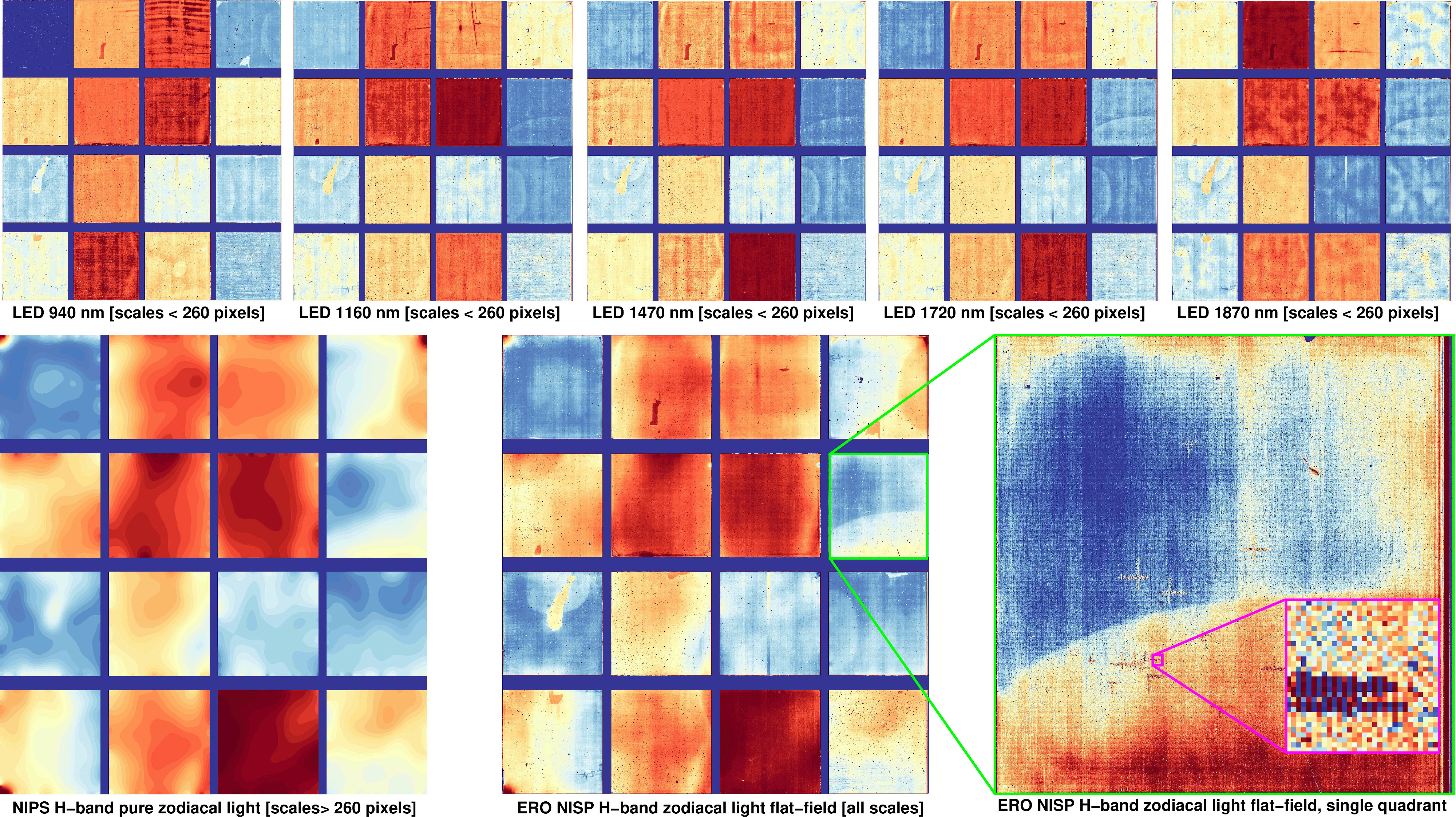}
	\caption{\ac{ERO} NISP zodiacal light flat-field.  This combines the large scales ($>\ang{;;80}$) derived solely from zodiacal light observed in any NISP exposures, with the small and medium scales ($<\ang{;;80}$) coming from specific \ac{LED} internal calibration images tailored to each NISP band (only three \acp{LED} being relevant for this purpose: 1160\,nm, 1470\,nm, and 1720\,nm for \YE, \JE, \HE, respectively). The arbitrary colour scale is used to fully explore the intensity range within each image. The top row displays the evolution of \{gain $\times$ quantum efficiency\} across all 16 NISP detectors at specific wavelengths, highlighting a large-scale pattern that is, however, omitted in the final flat-field in preference to the pattern emanating from the zodiacal light image (bottom left). The completed \ac{ERO} NISP zodiacal flat-field (bottom centre, for \HE) encapsulates all scales present in the \Euclid signal, ranging from pixel-to-pixel sensitivity variations and low-to-medium-sized features (bottom right, showing a detector of $\ang{;;612}\times\ang{;;612}$, with a zoom on a 30 by 30 pixels area showing the pixel-to-pixel sensitivity differences), extending to the entire \ac{FOV} scale. The relative colour range from dark blue to deep red is an intensity ratio of 12\% on average across the full mosaic panels and 5\% for the variations across the single detector on the lower right.
\label{fig:ZodiFlatNISP}}
\end{figure*}

NISP uses the same flat-field approach as VIS, combining zodiacal light and \acp{LED} to correct image variations. This section highlights the differences between the two instruments, focusing on how we used light sources and on the adjusted processing methods for NISP's \ac{NIR} detectors compared to VIS's optical detectors.

The sky background in the NISP bands is roughly the same as in VIS, about 22.3\,mag\,arcsec$^{-2}$, but a NISP pixel covers 9 times the area of a VIS pixel (\ang{;;0.3} versus \ang{;;0.1} per pixel). Additionally, the single exposure integration time for NISP is about 1/5th of that for VIS. As a result, the zodiacal light signal per pixel is stronger in NISP, around 60\,ADU in total per exposure. The associated photon noise (approximately 11 electrons) surpasses the readout noise (3.1\,ADU, equivalent to 6.2 electrons), providing robust statistics for analysing the zodiacal light background and instrument-induced structures.

For creating the NISP zodiacal light flat-field, we selected a reference field free of extended sources located at the north ecliptic pole, featuring far less stars than the south ecliptic pole field adopted for VIS nearly on the line of sight of the Large Magellanic Cloud (stellar density was lesser in the \IE-band, while the VIS resolution kept the footprint of stars limited). This choice was based on nearly 100 dithered frames captured over 3 days in early September 2023, with the same integration time as that used for the \ac{ERO} programme. Despite its proximity to the Galactic plane, as outlined in the mission plan \citep{Scaramella-EP1}, this area is one of the three Euclid Deep Fields and is notably free from strong Galactic cirrus emission. The compilation of exposures from many different pointings ensures that the median stack effectively eliminates any isolated contamination.

For NISP, each photometric band is matched to a specific \ac{LED}, as illustrated in \cref{fig:ZodiFlatNISP}. Detailed examination of these high-\ac{SNR} frames across the three bands revealed the necessity to include significantly larger physical scales in the final flat-field than is done for VIS, due to the existence of distinct features that span hundreds of pixels. Capturing these extended structures at high \ac{SNR} was crucial for achieving effective flat-fielding. Consequently, the crossover physical scale selected for NISP between the \ac{LED} flat-field and the pure zodiacal light flat-field is \ang{;;80}. Adjustment of this scale was approached with precision to avoid artificial amplification of any structure that could be present in both input elements. This was achieved by flat-fielding the input images from the Deep Field North and meticulously examining their uniformity in the most sensitive areas of the NISP mosaic. The final \ac{ERO} NISP flat-field is normalised to the average mode across all 16 detectors, resulting in detrended images that maintain characteristics similar to those of the raw data.

\subsubsection{Row correlated noise correction}

\begin{figure}
	\includegraphics[width=\linewidth]{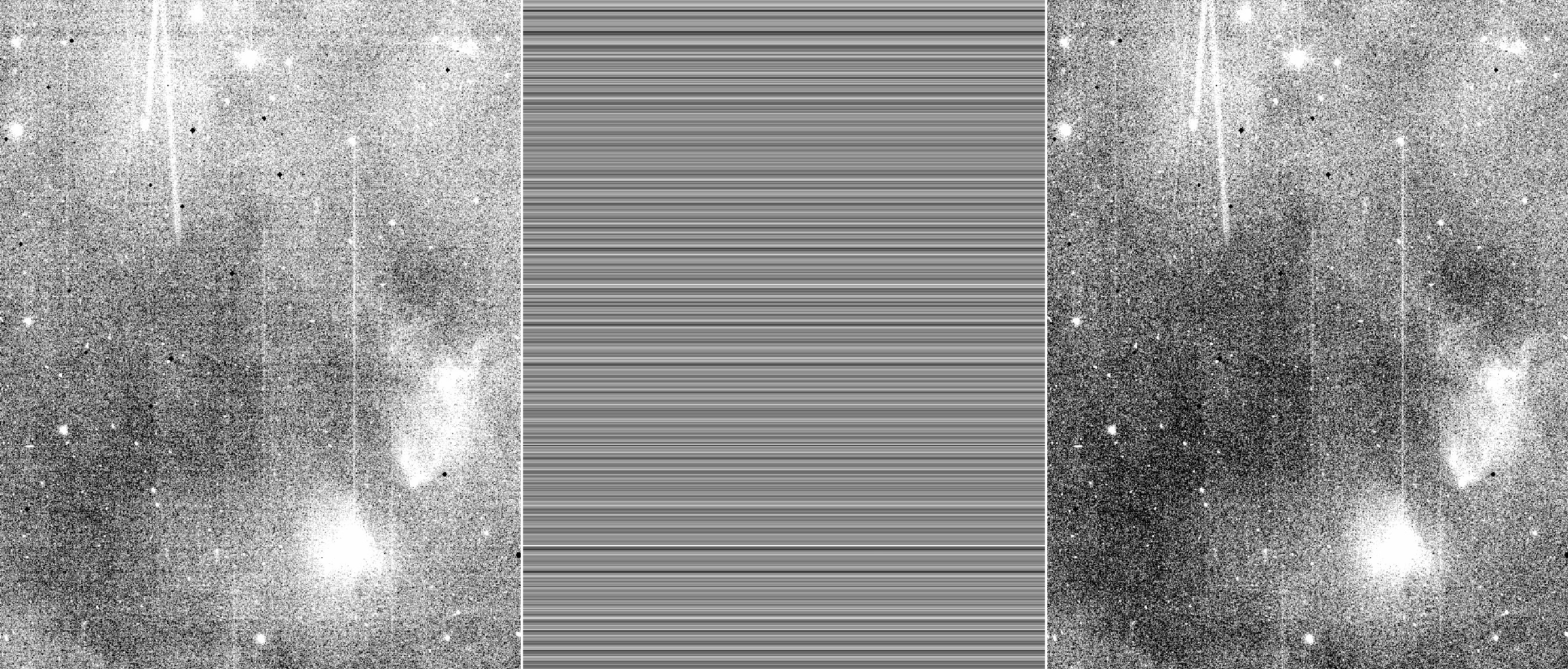}
	\caption{\textit{Left}: in the original image (900\,px$\times$1200\,px area of a single NISP detector), horizontal striping, indicative of row-correlated noise, is easily noticeable. \textit{Middle}: a noise model is constructed spanning the image's width to specifically target and quantify this pattern of noise. \textit{Right}: in the corrected frame, the row-correlated noise has been mitigated, demonstrating that this procedure effectively addresses the noise issue without substantially impacting the extended emission in the image. Residual image persistence caused by NISP spectra generates an apparent column-correlated noise.
\label{fig:RowNoise}}
\end{figure}

Upon visual inspection of the processed images, a faint residual noise is observed, exhibiting correlation across entire rows with variations around a single ADU (measured RMS of 0.7\,ADU). This detector phenomenon can be modelled and subtracted by building a median of all rows. To avoid affecting extended emission, structures beyond a 15-pixel scale are first removed from the resulting vector before subtraction from each column of the image. This correction is illustrated in \cref{fig:RowNoise}. While the fundamental noise properties remain unchanged, the data appear cleaner after correction. 

\subsubsection{High-fluence nonlinearity}

The NISP \ac{H2RG} sensors exhibit nonlinearity at high fluences, a characteristic whose details are unique to each pixel. Their full-well capacity is about 100\,000 electrons, equivalent to 50\,000\,ADU given NISP's average gain of 2 electrons per ADU. Nonlinearity typically becomes noticeable around 60\,000 electrons, corresponding to 30\,000\,ADU for NISP. Since \ac{ERO} science primarily deals with faint sources within the first few thousand ADUs, high-fluence nonlinearity is generally negligible. However, caution is warranted when measuring high fluences above 30\,000\,ADU, such as during photometric calibration using 2MASS \citep{cutri2003} stars.

There are also low-fluence nonlinearities, particularly from charge-persistence effects that we correct for empirically (\cref{sec:persistence}), and from count-rate nonlinearity \citep{biesiadzinski2011} that is related to persistence and not explicitly corrected for by the \ac{ERO} pipeline.

\subsubsection{Inter-pixel capacitance}

\ac{H2RG} detectors exhibit inter-pixel capacitance \citep{LeGraet22} resulting in charge transfer or `cross-talk' from the pixel where the charge initially accumulates to neighbouring pixels. This phenomenon increases the footprint of cosmic rays and hot pixels across a larger area of the image, and also the core of the \ac{PSF}, and can be considered another source of nonlinearity. However, since \ac{ERO} science primarily utilises NISP as a photometer (with apertures covering multiple pixels to measure the flux of galaxies and stars), we ignore this effect. Nonetheless, it does contribute to the overall photometry error budget.

\subsubsection{Cosmic rays}

We were unable to implement a robust \ac{CR} identification solution for NISP in time for this data release. With a pixel sampling of \hbox{\ang{;;0.3}} leading to an \ac{FWHM} of 1.6 pixels, distinguishing cosmic rays from stars becomes challenging because they often appear similar. Tools like {\tt deepCR}, effective for VIS data, are not suitable, due to inadequate training. The NISP images are accompanied by a flag map, where pixels likely affected by \acp{CR} are identified. However, we discovered that applying this mask would lead to the discarding of an excessive number of pixels, a problem reminiscent of the issue encountered with the bad-pixel mask versus sky coverage. As a result, we did not correct or flag cosmic rays and instead relied on iterative sigma-clipping methods to reject them when generating the science stacks.

For \ac{ERO} data employing a single \ac{ROS} with only four dithered exposures per band (see \cref{table:obsspecs}), a fraction of the \ac{FOV} situated at the intersection of the gaps between four detectors -- totalling nine areas within the main area of nominal \ac{SNR} -- receives astronomical signal only once. Consequently, these nine small rectangular areas still contain residual cosmic rays in the final stacks. However, by combining information from other \Euclid bands at the catalogue level, we can flag and reject these residual transients. Since they are unlikely to affect all four single visits on the sky across the three \ac{NIR} bands, this is achieved through the utilisation of a $\chi^2$ detection image (see \cref{sec:cats}).

\section{\label{sec:astrometry}Astrometric calibration}

\subsection{Initial astrometric solution}

In the context of processing the \ac{ERO} images for integration into the pipeline, we noted that the raw FITS headers only provide the central coordinates of the observed fields. To establish a refined astrometric framework, a crucial initial step involved anchoring a minimum of one quadrant for VIS or one detector for NISP to the \ac{WCS}. This key task was accomplished through the deployment of {\tt Astrometry.net}, operating in a local environment and utilising the most comprehensive index derived from \textit{Gaia}-DR3.

Advancing to the subsequent stage that entails the construction of global mosaic astrometry for both VIS and NISP instruments, the process is notably facilitated by the preliminary calibration of all quadrants or detectors via {\tt Astrometry.net}. In this vein, custom catalogues were generated for certain \ac{ERO} fields where stellar density was compromised, attributable to factors such as high Galactic latitudes (for instance, the Fornax field) or significant attenuation, such as in Messier 78. The catalogues were sourced from the Dark Energy Survey Instrument Legacy Imaging Surveys \citep{Dey2019} and the Unwise Catalogue \citep{Schlafly2019}, and were tuned for the small angular scale of \Euclid VIS quadrants by selecting quadrangles of stars with scales a factor of $\sqrt{2}$ and $2$ smaller than the standard \textit{Gaia}-based catalogues.

\subsection{Global image astrometry}

Astrometric calibration was performed using an enhanced version (V2.13) of {\tt SCAMP} \citep{Bertin2006}, utilising catalogues generated from processed images initially calibrated with {\tt Astrometry.net}. In essence, {\tt SCAMP} calculates a global astrometric solution by minimising, in the least-square sense, the mutual differences of the reconstructed source positions in all overlapping exposures, plus the differences in position with respect to an arbitrary reference catalogue. The reference catalogue employed for calibrating \ac{ERO} data is \textit{Gaia}-DR3 \citep{GaiaDR3}.

In {\tt SCAMP}, geometric distortions represent deviations from a fiducial projection that is assumed to be tangential (gnomonic). 
At any position in the focal plane, {\tt SCAMP} models the geometric distortions for a given instrument and filter as the sum of two polynomials: a chip-dependent constant polynomial; and a mosaic-wide, exposure-dependent polynomial of low degree. The VIS and NISP focal planes exhibit minimal geometric distortions; we adopted a third-degree polynomial for the chip-dependent component and a first-degree polynomial for the exposure-dependent component, since we found them sufficient to calibrate astrometry to the milliarcsecond level. VIS and NISP were calibrated separately, because no gain in precision was achieved with a combined astrometric run, with the relatively coarse sampling of NISP remaining the limiting factor.

A considerable limitation of {\tt SCAMP} is its inability to directly incorporate proper motions within the calibration process. As a result, for observation and reference data sets spanning different epochs, the calibrated coordinates must be adjusted in a subsequent step \citep{Bouy2013}.
\ac{ERO} exposures are captured within a brief period, thereby sharing a common epoch, $t_{\rm{ERO}} = 2023.7$.
However, there is a substantial temporal gap when compared to the \textit{Gaia}-DR3 epoch, $t_{\rm{DR3}} = 2016.0$.
{\tt SCAMP} corrects for the proper motion vector $\vec{\mu}_{*\rm DR3}$ of each reference \textit{Gaia} star by actualising its celestial coordinates using

\be
\label{eq:propcorr}
\begin{pmatrix}
\alpha \cos \delta\\
\delta
\end{pmatrix}_{\rm ERO}
=
\begin{pmatrix}
\alpha\cos \delta\\
\delta
\end{pmatrix}_{\rm DR3}
+ \kappa\,(t_{\rm ERO} - t_{\rm DR3})\, \vec{\mu}_{*\rm DR3}\;,
\ee
where
\be
\kappa = \frac{|\vec{\mu}_{*\rm DR3}|^2}{|\vec{\mu}_{*\rm DR3}|^2 + |\vec{\sigma}_{\mu_{*\rm DR3}}|^2}
\ee
is a Tikhonov regularisation factor that accounts for possibly large proper motion uncertainties; in practice $\kappa$ remains very close to 1 for a large majority of stars.

\begin{figure}
	\includegraphics[width=\linewidth]{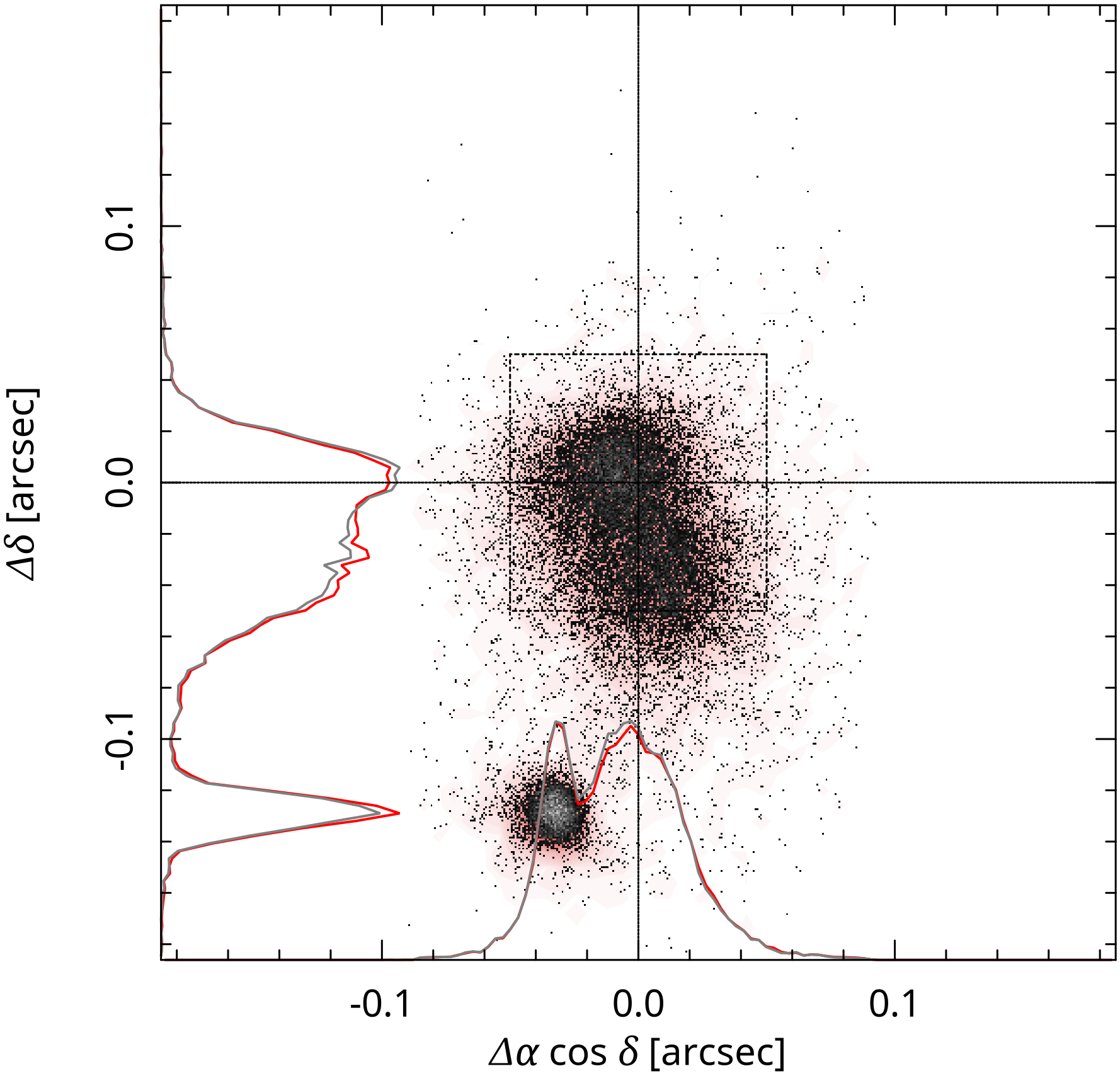}
	\caption{Post-calibration astrometric residuals along the RA and Dec axes for unsaturated \textit{Gaia}-DR3 stars in the NGC\,6397 field \citep{EROGalGCs}. Residuals are computed for each star as the difference between the calibrated coordinates on the individual VIS exposures and the corrected \textit{Gaia}-DR3 coordinates (see text). Grey points represent a subsample of detections with a \ac{SNR}$\,{>}\,100$ on \ac{ERO} exposures. 
    The plot reveals two distinct groups of stars, represented as a large cloud slightly offset from the centre and a small compact cloud clearly separated from the centre, both consisting of stars lacking proper motion data in the DR3 catalogue. The position offset of the smaller `cloud', located in the lower left, aligns well with previous estimates of the NGC\,6397 absolute proper motion \citep[$\vec{\mu}_* = (3.6, -17.3)\,{\rm mas\,yr}^{-1}$,][]{Kalirai2007}, over a period of 7.7 years. 
    A dashed square at the centre of the plot illustrates the size of the VIS pixel footprint for comparison. On this specific \ac{ERO} field the internal astrometric precision is 2.7\,mas RMS.
    }
    \label{fig:ngc6397_residuals}
\end{figure}

Although the \ac{ERO} projects were all conducted nearly perpendicular to the antisolar direction, we neglected the effect of parallax between the \textit{Gaia} and \Euclid observations that is below the milliarcsecond level for the vast majority of stars used in the calibration.
For simplicity, we also ignored correlations between the positional errors in right ascension and declination, as well as between positions and proper motions in the \textit{Gaia} sample.
The final positional uncertainties corrected in the astrometric reference sample are determined by adding in quadrature the proper motion uncertainties, scaled by the time difference.
A significant portion of the \textit{Gaia}-DR3 catalogue entries lack proper motion measurements; rather than discarding these entries outright, we found that it enhances the robustness of the solution for {\tt SCAMP} to assign zero proper motions to these stars, with an arbitrary proper motion uncertainty of 10\,mas\,year$^{-1}$ in both axes.
\Cref{fig:ngc6397_residuals} shows an example of astrometric residuals for the challenging NGC\,6397 \ac{ERO} field \citep{EROGalGCs}, where most globular cluster stars lack proper motion estimates in \textit{Gaia}-DR3.

Overall for VIS, the median internal astrometric precision (dispersion among calibrated coordinates) for sources with an \ac{SNR}$\,{>}\,100$ is 6\,mas RMS across the 17 \ac{ERO} fields in both axes.
This figure accounts for contributions from compact, yet resolved sources; when the selection is limited to point sources, the median internal dispersion decreases to 2\,mas RMS.
Additionally, the median external accuracy, defined as the RMS dispersion of the difference between the averaged calibrated coordinates over all overlapping exposures and the corrected \textit{Gaia}-DR3 coordinates, is 8\,mas RMS.
In comparison, the performance metrics for NISP are 15\,mas RMS for internal precision and 10\,mas RMS for external accuracy, respectively.
The astrometric performance of both VIS and NISP indicates that images are calibrated to a precision of 1/15th of a pixel across the entire \ac{FOV}.

\section{\label{sec:stacking}Resampling and stacking}

\begin{figure*}[hbtp!]
	\includegraphics[width=\linewidth]{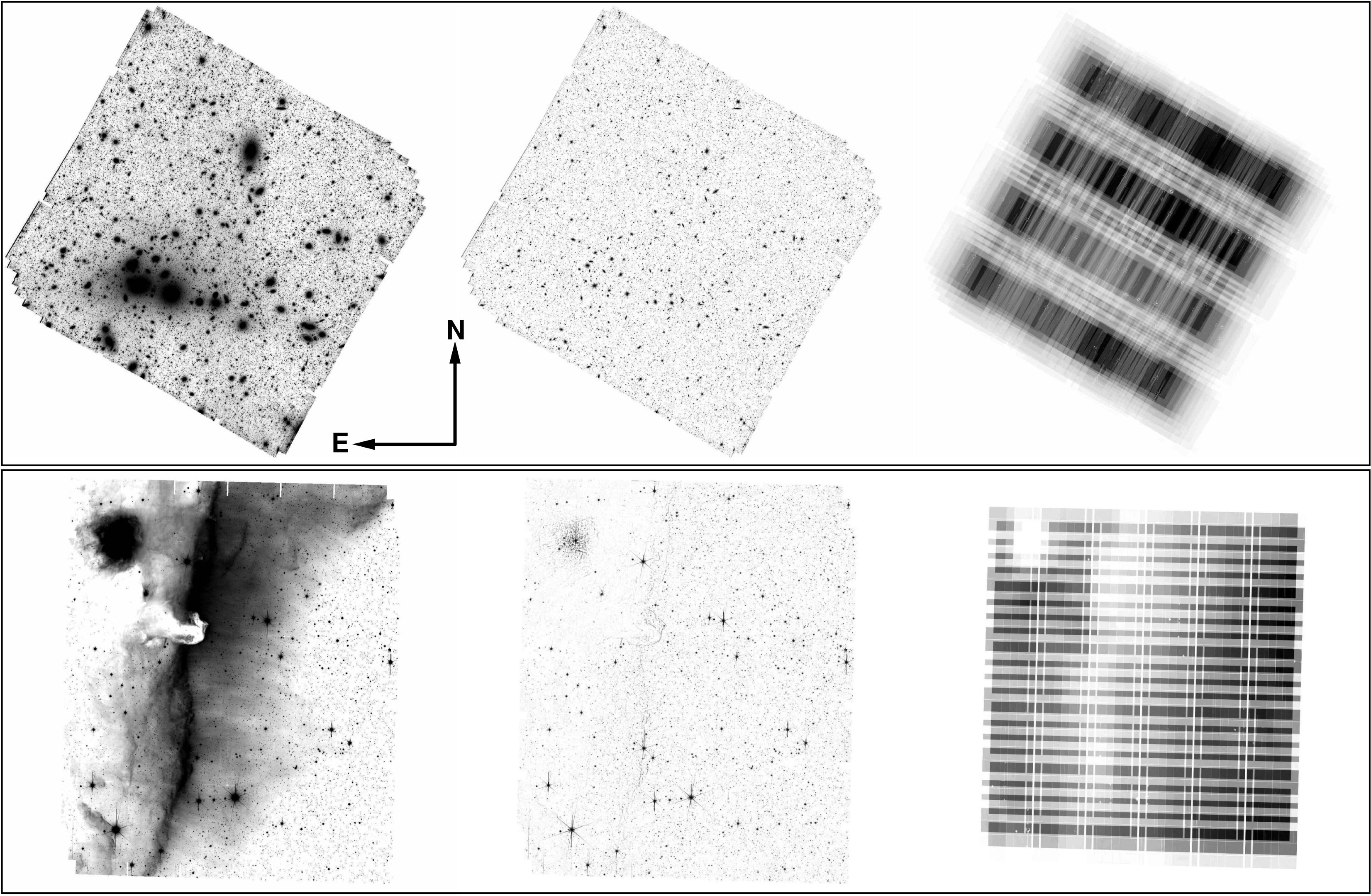}
    \caption{\textit{Top}: (left to right) \ac{ERO} Perseus cluster in the \HE-band (full \ac{FOV} of 0.70\,deg$^2$), showing the extended emission stack, compact-sources stack, and weight map (all in inverted scale). \textit{Bottom:} (left to right) \ac{ERO} Horsehead Nebula in the \IE-band (full \ac{FOV} of 0.59\,deg$^2$), showing the extended emission stack, compact-sources stack, and weight map (all in inverted scale). All \ac{ERO} fields get rotated to deliver a standard equatorial projection with north up and east to the left. The main difference is the suppression of extended emission in the compact-sources stack to optimise compact source photometry. The two flavours of stacking are motivated by the need to optimise the photometry for each class of objects (it is not recommended to use the extended-emission stacks for compact-source science). The weight maps echo the observing strategy, with a \ac{SNR} that can vary greatly across the image. This is particularly striking on the bottom right, based on a standard \Euclid observing sequence.
    \label{fig:Stacks}}
\end{figure*}

Image resampling and stacking represents a classic method for combining individual exposures onto a common pixel grid.
Resampling affects VIS and NISP exposures, both of which are significantly undersampled -- NISP even more so -- potentially leading to image smoothing and/or aliasing artefacts.
A non-destructive alternative for merging multiple observations is to conduct simultaneous measurements across all overlapping exposures, each on its own pixel grid \citep{Bertin2020}.
However, this method entails slower processing, yields larger and more complex end products, and complicates the handling of residual glitches.

Therefore, in the context of the initial \Euclid science efforts with the \ac{ERO} programme, we opted to adhere to the conventional method of resampling and stacking. The {\tt SWarp} software package \citep{Bertin2002}, is at the root of the two flavours of \ac{ERO} stacks that we use (\cref{fig:Stacks}), as now described.

\subsection{The compact-sources stack}
For the compact-sources stack we activated the background modelling and subtraction, removing extended emission signatures and thus yielding a background of zero ADU. We refer to this stack as the `\ac{ERO} compact-sources stack', optimised for the detection and photometry of compact sources. 

We resampled and stacked the images in a single {\tt SWarp} run, adopting a mesh size of \ang{;;6}, matched to both instruments -- with {\tt BACK\_SIZE} set to 64 pixels for VIS and 22 pixels for NISP -- to eliminate extended emission. This facilitates for example the detection of faint sources embedded in the halos of galaxies. We utilised the {\tt SWarp} option for clipped-mean stacking \citep{Gruen2014} to enhance the final\ac{SNR} also excluding \acp{CR} in NISP and occasional \ac{CR} residuals from the {\tt deepCR} operation for VIS. 

Prior of this step, a weight map was produced for each input image using {\tt WeightWatcher} \citep{Marmo2008}, based on the bad-pixel mask and the flat-field used for detrending the raw data. {\tt SWarp} compiles a weight map for the science stack from all input products, which is crucial for tracking the noise properties across the image and is a key element for source extraction (see \cref{sec:cats} for the creation of the catalogues). {\tt SWarp} also updates the average gain in the stack's FITS header ({\tt GAIN}) according to the weight map. For NISP, a ${\chi}^2$ detection image \citep{Szalay1999} is eventually generated by {\tt SWarp} based on the three \YE, \JE, \HE stacks.

\subsection{The extended-emission stack}
The extended-emission stack preserves extended emission (covering both high and low surface-brightness science) and is suited for both high and low surface-brightness objects. It is created by initially generating an individual {\tt SWarp}-ed image from each input image, with the background subtraction option deactivated to retain all scales. The {\tt FSCALASTRO\_TYPE} option is set to {\tt NONE} to ensure the preservation of the number of ADUs per pixel, which makes this stack inappropriate for the study of compact sources. 

Despite minimal variations from one exposure to the next, the true zodiacal background for each image is measured through an optimisation step to a precision of 0.01\,ADU to facilitate piston correction during stacking (the background is steady at the sub-ADU level from exposure to exposure due to the stable environment). This is achieved using a dedicated \ac{CFHT}-MegaCam {\tt Elixir}-LSB pipeline with a sigma-clipped mean function, developed for the Next Generation Virgo Survey \citep{Ferrarese2012}. The median value of the background levels from the input images is reintegrated into the final extended-emission science stack, aiding in the tracking of noise properties. Sigma-clipping is set to be more stringent for NISP than VIS to eliminate residual persistence. The weight map produced by {\tt SWarp} for the compact-sources stack serves as a proxy for the extended-emission stack, and the FITS {\tt GAIN} keyword is duplicated from the compact-sources stack to maintain consistency.

\subsection{Resampling}
The resampling step included the adjustment of flux scaling to achieve our target photometric \ac{ZP}, as outlined in the preceding section, utilising {\tt SWarp}'s {\tt FSCALE\_DEFAULT} option.
A crucial decision in this process is the selection of the interpolator.
Given the low number of overlapping exposures ($\le 4$), interpolating functions with support smaller than the destination pixel, such as those used by Drizzle \citep{Fruchter2002}, are impractical here.

For VIS we opted for the {\tt Lanczos3} kernel as the interpolant so as to preserve sharpness as much as possible -- a requirement driven by the \ac{ERO} science -- and to minimise correlations and moir\'e effects in the background noise. Through this process, the VIS \ac{FWHM} changes from \ang{;;0.136} in the unresampled image to \ang{;;0.158} post-resampling and stacking, as measured by {\tt PSFex} \citep{Bertin2011}. The VIS channel is significantly undersampled, and the bandwidth-preserving quality of {\tt Lanczos3} interpolation is not immune to aliasing. Consequently, the precision of certain measurements on VIS image stacks, such as \ac{PSF} fitting, may be compromised by aliasing in the brightest sources. However, in our tests we found that the slight improvement in resolution over other resampling methods -- for example a bilinear kernel -- in the resulting stacks outweighed the negative effect of aliasing residuals, making it a suitable compromise for \ac{ERO} science.

\begin{figure}[th]
    \centering
    \includegraphics[width=1.0\hsize]{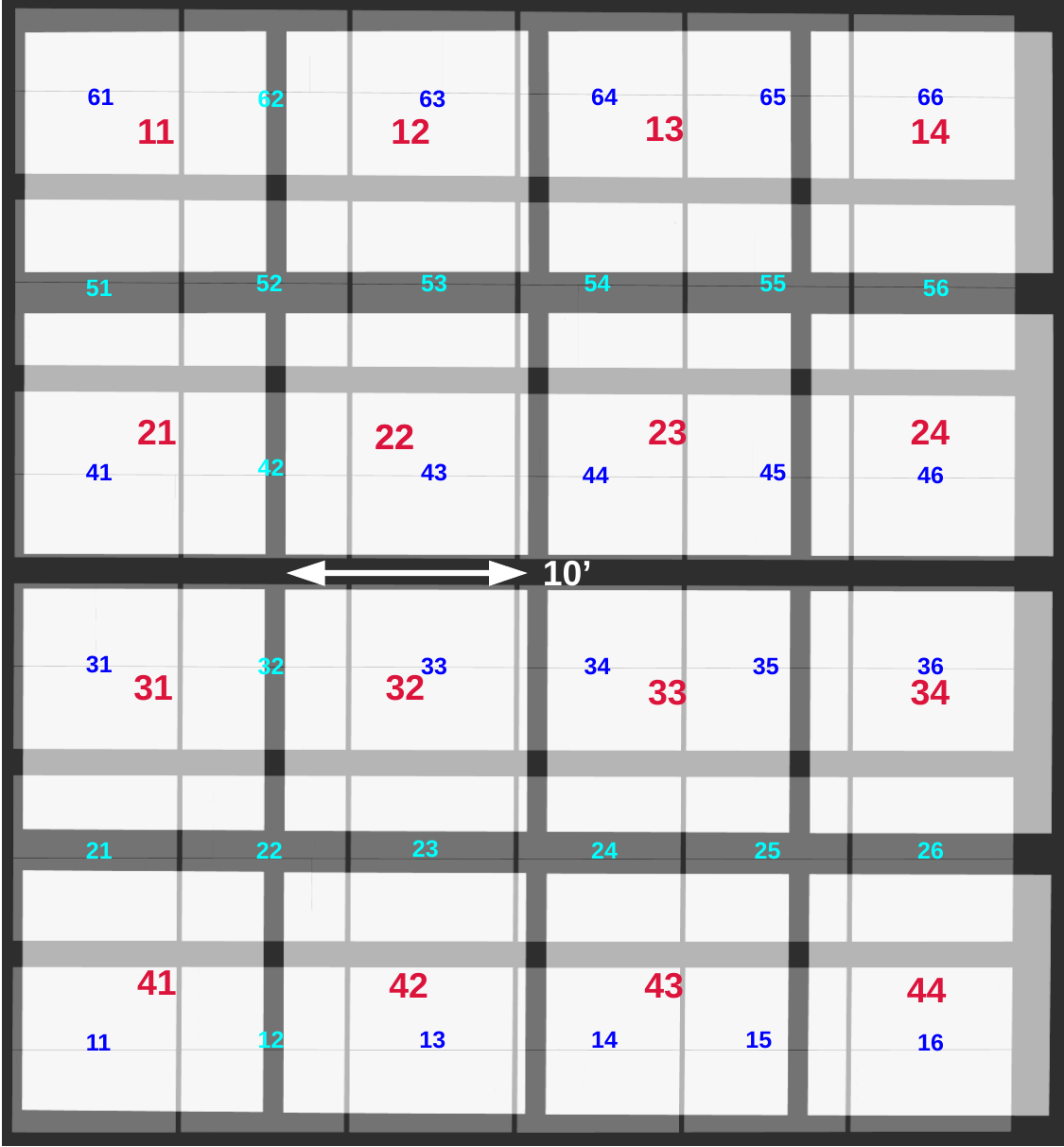}
    \caption{Common \ac{FOV} of VIS and NISP, generated from two simultaneously taken and VIS and NISP images, with common astrometry applied. 
    Small blue numbers refer to VIS and large red numbers to NISP detector IDs, with interchip gaps being evident. The VIS detectors show an additional thin horizontal gap from the charge-injection lines. The respective spatial and angular offsets between both instruments are \mbox{\ang{;;52.5}} and \ang{0.078}. Figure from \citet{EuclidSkyOverview}.}
    \label{fig:vis_nisp_fov}
\end{figure}

For NISP images, employing Lanczos interpolation would lead to excessive artefacts due to the more pronounced undersampling and the prevalence of cosmic rays predominantly affecting individual pixels. Consequently, we used bilinear interpolation instead. The NISP \ac{FWHM} then varies from \ang{;;0.42} to \ang{;;0.49} on average across the \YE, \JE, and \HE bands, as measured by {\tt PSFex}.

The pixel-to-pixel correlations resulting from resampling affect the amplitude of the background noise measured by {\tt SourceExtractor}.
Utilising {\tt SkyMaker v4} simulations \citep{Bertin2009a, Carassou2017}, we estimate that {\tt SourceExtractor} underestimates the RMS amplitude of the background noise on larger scales by a factor of approximately 1.32 for the VIS channel and 1.69 for the NISP. This assessment is based on the assumption that the input readout and photon noise are perfectly white, and that an infinite number of images, uniformly distributed in position, contribute to the stack.

\subsection{Coaddition}
The culmination of stack processing that impacts the pixels involved creating a common-\ac{FOV} mask for both the VIS and NISP instruments, retaining only pixels containing valid sky data in both stacks. Although VIS and NISP share a common \ac{FOV} as shown on \cref{fig:vis_nisp_fov}, the differing geometries of their detector mosaics and the size of the gaps result in distinct final stack outlines for each instrument. The binary mask aligns both instruments to a uniform standard. It is applied not only to the stacks but also to their respective weight maps. 

The process does not attempt to create clean edges around the images; instead, the final outline is determined solely by the combination of the dithering pattern, the sky orientation from L2 at the time of observation, and the footprint of each mosaic. In the FITS files, masked pixels are assigned the value {\tt NaN}.
In the final stage of processing, crucial FITS keywords are adjusted to accurately reflect the characteristics of the data (e.g. {\tt SATURATE}, {\tt GAIN}), alongside other miscellaneous information such as the origin of the data, and details about the pipeline and its operation.

\section{\label{sec:photometry}Photometric calibration}

\begin{figure*}[t]
\includegraphics[angle=0,width=1.0\hsize]{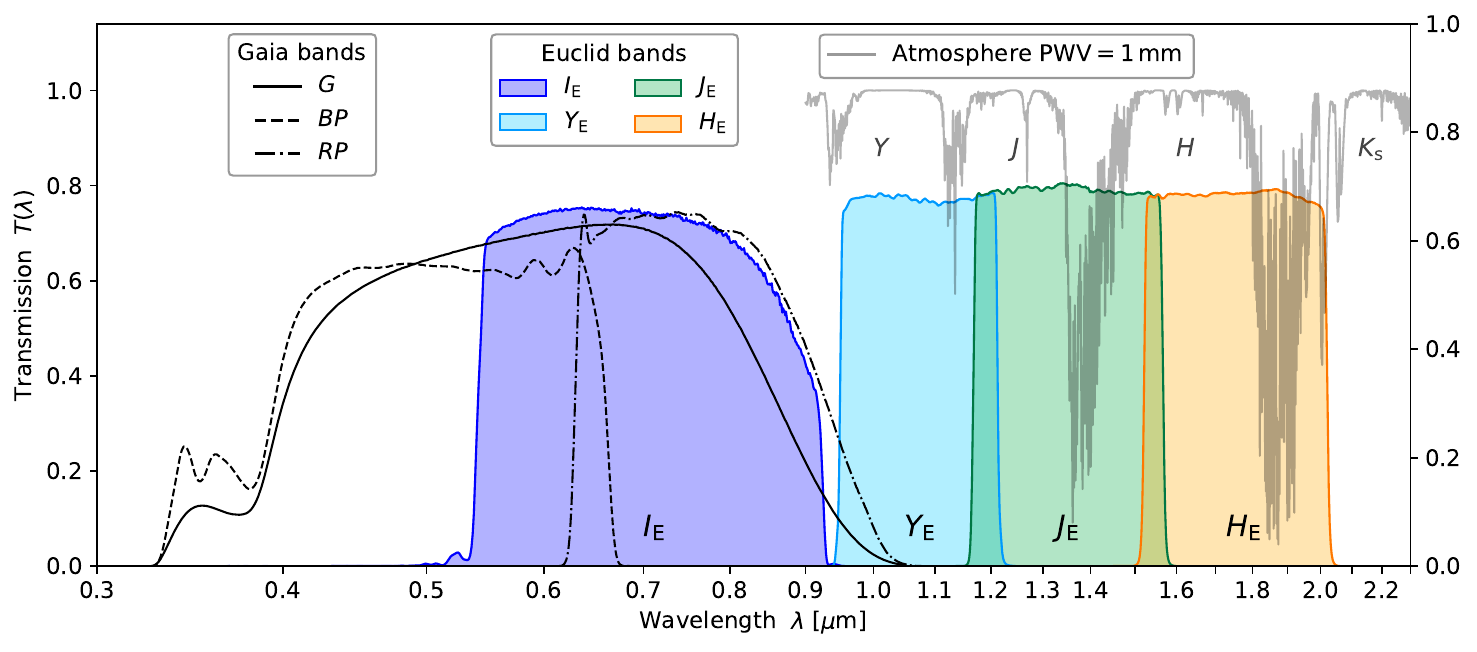}
\caption{Spectral response of \Euclid's wide imaging bands. For comparison we show the \textit{Gaia} bands that we used -- among others -- for photometric calibration of the VIS \IE-band, and the atmospheric transmission that limits the ground-based $Y$, $J$, $H$, and $K_{\rm s}$ bands, for a precipitable water vapour level of 1.0\,mm from \cite{rothman2013}. Considerable colour terms may arise for individual sources when transforming fluxes in the wide \Euclid bands to fluxes in other photometric systems.}
\label{fig:euclid_passband_comparison}
\end{figure*}

\begin{figure*}
\includegraphics[clip=,trim=0 0 65 0, height=0.31\textwidth]{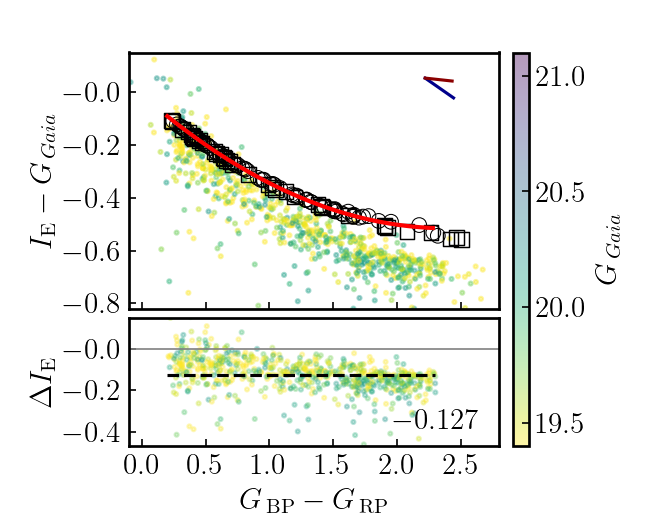}
\includegraphics[clip=,trim=0 0 65 0, height=0.31\textwidth]{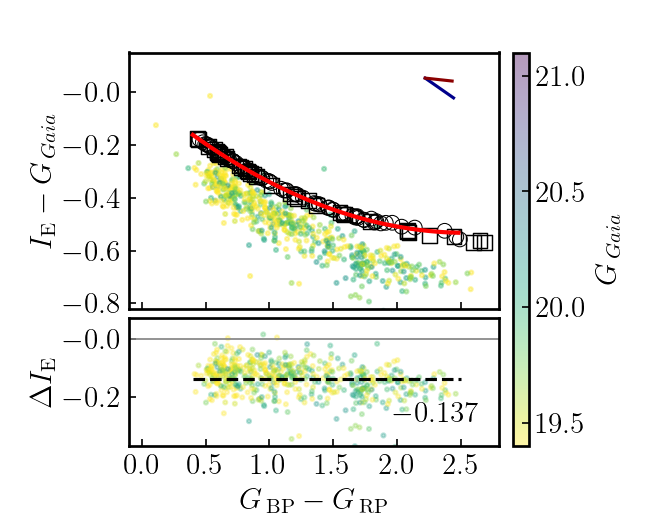}
\includegraphics[clip=,trim=0 0 0 0, height=0.31\textwidth]{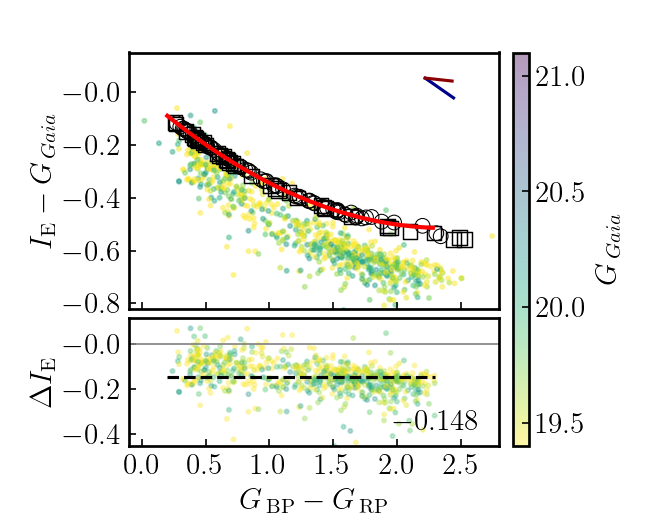}
\caption[]{Calibration of $\IE$ against \textit{Gaia}. The upper panels display the colour transformation between \textit{Gaia}-DR3 photometry (converted to the AB magnitude system) and $\IE$ for Milky-Way stars in the ERO fields towards Dorado (\textit{left panel}), Perseus (\textit{middle}) and NGC\,2403 (\textit{right}); in the lower sub-panels, the residual difference between the predicted and observed $\IE$ photometry is plotted. The coloured symbols represent observations, as determined from stacked images using photometry in a large aperture and the pre-launch \ac{ZP}, that is a stack \ac{ZP} of 30 as described in Sect.~\ref{sec:photometry}. An average foreground extinction is applied to the synthetic photometry, as adequate for each line of sight; the empirical data have not been dereddened. Two extinction vectors are shown in each figure, both for $A_V=0.5$, with one appropriate for the bluest stars and one for the reddest ones. Coincidentally the extinction vector is almost aligned with the colour-transformation curve at all colours; as a result extinction corrections of up to $A_V=1$ change $I_{\rm E,obs} - I_{\rm E,pred}$ by less than 0.01\,magnitude. }
\label{fig:Gaia2IE}
\end{figure*}

The initial flux calibration of the \ac{ERO} data is based on pre-launch instrument and telescope parameters, and it is followed by validation against external catalogues. The pre-launch \acp{ZP}, denoted as $\mathrm{ZP}_0$, are documented for NISP in \citet{Schirmer-EP18} and calculated similarly for VIS. Their respective values are 25.72, 25.04, 25.26, and 25.21 for $\IE$, $\YE$, $\JE$, and $\HE$, representing the AB magnitudes that correspond to a signal of 1\,e$^-$\,s$^{-1}$, for a source with a frequency-flat \ac{SED}. The \acp{ZP} in user-ready  \ac{ERO} data correspond to the AB magnitudes for 1 post-processing ADU. Their initial values depend on the gain $g$ and exposure time $\tau$ for each detector. For VIS, with a typical $g=3.5$\,electrons\,ADU$^{-1}$ and an exposure time $\tau=566$\,s, 1 initial ADU corresponds to a magnitude of 
$25.72 - 2.5 \log_{10}(g/\tau) =31.24$.
For the \ac{ERO} pipeline we adopted an arbitrary $\mathrm{ZP}=30$ for all images at this stage; the pixel values were rescaled accordingly, with a corresponding adjustment of relevant image-header keywords such as the gain and saturation values. 

The stacking procedure, which typically averages four images, has no effect on the \ac{ZP} but does affect the effective gain. The gain is not constant over a stack, for instance because the number of input images available per output pixel varies as a result of gaps between detectors and masked pixels. An effective gain is computed by {\tt Swarp} and stored in the image header. {\tt Swarp} also provides a weight map that incorporates information on the number of images used at each pixel. For photometric measurements, 
{\tt SourceExtractor} automatically adjusts the effective pixel gain using the input weight map (with $\text{WEIGHT\_GAIN} = \text{Y}$, the default), assuming that the pixel gain scales with the weight.

The initial photometric calibration of the stacks was verified against external catalogues to adjust \acp{ZP} where necessary. The availability of external catalogues for this purpose varies depending on the observed area. The all-sky \textit{Gaia} data are adequate for the validation of $\IE$ photometry (see \cref{fig:euclid_passband_comparison}). Additionally, data from the \ac{DES} were used \citep[their DR2;][]{Abbott+2021}, as well as Pan-STARRS 3pi Survey DR2 data from MAST \citep[PS1;][]{Chambers2016}; individual \ac{ERO} projects also made use of data from 
other surveys, as available. The process is illustrated below for the \acp{FOV} of \ac{ERO} Dorado, which overlaps with \ac{DES} and \ac{ERO} Perseus, which overlaps with PS1.

The validation procedure involved comparing the magnitudes of non-saturated stars with those predicted from external catalogues, using transformations calculated from tables of synthetic stellar photometry. Initially, we identified the types of stars present within the relevant magnitude range along the line of sight towards the \ac{ERO} field. This identification relied on the Besan\c{c}on model of the Galaxy \citep{Robin+2003,Czekaj2012,Lagarde+2021},\footnote{\url{https://model.obs-besancon.fr}} which also underpinned the pre-launch sky simulations by the Euclid Consortium \citep{EP-Serrano}. VIS saturates near $\IE=18.5$\,AB\,mag. For $\IE$ magnitudes between 18.5 and 24, main sequence stars overwhelmingly predominate. Stars near the turn-off, being intrinsically brighter, are generally found in this magnitude range when they reside far out in the halo, thus being metal-poor and $\alpha$-enhanced; as we move to intrinsically redder and fainter stars along the main sequence, we encounter more disc dwarf stars with typically Solar metallicities and $[\alpha/\mathrm{Fe}]$ ratios. This distribution was taken into account when selecting relevant stellar spectra for synthetic photometry,
thus reducing the dispersion in synthetic colour-colour diagrams compared to combining stars of all types. The selection can also aid in calculating suitable colour-dependent extinction coefficients.

Two libraries of stellar \acp{SED} were employed: (i) a collection based on the semi-empirical BaSeL 2.2 library by \citet{Lastennet+2002}, corresponding to the library described in \citet{EP-Serrano}; and (ii) a collection of theoretical stellar spectra generated with the \texttt{Phoenix} stellar atmosphere and radiative transfer code by \citet{Husser+2013}, previously utilised by \citet{Powalka+2016_fiducials} for a similar objective. These libraries are referred to as {\tt SSED} and {\tt TSED}, respectively, with only the latter allowing for an explicit consideration of the trend in $[\alpha/\mathrm{Fe}]$. The selection process is demonstrated for two \ac{ERO} fields in Appendix~\ref{apdx:DoradoFluxcal}.

\begin{figure*}
\includegraphics[trim=190 90 195 75, width=\linewidth]{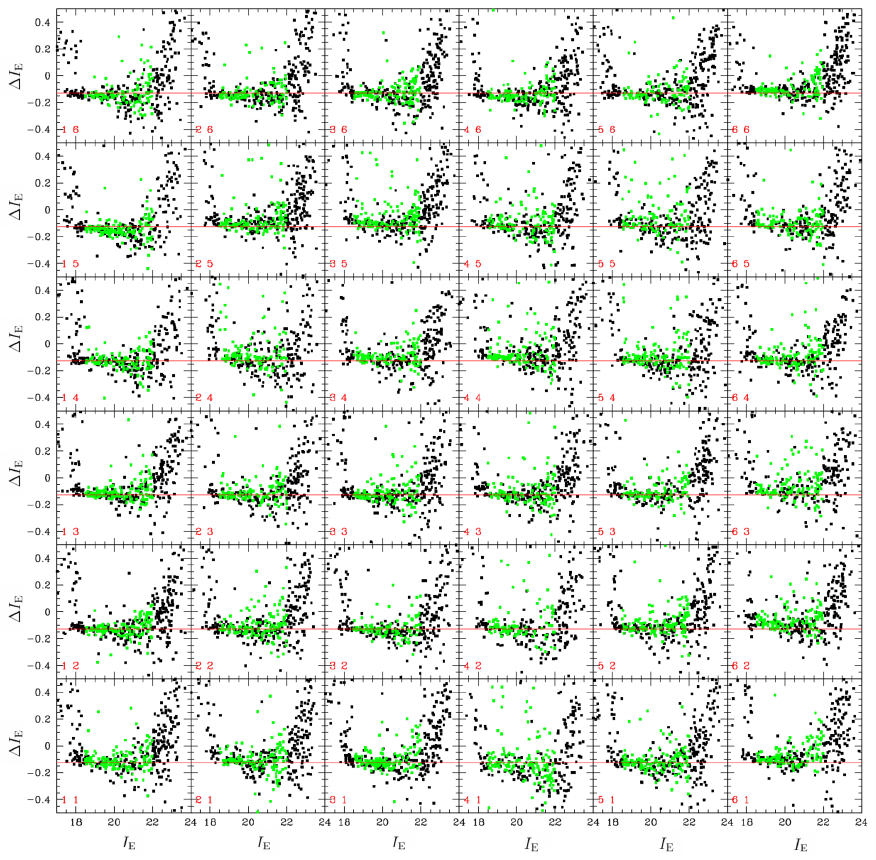}
    \caption{Illustration of the $\IE$ magnitude offset with respect to pre-launch estimates of the \ac{ZP}, per detector, using one exposure on the Perseus galaxy cluster. The horizontal line indicates the average offset, while the green symbols represent stars with a \textit{Gaia} colour of $(G_{\rm BP}-G_{\rm RP})<1.8$. These stars were specifically used to calculate the average offset, highlighting the variations across different detectors and emphasising the impact of stellar colours on the observed magnitudes.
    \label{fig:photcal_perDet}}
\end{figure*}

\subsection{VIS photometric calibration}
\label{subsec:VISphotcal}

\textit{Gaia} magnitudes are particularly well-suited for predicting $\IE$ magnitudes due to their accuracy and uniformity across the sky, as well as the low dependence of the theoretical transformation from \textit{Gaia} passbands to $\IE$ on the adopted \ac{SED} models. This compatibility is evident in Fig.~\ref{fig:euclid_passband_comparison} and in particular \cref{fig:Gaia2IE}, where the synthetic photometry from both the semi-empirical and theoretical libraries shows good overlap across a wide range of colours.\footnote{The colours of the models of \citet{Pickles1998} also align with those shown in \cref{fig:Gaia2IE}} Coincidentally the extinction vectors, which are colour-dependent in the broad $\IE$ filter, are essentially tangent to the expected stellar locus in plots of $\IE - G_{Gaia}$ versus $G_\mathrm{BP}-G_\mathrm{RP}$ at all relevant colours. The empirical data points in \cref{fig:Gaia2IE} were derived from $\IE$ aperture photometry in an aperture size of \ang{;;4.8} radius. According to our \ac{PSF} models, this size encapsulates more than 98\% of the light (\cref{tab:enclosedenergy}). We corrected for the small missing fraction when needing to accurately express the true brightness of observed stars.

The \textit{Gaia} data were acquired through a cross-match with \textit{Gaia}-DR3 using the CDS\footnote{Centre de Donn\'ees astronomiques de Strasbourg.} cross-match tool \citep{Boch+2012,Pineau2020} accessible via the {\tt Topcat} software \citep{Taylor2005}; the passbands for synthetic photometry were also based on \textit{Gaia}-DR3.\footnote{\textit{Gaia} transmission curves were obtained via the SVO service (cf. note \,\ref{footn:svo}) in September 2023 and verified to match those described at \url{https://www.cosmos.esa.int/web/gaia/dr3-passbands}} \textit{Gaia}'s photometry was initially presented in the Vega magnitude system. Our conversion from Vega to AB magnitudes utilises the model Vega spectrum sourced from the SVO filter transmission service\footnote{\url{http://svo2.cab.inta-csic.es/theory/fps/} \label{footn:svo}} \citep{Rodrigo2012,Rodrigo2020} in September 2023.\footnote{The SVO Vega spectrum was compared to various Vega model versions available at the Space Telescope Science Institute (\url{https://ssb.stsci.edu/cdbs/calspec/}), and was found to be identical to {\tt alpha\_lyr\_stis\_010} at wavelengths below $1\,\micron$ and within a few mmag of {\tt alpha\_lyr\_stis\_011} at longer wavelengths.}

\begin{figure*}
\begin{center}
\includegraphics[clip=,trim=150 0 100 0, height=0.30\textwidth]{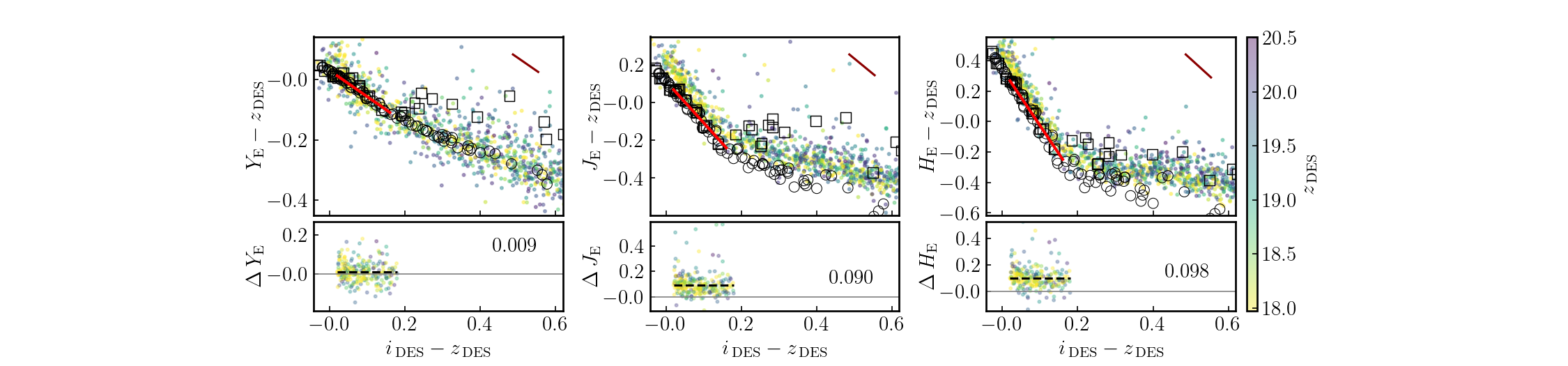}\\
\includegraphics[clip=,trim=150 0 100 0, height=0.30\textwidth]{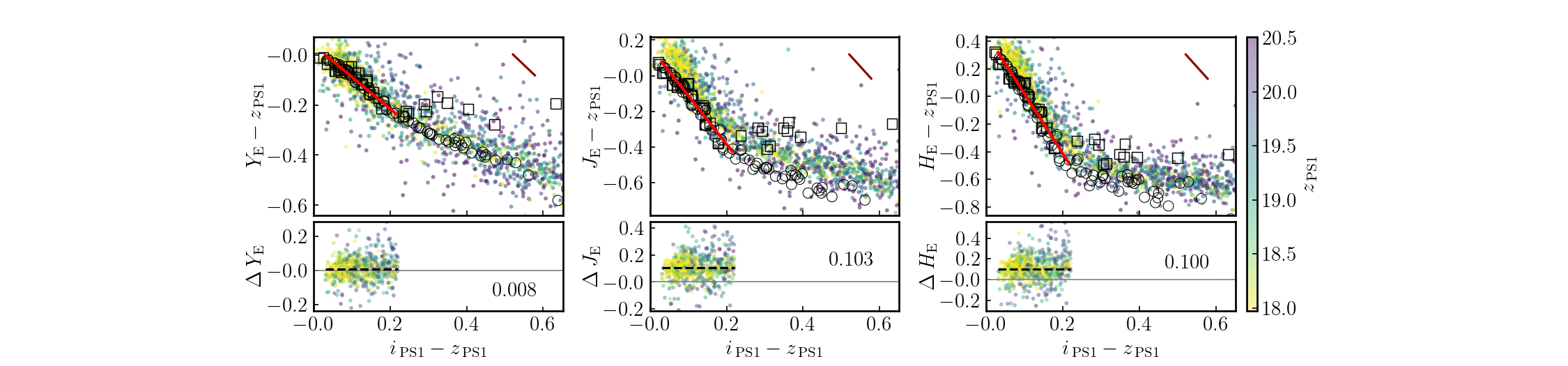}
\end{center}
\caption[]{Colour-transformation between \ac{DES} and \Euclid NISP photometry for Dorado (top), and between PS1 and \Euclid NISP photometry for Perseus (bottom). The figure setup is as in \cref{fig:Gaia2IE}. Extinction is applied to the synthetic magnitudes ($A_V=0.05$\,mag in the case of Dorado, $A_V=0.45$\,mag in the case of Perseus), and a typical extinction vector for $A_V=0.5$\,mag is also shown.}
\label{fig:EXTvsNISP}
\end{figure*}

The residuals observed vary with brightness and colour (\cref{fig:Gaia2IE}). This comparison was extended across several \ac{ERO} projects and we also employed PS1 as a reference in place of \textit{Gaia}, yielding similar outcomes. Factors contributing to these discrepancies include  nonlinearity within the VIS instrument, the \ac{PSF} dependence on magnitude known as the brighter-fatter effect \citep{antilogus2014}, possible changes of the in-flight passbands compared to pre-launch measurements, and the fact that flux calibration in broad filters inevitably has a colour dependence that a flat-field based on a single light source (zodiacal light in our case) cannot capture. Further analysis will be needed to disentangle the effects of brightness and colour in the ERO data and remove the residual trend. An average offset was calculated for \ac{ERO} data using stars within the range of $18.5\leqslant G_{Gaia} \leqslant 20$ (after conversion to AB magnitudes), amounting to $-0.13$\,mag, leading to the 30.13 \ac{ZP} reported in all final VIS stack FITS headers. While the residuals for most \ac{ERO} projects are centred around this value, some exhibit deviations as large as 0.05\,mag.

To further investigate the observed discrepancy between observed $\IE$ and the prediction based on Gaia, as well as the dispersion in that comparison, we analysed the residuals by segregating them according to each detector in the VIS mosaic. \Cref{fig:photcal_perDet} depicts such an analysis. The horizontal red line represents the average \ac{ZP} offset for the entire mosaic. It is observed that the \acp{ZP} vary substantially across different detectors. An examination of the 16 individual images reveals that in some cases the \ac{ZP} is not uniform even within a single detector. This variation is likely attributable to drift among the four amplifiers used in each detector. We further find a lack of uniformity from one image to the next, underscoring the complexity of achieving consistent photometric calibration across the mosaic. Calibrations and observations conducted during the \ac{PV} phase (concurrent with the ERO observations) and described in \cite{EuclidSkyOverview} demonstrate that, with corrections in place for effects not considered here (including the brighter-fatter effect, optical ghosts, non-linearity, and the illumination correction), the Euclid surveys meet the relative photometric requirement of better than $1\%$.

Another factor contributing to spatial and temporal photometric variations is the contamination of optical surfaces with water ice from molecular outgassing \citep{Schirmer-EP29}. It is known that \Euclid experiences outgassing effects in particular in the VIS optical path, resulting in time-variable throughput losses that may also be spatially variable. A thermal decontamination of the affected mirror was successfully undertaken in March 2024 to restore the transmission to immediate post-launch values. Directly calibrating VIS images taken at different times against \textit{Gaia} removes at least the global flux-scale offsets.

\subsection{NISP photometric calibration}
\label{subsec:NISPphotcal}

An initial evaluation of the fluxes from non-saturated NISP point sources was conducted using 2MASS photometry \citep{cutri2003}, following the linear transformation equations detailed in appendix~C of \citet{Schirmer-EP18}. A limitation of this approach is the restricted dynamic range for comparison, spanning only about 1 magnitude. Specifically, the faintest sources in 2MASS approach the saturation limit in NISP. There is a broader overlap with more recent \ac{NIR} surveys that reach depths between those of 2MASS and \Euclid; these surveys typically base their calibration on 2MASS, offering a more extended range for comparison and potentially enhancing the accuracy of the NISP flux comparison.

\begin{figure*}
	\includegraphics[width=\linewidth]{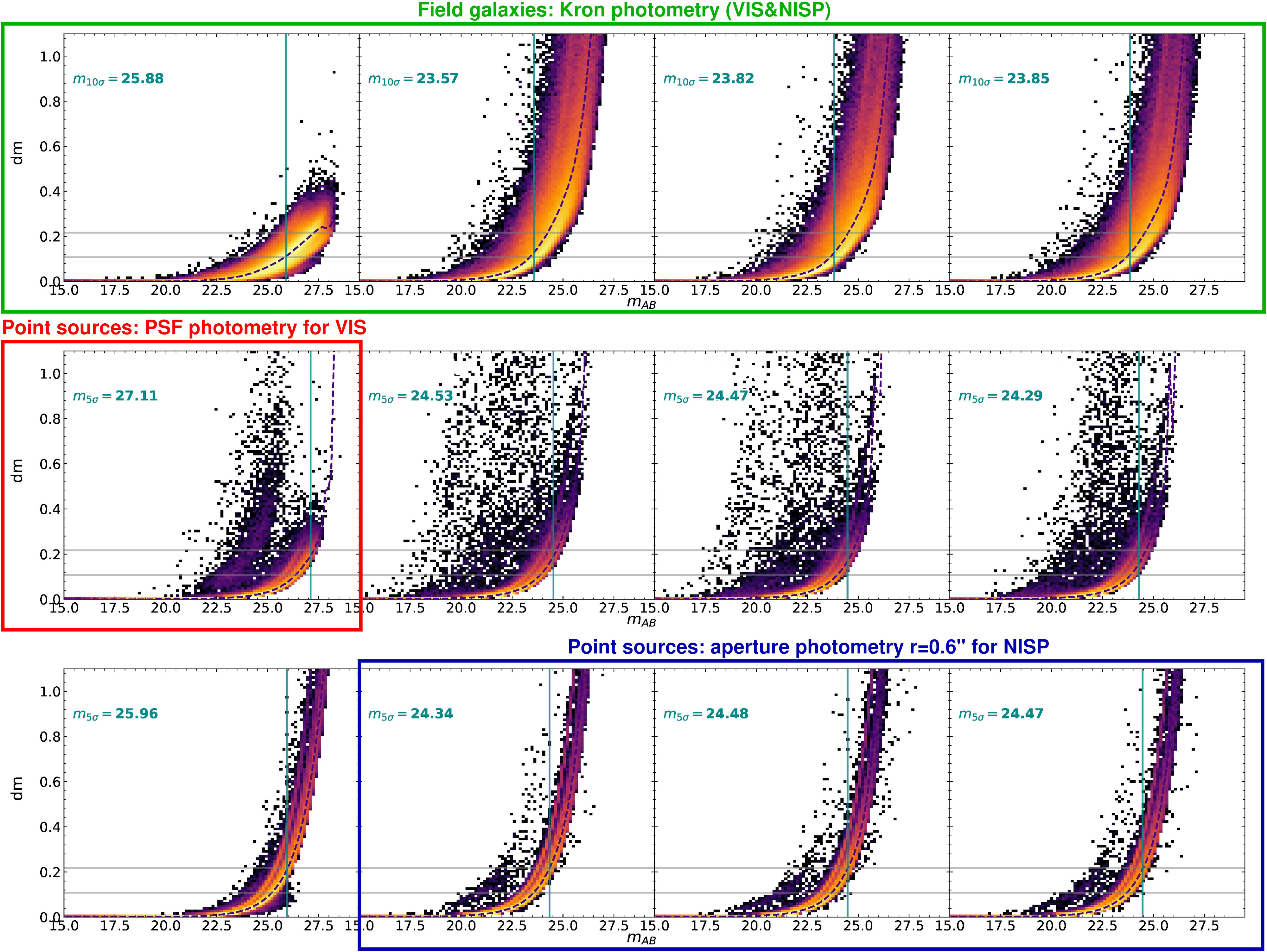}
    \caption{Depth metrics  for evaluating the \ac{ERO} catalogues' performance, based on the photometric measurement error (or `dm') reported by {\tt SourceExtractor}. The plot order from left to right is \IE, \YE, \JE, and \HE. \textit{Top row}: {\tt SourceExtractor}'s {\tt MAG\_AUTO} (Kron magnitude) provides the most accurate estimate of the total magnitude for galaxies in both VIS and NISP, using a conservative threshold of ${\rm {\tt CLASS\_STAR}}=0.65$ for star-galaxy separation (statistics from the \ac{ERO} Abell\,2390 field).
    \textit{Middle row}: For stars, the VIS resolution facilitates effective separation of objects using {\tt SourceExtractor}'s \ac{PSF}-fitting photometry ({\tt MAG\_PSF}). However, this method is less effective with NISP, which tends to blend objects because of its coarser pixel scale, as evidenced by the scatter in comparison to VIS results; it also underestimates the photometric error on compact sources compared to aperture photometry (statistics from the \ac{ERO} NGC\,6397 field, also used on the bottom plot).
    \textit{Bottom row}: A \ang{;;1.2} diameter aperture is used for all objects in NISP ({\tt SourceExtractor}'s {\tt MAG\_APER}), with the fitting tool selecting the deepest part of the distribution on the right, primarily composed of stars. An aperture correction is then applied to the numbers derived from such analyses, based on our analysis of enclosed energy (offsets: $\YE=-0.130$, $\JE=-0.167$, and $\HE=-0.249$; refer to \cref{tab:aperoffset} for the aperture corrected values for NGC\,6397 here).
    \label{fig:DepthMetric}}
\end{figure*}

The top row of \cref{fig:EXTvsNISP} presents a comparison with photometry from the \ac{DES} along the line of sight towards Dorado. Notably, for cool stars, significant discrepancies are seen between the synthetic photometry produced by different spectral libraries. The {\tt SSED} library exhibits a problematic junction between the optical and \ac{NIR} wavelength ranges, resulting in a step-like anomaly near 1\,\micron~in some spectra. Conversely, the {\tt TSED} library demonstrates more consistent behaviour; however, like other theoretical libraries, it does not perfectly align with empirical \acp{SED} below approximately 4500\,K, as noted by \citet{Lancon+2021}. Near the hook that appears in the empirical stellar sequence in some of the colour-colour diagrams, the {\tt TSED} library shows that the expected behaviour depends strongly on metallicity and surface gravity, adding uncertainty unless detailed stellar parameters are exploited. Consequently, transformation equations between systems are derived from, and applied exclusively to warmer stars, where model-dependence is minimal. Extinction effects are included but negligible in these colour planes towards Dorado, compared to differences between models.

To determine if systematic uncertainties in \ac{DES} colours (represented on the $x$-axis) contribute to the observed discrepancy between predicted and observed NISP magnitudes, comparisons were also made using $(i_{\mathrm{DES}} - y_{\mathrm{DES}})$ and $(y_{\mathrm{DES}} - z_{\mathrm{DES}})$. The residuals were consistent across these comparisons, and showed an offset of  $-0.05$ in $\YE$, and of approximately $+0.1$ in $\JE$ and $\HE$. 

The lower row of \cref{fig:EXTvsNISP} considers the data of the Perseus \ac{ERO} project, and provides a comparison between NISP photometry and PS1 data. In this region, extinction plays a more significant role with typical values around $E(B-V)= 0.15$\,mag, exhibiting spatial variability by approximately a factor of 1.8 across the field \citep{EROPerseusDGs}. Furthermore, the extinction varies with distance for stars within the Galactic disc. This contributes to dispersion but mostly at the red end of the stellar colour distribution, which we avoid in the calibration procedure. The most significant effect is seen for \HE, where an increase of the adopted $A_V$ by 0.1\,mag decreases the offset between predicted and observed \IE by 0.011\,mag. With $A_V=3.1 E(B-V) = 0.45$\,mag, the average deviation between predicted and observed magnitudes in Perseus is negligible for \YE, and approximately $+0.1$\,mag for \JE and \HE. This analysis was supplemented with a consistency check using galaxy magnitudes, converted to \HE from the 2MASS galaxy catalogue, yielding similar discrepancies. Because the average offset values were finalised after the images for first \ac{ERO} data-release were frozen for science production and distribution as part of the first \ac{ERO} data release presented in this paper, the photometric zero points in those first-release NISP image FITS keywords are not updated;\footnote{All the \ac{ERO} teams were informed, however.} we recommend to subtract 0.1\,mag from the header \acp{ZP} of the $\JE$ and $\HE$ stacked images, and to subtract 0.1\,mag to all $\JE$ and $\HE$ magnitudes in the ERO catalogues delivered with this first release. The next \ac{ERO} public data release will integrate this adjustment. For consistency in the first ERO public data release, \cref{tab:summaryall} summaries the dataset's general properties, listing the zero points as found in the FITS headers and adopted for catalog production.

\begin{table*} [htbp!]
\begin{center}
    \caption{Aperture-magnitude corrections for the \ac{ERO} catalogues. These values should be applied to the catalogue entries for the relevant {\tt MAG\_APER} (from the second to the tenth listed here, with the first not being relevant).}
    \label{tab:aperoffset}
    \centering
    \begin{tabular}{lccccccccc}
    \hline\hline
    \noalign{\vskip 1pt}
    Catalogue aperture & 2  & 3 & 4 & 5 & 6 & 7 & 8 & 9 & 10\\
    \hline
    \noalign{\vskip 2pt}
    Aperture diameter [arcsec]  & 2.0 & 0.3 & 0.6 & 1.2 & 2.4 & 4.8 & 9.6 & 5.85 & 8\\
    VIS diameter [pixels  & 20  & 3 & 6 & 12  & 24 & 48  & 96 & 58.5 & 80\\
    NISP diameter [pixels]  & 6.7 & 1 & 2 & 4 & 8 & 16 & 32 & 19.5 & 26.7   \\ 
    \hline
    \noalign{\vskip 1pt}
    \IE offset & $-0.039$ & $-0.604$ & $-0.265$ & $-0.120$ & $-0.062$ & $-0.034$ & $-0.017$ & $-0.013$ & $-0.009 $\\
    \YE offset& $-0.050$ & $-1.748$& $-0.418$ & $-0.130$ & $-0.076$ & $-0.043$ & $-0.025$ & $-0.021$ & $-0.016 $\\
    \JE offset& $-0.062$ & $-1.827$& $-0.520$ & $-0.167$ & $-0.098$ & $-0.053$ & $-0.030$ & $-0.025$ & $-0.019 $\\
    \HE offset& $-0.070$ & $-1.962$& $-0.653$ & $-0.249$ & $-0.120$ & $-0.059$ & $-0.032$ & $-0.027$ & $-0.020 $\\
    \hline
    \end{tabular}
\end{center}
\end{table*}

Overall, the derived \acp{ZP} align closely with pre-launch expectations, showing variations within approximately 0.1\,mag across different \ac{ERO} projects.  Such variations are entirely plausible, considering that the \ac{ERO} pipeline has not yet undergone nonlinearity calibration or photometric flat-fielding adjustments. Furthermore, the absolute quantum efficiency of the NISP detectors carries an uncertainty of about 5\%, due to an entanglement with the detector gain \citep{Waczynski16,secroun2018} among other factors. Finally, the definition of the total stellar flux, encoded in aperture corrections, may differ between surveys. Subsequent analysis of photometry in the three NISP bands within the \ac{ERO} data framework did not uncover any significant inconsistencies in this calibration, staying within the broader uncertainties inherent to the \ac{ERO} pipeline process.

In summary, the uncertainty of the absolute flux calibration of the VIS data after calibration against \textit{Gaia} fluxes, is better than 1\% on average, with residual trends versus colour and brightness with joint amplitude of the order of 10\%.  In NISP, the absolute calibration is limited to about 10\% by the stronger model-dependence of the colour-transformations to reference data, and a limited calibration set in the early months of observations.

\section{\label{sec:cats}Compact source catalogues and general performance of the \texorpdfstring{\ac{ERO}}{ERO} data set}

\subsection{PSF modelling}

For photometric analyses, a \ac{PSF} model is essential for each image stack.
The \ac{PSF} model is computed using {\tt PSFEx} that employs a super-resolution algorithm along with a simple regularisation scheme to map the \ac{PSF} at sub-pixel resolution. To minimise the contribution from spurious Nyquist modes, we choose a \ac{PSF} sampling step of 0.45 pixels.

The \Euclid \ac{PSF} shows remarkable homogeneity across the \ac{FOV}, aligning with the stringent requirements set for the instrument.
A second-degree polynomial adequately captures the PSF's subtle variations across the entire image.
As anticipated, the precision of the \ac{PSF} models derived from the stacks is limited by aliasing and image resampling, with these effects being particularly pronounced in the NISP data.

Despite these challenges, the typical $\chi^2$ per degree-of-freedom value for \ac{PSF} models fitted on bright point sources in VIS stacks remains at a reasonable level of 1.7. However, for the NISP stacks, this value escalates to around 10, indicating a higher level of deviation from the ideal \ac{PSF} model, primarily due to undersampling.

\begin{table*}[ht!]
\begin{center}
    \caption{General median properties of the 17 \Euclid \ac{ERO} fields, with a 0.60\,deg$^2$ sky coverage for single \ac{ROS} observations. \Cref{table:bigtable1} and \cref{table:bigtable2} list all properties per ERO field. All magnitudes are instrumental and in the AB system. The range of the sky background reflects the ecliptic latitudes covered across the ERO fields.}
    \label{tab:summaryall}
    \centering
    \begin{tabular}{lccccl}
    \hline
    & & & & & \\[-9pt]
    Band & \IE & \YE & \JE & \HE & Unit / note \\
    \hline
    & & & & & \\[-8pt]
    \ac{FWHM} & 0.157 & 0.477 & 0.486 & 0.492 & arcsec \\
    \ac{FWHM} & 1.57 & 1.59  & 1.62 & 1.64 & pixel \\
    Pixel sampling & 0.1 & 0.3 & 0.3 & 0.3 & arcsec\,pixel$^{-1}$\\
    Depth 5\,$\sigma$ stars & 27.12 & \textsl{24.45} & \textsl{24.60} & \textsl{24.52} &mag (PSF or \textsl{aperture} photometry)\\
    Depth 10\,$\sigma$ galaxies & 25.29 & 22.97 & 23.22 & 23.31 &mag (Kron magnitude) \\
    Limiting surface brightness & 29.9 & 28.2 & 28.4 & 28.4 & mag\,arcsec$^{-2}$, $\ang{;;10}\times\ang{;;10}$ scale at 1\,$\sigma$\\
    Sky background & 22.3 & 22.1 & 22.2 & 22.3 & mag\,arcsec$^{-2}$, range: [-0.5, +0.5]\\
    Photometric zero point & 30.13 & 30.0 & 30.0 & 30.0 & mag, ADU (first ERO public data release)\\ 
    Astrometry (internal) & 5.1 & 14.9 & 14.9 & 14.9 & mas \\
    Saturation & 20\,900 & 141\,000 & 107\,000 & 111\,000 & ADU\\ 
    Total number of sources &11\,359\,274 &5\,257\,808 & 5\,257\,808 &5\,257\,808 & All EROs (common detection on NISP) \\
    \hline
    \end{tabular}
\end{center}
\end{table*}

\subsection{Production of the science validation catalogues}

Rich photometric catalogues were produced using {\tt SourceExtractor} on both the VIS and NISP compact-sources stacks, utilising inputs such as \acp{PSF} derived by {\tt PSFex}, the weight maps, the NISP ${\chi}^2$ detection image created by {\tt SWarp}, and the updated photometric \acp{ZP} as previously described. Source extraction for the VIS \IE-band catalogue was performed in stand-alone mode, benefiting from its higher depth and resolution. The deep nominal detection image produced by combining the three NISP bands led to a uniform source extraction approach across all three bands and a consistent segmentation map from the three {\tt SourceExtractor} runs for NISP for a given field. The \ac{ERO} collection primarily includes 34 stacks and 17 weight maps and catalogues for VIS, together with 102 stacks and 51 weight maps and catalogues for NISP.

The \ac{ERO} catalogues include model-fitting measurements (e.g. {\tt SourceExtractor}'s {\tt MODEL} / {\tt SPHEROID} / {\tt DISK} / {\tt POINTSOURCE}), totalling 163 requested parameters, with several generating multiple columns in the photometric catalogue. Consequently, the \ac{ERO} science-validation catalogues feature a total of 363 parameters per identified source. \Cref{table:erocatalogparams} in \Cref{apdx:EROCatalogParameters}  lists all input parameters along with their descriptions and units. 

{\tt SourceExtractor} was configured with a detection and analysis threshold of $1.5\,\sigma$, bringing a comprehensive census of compact sources (stars and galaxies) within the images. The internal background subtraction utilised a mesh size of \ang{;;6.4}, mirroring the precise setting employed by {\tt SWarp} for the stacks. The catalogues provide a series of 10 aperture photometry measurements ({\tt MAG\_APER}). For VIS (\ang{;;0.1}\,pixel$^{-1}$) the diameters of the apertures in pixels are {20, 3, 6, 12, 24, 48, 96, 58.5, and 80}, spanning from the 2nd to the 10th aperture. For NISP, these pixel values are adjusted due to the different pixel scale of \ang{;;0.3}\,pixel$^{-1}$. The initial aperture should be disregarded because it was an exploratory measure based on \ac{FWHM} that did not yield useful insights. The 9th and 10th apertures align with certain apertures used in the \ac{DES} official products (\ang{;;5.8437}, and \ang{;;8}). The magnitude offsets that must be applied to the catalogues for each aperture to obtain the total magnitude are detailed per \Euclid band in \cref{tab:aperoffset}, stemming from an analysis of encircled energy in the \ac{PSF} discussed later in \cref{sec:extendedPSF}.

While individual \ac{ERO} teams typically developed photometric catalogues tailored to their specific scientific objectives, a collaborative effort among the six teams resulted in the establishment of a general format designed primarily for compact-sources science. These catalogues are both comprehensive and versatile, facilitating a wide spectrum of scientific research. Accompanying the stack images, these catalogues represent a crucial component of the \ac{ERO} public data release, enabling diverse astronomical studies.

These catalogues also served to assess the performance attained in each \ac{ERO} field as outlined in \cref{table:bigtable2}, leading to their designation as science-validation catalogues. To reproduce and verify the work, miscellaneous configuration files and derived data products from the {\tt AstrOmatic} tools ({\tt PSFex}, {\tt SourceExtractor}) used to produce the components of the final catalogues are provided as well. Among these data products is the segmentation map generated by {\tt SourceExtractor}, offering a comprehensive suite of resources for in-depth analysis and validation of the \ac{ERO} data-processing methodology.

In the first \ac{ERO} data release, there is no catalogue specifically dedicated to the photometry of extended sources (extension larger than \ang{;;20}). The creation of such a catalogue, based on the extended-emission stacks, is scheduled for inclusion in the next \ac{ERO} public data release.

\subsection{Performance summary of the ERO data set}
 
We used generic depth metrics widely adopted in the field to reflect the performance of the \ac{ERO} for compact sources based on the science-validation catalogues. The Kron magnitude, {\tt MAG\_AUTO}, by {\tt SourceExtractor} is a robust estimator for the total magnitude of extended sources. To ensure the estimated depth is based exclusively on galaxies, a strict {\tt CLASS\_STAR} value of 0.65 was used for the star-galaxy separation criteria by {\tt SourceExtractor}. We adopted a 10\,$\sigma$ detection limit above the background noise to derive meaningful physical parameters. For point sources, we first investigated PSF photometry for both VIS and NISP, exploiting the PSF produced by {\tt PSFex} in {\tt SourceExtractor}. The superior resolution of VIS leads to a clear separation of compact sources and galaxies, without the use of the {\tt CLASS\_STAR} criteria, and the depth is derived at the 5\,$\sigma$ level for solid detection and meaningful photometric measurement. This metric is not adequate for NISP due to the coarse sampling, which does not allow for an effective separation of stars and galaxies (even with {\tt CLASS\_STAR} $>0.9$). Instead, for robustness, we adopted a large photometric aperture of \ang{;;1.2} diameter (4 pixels) and derived the depth on the deepest edge of the population, ensuring we are probing only the point sources. \Cref{fig:DepthMetric} illustrates these depth metrics adopted for galaxies and point sources. The \ac{LSB} metric and performance of the ERO data set is described in \cref{sec:LSBmetric}.
\cref{table:bigtable1,table:bigtable2} in \cref{sec:datasummary} detail the essential characteristics of each \ac{ERO} field, with \cref{tab:summaryall} summarising the principal attributes (sky coverage, etc.) and performance metrics across all 17 fields.

\section{\label{sec:lsb}Performance for extended emission science}

\subsection{The \Euclid optical design and the ERO extended-emission science}

The selection of \Euclid's Korsch optical design \citep{korsch1977, Bougoin2018} was motivated by the goal of achieving an optimal \ac{PSF} for the primary gravitational lensing probe, focusing on maximising the energy concentration within the PSF's core across a broad \ac{FOV}. This optical design is instrumental in addressing spherical aberration, coma, astigmatism, and field curvature, and supporting a wide \ac{FOV} while reducing stray light. The effective minimisation of scattered light -- a direct result of the optical design complemented by advanced baffling on the instrument platform \citep{Gaspar2016} -- underscores the spacecraft and its instruments' exceptional capability to investigate the \ac{LSB} Universe. While this potential was not initially anticipated in \Euclid's science objectives, subsequent modelling and simulations have highlighted its significance \citep{Scaramella-EP1, Borlaff-EP16}. The in-flight science performance concerning the detection of extended emission is assessed in this section through detailed analysis of real data from \ac{ERO} projects that aim to explore \Euclid's capabilities for \ac{LSB} science in particular to their fullest extent \citep{EROLensData, EROPerseusOverview, ERONearbyGals, EROPerseusICL, EROPerseusDGs, EROFornaxGCs}.

The critical factor in assessing a telescope's ability to investigate ultra-faint extended emission, or faint contrasts against a noisy background, is the characterisation of its extended \ac{PSF}, as discussed in various works by \cite{Mihos2005}, \cite{Abraham2014}, \cite{Sandin2014}, \cite{Watkins2016}, \cite{InfanteSainz2020}, and \cite{Liu2022}. The extended \ac{PSF} quantifies the distribution of energy within the PSF's wings with respect to its core, often detectable up to tens of arcminutes away. The relevance to \ac{LSB} science is clear: a \ac{PSF} with significant energy in its extended wings, manifesting as diffuse light, will inherently compromise the telescope's capability to discern genuine signals of this nature. This section is dedicated to validating the premise that, for \ac{LSB} science at its limits, \Euclid data require no correction for the extended \ac{PSF} for VIS and NISP.

\begin{figure}
	\includegraphics[width=\linewidth]{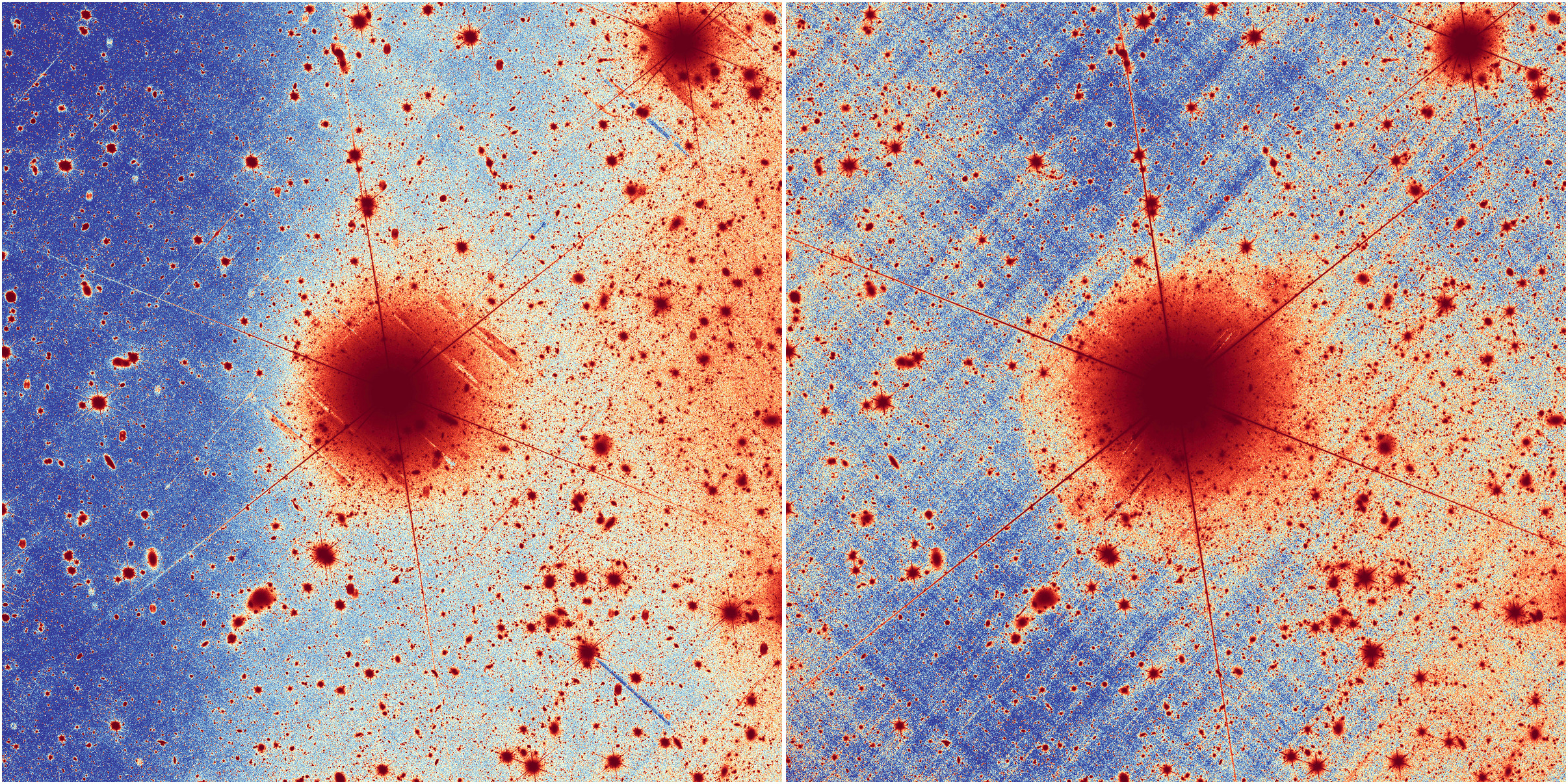}
    \caption{HD\,1973, a 5th magnitude star in the \HE-band ($\IE=6.3, \YE=5.5, \JE=5.3, \HE=4.9)$, is featured on a deep (3 ROS) \ac{ERO} stack in the \IE-band (left) and the \HE-band (right) within a $\ang{;23;}\times\ang{;23;}$ field. The extended halo of VIS (left) concludes smoothly, exhibiting no detectable signal beyond a \ang{;5;} radius. Conversely, the NISP extended halo (right) terminates more abruptly -- a characteristic of its refractive optics -- yet also without a detectable signal past a \ang{;5;} radius. The diffraction spikes for VIS extend outward to approximately \ang{;12;}, whereas for NISP in \HE-band they reach out to about \ang{;20;}. The gradient observed in the background of the VIS image is attributed to a true \ac{LSB} signal from Galactic cirrus intersecting the line of sight, demonstrating \Euclid's exceptional ability to discern faint contrasts at high \ac{SNR} across a broad field.
    \label{fig:VBPPhoenicis}}
\end{figure}

\begin{figure}
	\includegraphics[width=\linewidth]{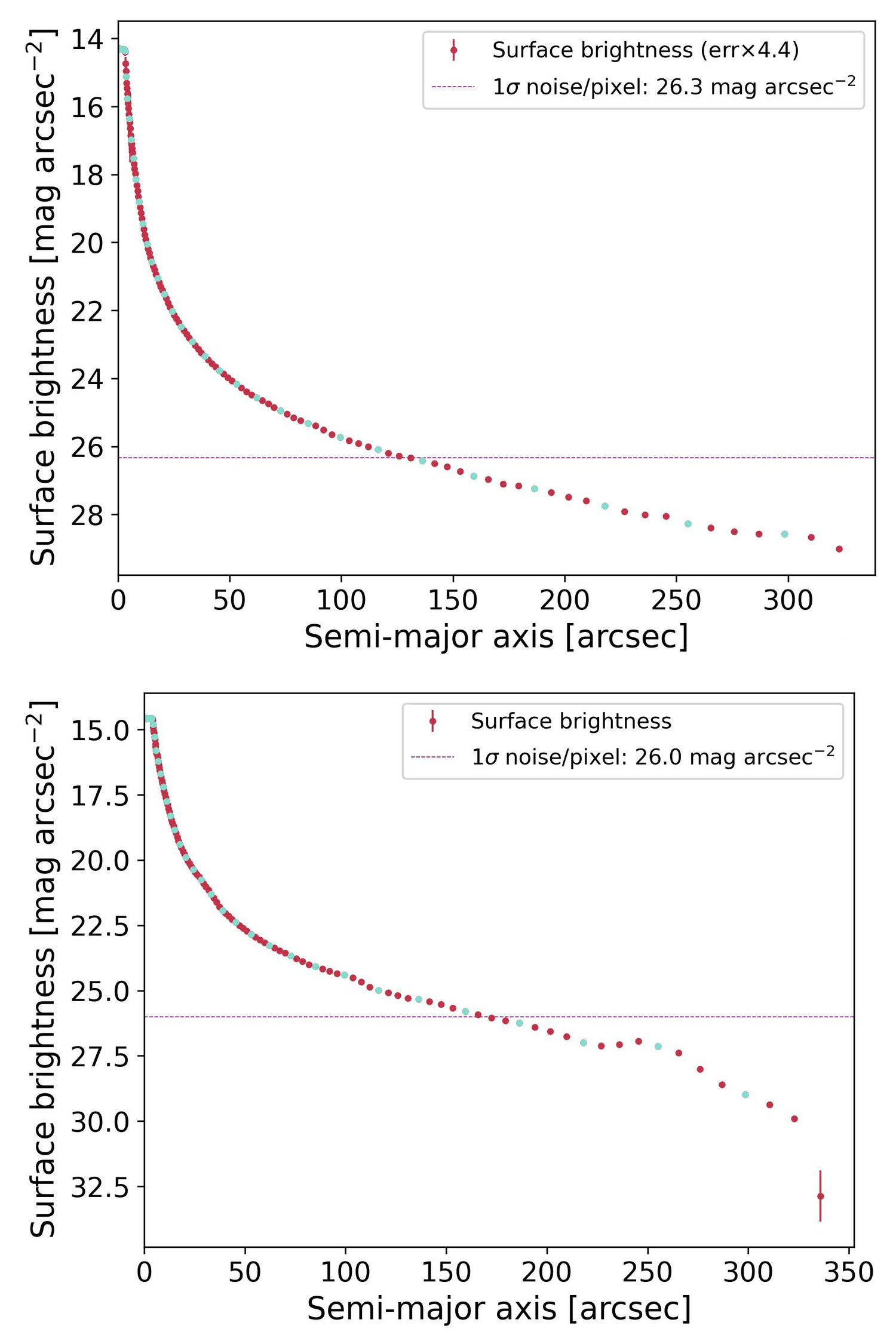}
    \caption{Radial profile of HD\,1973 on a deep (3 ROS) \ac{ERO} stack in the \IE-band (top) and the \YE-band (bottom) reveals distinct characteristics of the VIS and NISP instruments. For VIS, the extended halo concludes at a radius of \ang{;;290}, with {\tt AutoProf} not detecting any signal beyond this point. In contrast, the NISP extended halo exhibits a notable excess of light at a radius of \ang{;;260}, indicative of its refractive optics (this feature is observable as a distinct arc in \cref{fig:VBPPhoenicis}), and terminates shortly thereafter at a radius of \ang{;;320}. Cyan and red points refer to the colours of isophotes in an associated {\tt AutoProf} visual, which help guide the eye when comparing plots.
    \label{fig:AutoProfBlindingStar}}
\end{figure}

\subsection{The observational \Euclid extended PSF}
\label{sec:extendedPSF}
\subsubsection{Encompassed energy of the \Euclid extended PSF}

In analysing the extended \ac{PSF} within the \ac{ERO} data to estimate encircled energy up to a certain radius, we focused on the brightest star in the \ac{ERO} fields, HD\,1973 (V$^{*}$BP Phoenicis, with AB magnitudes of 5.1 and 4.7 in the 2MASS $J$ and $H$ bands, see \cref{fig:VBPPhoenicis}) near Abell\,2764 \citep{EROLensData}. This star shows a prominent extended halo and distinct diffraction spikes that extend to large radii. Utilising {\tt AutoProf}, a tool originally designed for galaxies, we conducted isophotal photometry on the \ac{PSF} halo to obtain radial profiles in all four bands. By enforcing circular isophotes, we were able to extract a reliable signal up to \ang{;5.0;} for VIS and \ang{;5.2;} for NISP, ceasing at a surface brightness of 28.5\,mag\,arcsec$^{-2}$ for VIS and 30.0\,mag\,arcsec$^{-2}$ for NISP. Despite the star's luminosity and the observation's depth (three times the standard \Euclid depth), we could not identify any signal beyond these extents and concluded we did reach the full extent of the extended PSF. HD\,1973 approaches a magnitude close to \Euclid's threshold for avoiding such observations \citep[4\,AB\,mag in each band, see][]{Scaramella-EP1}, an indication that the telescope will have rare encounters with such stars in the \ac{EWS} during its mission: the \ac{EWS} is composed of 27\,571 pointings, with only 293 of them (1\%) featuring a star $\IE<5$, and 1350 (5\%) with a star $\HE<5$.

\begin{figure*}[hbtp!]
	\includegraphics[width=\linewidth]{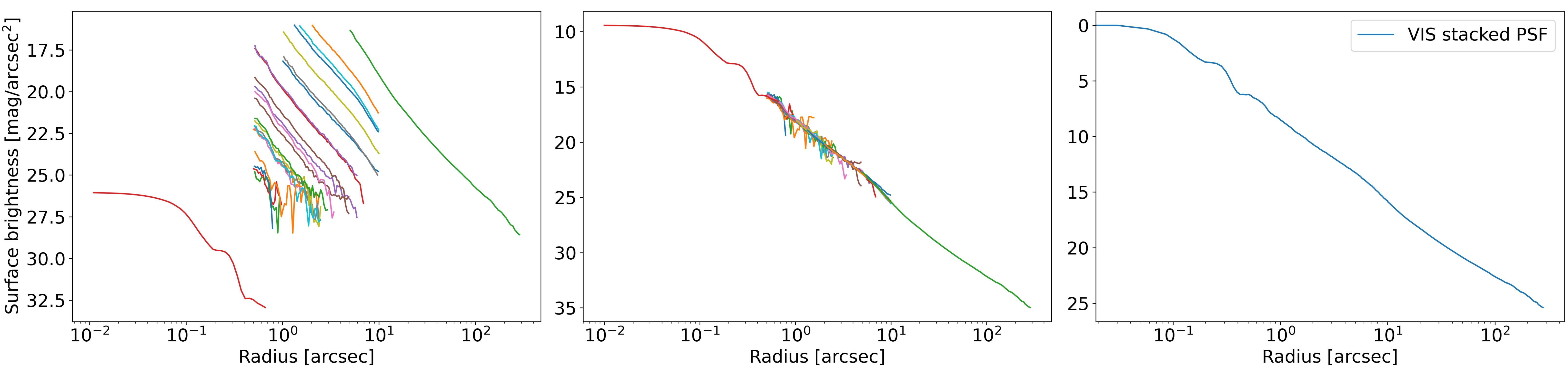}
    \caption{\textit{Left panel}: We analysed radial profiles of stars of varying brightness, clipping the radius range to balance their contribution against the high-quality radial profile of HD\,1973 (shown in green, upper right) and the high-\ac{SNR} {\tt PSFex} radial profile at the \ac{PSF} core based on thousands of bright unsaturated stars (depicted in red, lower left). This approach ensured that the aggregate profile leverages both the detailed observation of a singularly bright star and the precision afforded by {\tt PSFex} in the PSF's core regions.
    \textit{Middle panel}: The optimisation step involved aligning all profiles to achieve coherence across the data set. This optimisation process is critical for synthesising a unified radial profile from the disparate data points provided by stars of different magnitudes and the detailed profiles of HD\,1973 and the \ac{PSF} core.
    \textit{Right panel}: The final VIS extended PSF, with the profile normalised to a base surface brightness of $\mu_0=0$ (corresponding in that case to a $\IE=3.1$ mag star). This standardised profile serves as a comprehensive model of the VIS PSF's behaviour across a wide range of radii.
    \label{fig:VISStackedPSF}}
\end{figure*}

This star alone would be sufficient to characterise the extended PSF, except that the detector saturates in the brightest central regions of the stellar image (top left of the two profiles of \cref{fig:AutoProfBlindingStar}). This information is needed to anchor the energy found in the extended halo to the energy found in the core of the \ac{PSF} \citep{InfanteSainz2020, Liu2022}. To extend the \ac{PSF} brightness profile into the inner regions, as illustrated in \cref{fig:VISStackedPSF}, we extracted from single non-resampled frames a number of intermediate-brightness saturated stars, and some unsaturated stars, again using {\tt AutoProf} to determine the profile of the \ac{PSF} halo. For each profile (23 in total in each band) we manually determined the region in which the profile was robust by identifying clear features of saturation (a flattened profile in the inner regions) and the noise limit in the outskirts (that is high variance surface-brightness measurements).

For the key inner few pixels, we produced a pixel map with {\tt PSFex} at 10$\times$ the native resolution, then converted to a radial profile using {\tt AutoProf}. {\tt PSFex} builds a complex polynomial model of the \ac{PSF} through the analyses of thousands of bright unsaturated stars across a single non-resampled frame here, and captures key features of the Airy disc in all four bands in the \ac{ERO} data. 

This strategy averages out from the core out to a radius of \ang{;;10} possible second-order effects related to our broad filters versus various spectral types of stars. All of the profiles for the \IE-band can be seen in the left panel of \cref{fig:VISStackedPSF} where the stars have not yet been shifted from their original extraction. Using each profile only in the region where it was most robust, we performed an optimisation to rescale each profile until it aligned with the profile in star HD\,1973. An example of this alignment for the \IE-band can be seen in the middle panel of \cref{fig:VISStackedPSF} where all the profiles are now aligned. We note that some profiles have no overlap with star HD\,1973, however by simultaneously aligning all the profiles we could bridge the gap from \ang{;;300} down to the sub-pixel level at the very core of the \ac{PSF}. As a final step, we produced a collapsed profile which was the median of all profiles that contribute in a given region. This can be seen for the \IE-band in the right panel of \cref{fig:VISStackedPSF} which spans over four orders of magnitude in radius; the profile is normalised to $\mu_0=0$ by convention.

\begin{figure}
\centering
	\includegraphics[width=1.0\linewidth]{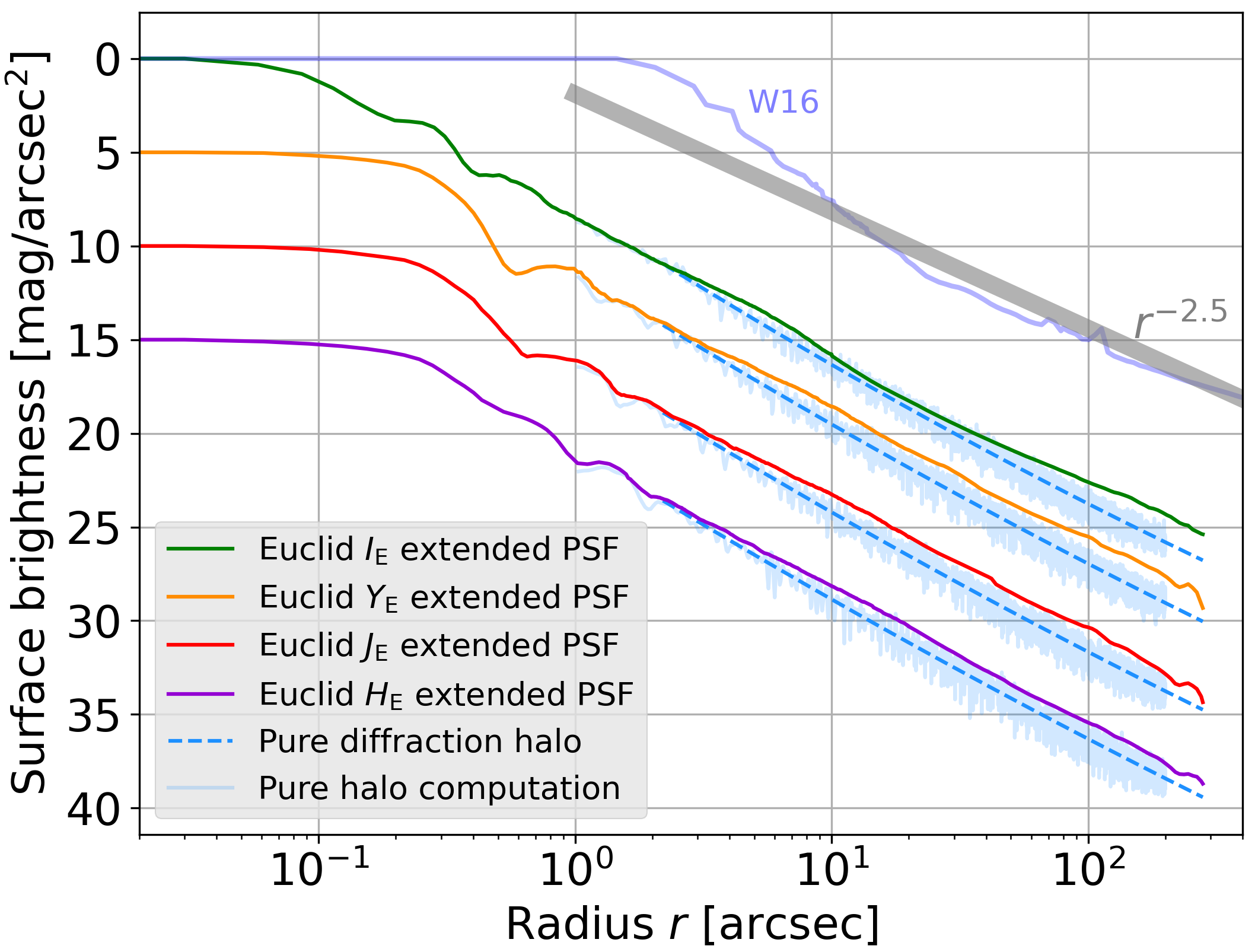}
    \caption{\Euclid's extended \ac{PSF} across the four bands is depicted with solid lines representing the stacked \ac{PSF}, as detailed in \cref{fig:VISStackedPSF}, derived from \ac{ERO} data. The \IE-band is normalised to $\mu_0=0$, while the NISP \YE, \JE, and \HE bands are visually separated by respective offsets of +5, +10, and +15 vertical units for clarity. The total magnitude of a star corresponding to each of the four profiles anchored at $\mu_0=0$ is $\IE=3.1, \YE=1.6, \JE=1.5$, and $\HE=1.4$ (it takes a brighter star in NISP to reach the same $\mu_0=0$ as VIS, which concentrates more energy in the core of the PSF, while the longer the wavelength, the poorer it gets in the \ac{NIR}). The simulation of the diffraction halo (\cref{apdx:OpticalModel}) is presented in pale blue, where the pale area represents the band-specific noisy computation profile and the dashed line indicates a second-order polynomial fit (an $r\,^{-3}$ power law). This fit is aligned with the \ac{PSF} at \ang{;;2} radius. The extended \ac{PSF} concludes at \ang{;5;} radius, and its average slope shows only slight deviation from the theoretical model of a pure diffraction halo. This consistency underscores the exceptional quality of the optics, which appear to be free from particulate contamination. The three \ac{NIR} extended \acp{PSF} are similar (constant offset between the purple, red, and orange profiles), but clearly contain more energy at large radii than the VIS extended PSF (the artificial 5\,mag offset between the green and orange lines here quickly diminishes with radius). We represent in grey the median behaviour of all best wide-field \ac{LSB} observatories reviewed by \cite{Liu2022} with an $r\,^{-2.5}$ power law, shifted up to match our $\mu_0=0$, as originally done by \cite{Watkins2016}. The specific extended $V$-band \ac{PSF} of the Burrell Schmidt telescope is reproduced here (W16), as published in \cite{Watkins2016}, limited to the range of this plot although it extends to a radius of \ang{1;;}, ending at a surface brightness of 22.2\,mag\,arcsec$^{-2}$. This demonstrates that \Euclid excels in minimising scattering of light, suppressing the extended wings better than the best ground-based optical telescopes by 8\,magnitudes, and 6\,magnitudes in the \ac{NIR}, thereby opening a new observational window on the Universe.
    \label{fig:ExtendedPSF}}
\end{figure}

\begin{figure}
	\includegraphics[width=\linewidth]{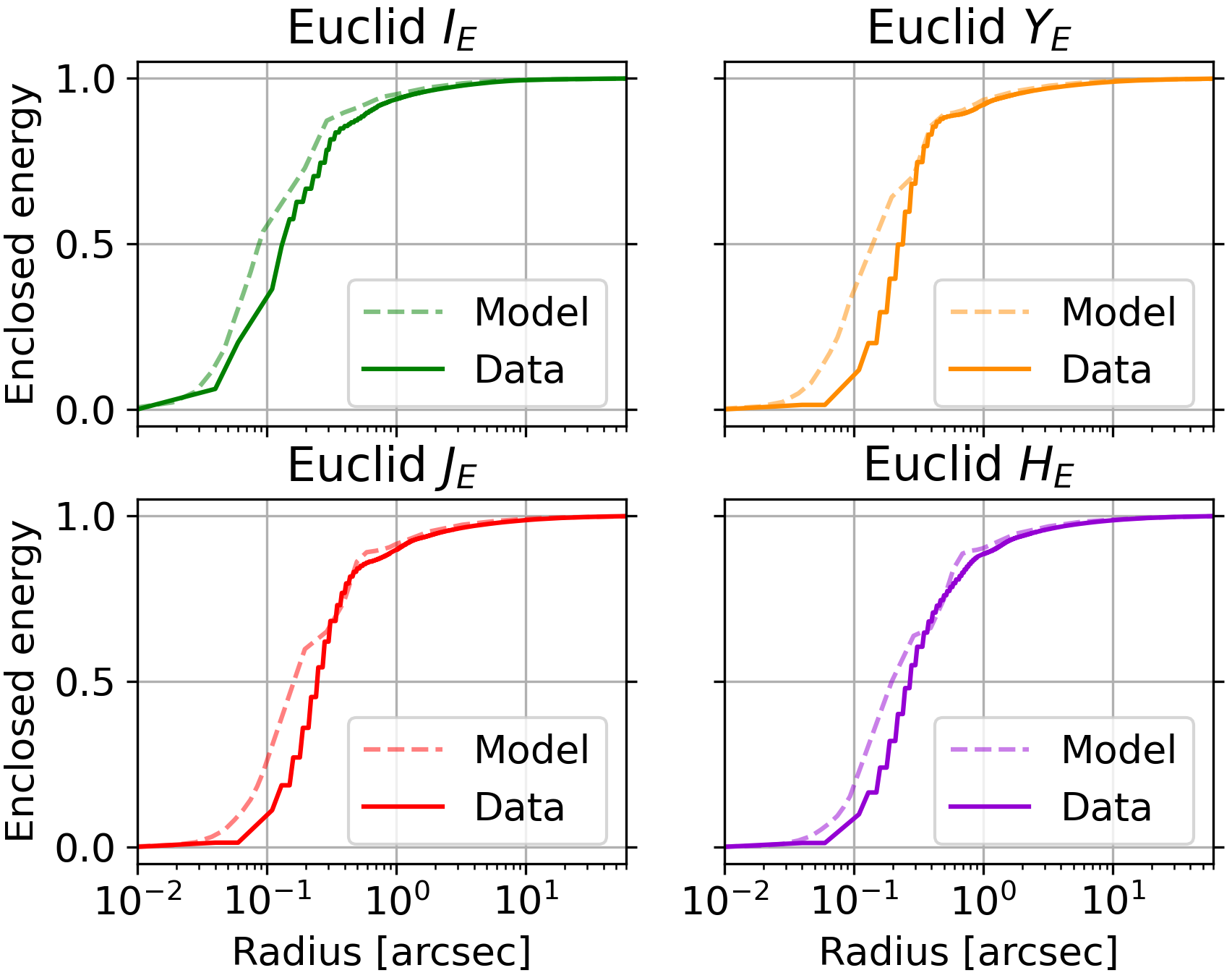}
    \caption{Comparison of the encircled energy in the inner part ($<\ang{;;10}$) of the extended \ac{PSF} between measured data and a simulation of perfect optics for all four \Euclid bands. The near match at the core, based on a {\tt PSFex} profile from our sky data, indicates our extended \ac{PSF} accurately reflects expected performance. The slight disagreement between the simple, aberration-free model and the data at a radius below \ang{;;1} ($10^0$) most likely stems from low-order aberrations introduced by optics. These aberrations tend to slightly broaden the core but have little to no effect on the extended PSF. The observation that the model-to-data discrepancy decreases with increasing wavelength tends to support this hypothesis, as optical aberrations are more pronounced at shorter wavelengths. It should be noted that the disagreement appears more pronounced due to the use of a log-log plot.
    \label{fig:EnclosedEnergy}}
\end{figure}

With this extended \ac{PSF} available for all \Euclid bands as shown in \cref{fig:ExtendedPSF}, we could then simply integrate the light to any radius and compare with the full integral. We find exceptional light enclosure for \Euclid, with the values listed in \cref{tab:enclosedenergy} exceeding the original design specifications. VIS and NISP exhibit similar performance with about 90\% of the encircled energy within a \ang{;;1} radius, 99\% at \ang{;;10}, and 100\% at \ang{;;300}.
In \cref{fig:EnclosedEnergy} we compare our results with our simulations described in \cref{apdx:OpticalModel}, extending to the \ac{PSF} core and verifying the accuracy of our data-based model. Given the basic assumptions of our theoretical \ac{PSF} model, including ideal optics without phase variations or defects, this outcome underscores the effectiveness of \Euclid's optical design and fabrication quality.

In \cref{fig:ExtendedPSF} simulations extending to larger radii are represented in blue. Across all four bands, there is a deviation from the extended wings of the Bessel-function characteristic of the Airy disc ($r^{-3}$ slope) at large radius, with our simulation assuming an ideal circular pupil. Nevertheless, the slope measured from \ac{ERO} data shows only a minor divergence from this ideal case of a pure diffraction halo and never reaches an $r^{-2.5}$ slope. Such a slope would suggest a profile influenced by defects, dust, and various aberrations at the millimetre scale on the primary mirror, as discussed by \cite{Racine96}. The observed dominance of optical diffraction over mirror surface roughness underscores the excellence of the manufacturing process and suggests that a negligible amount of particulates was deposited on the primary mirror during launch.

The extended \ac{PSF} does not exhibit any obvious effects from the thin nanometre-scale layer of ice that accumulated on the optics near the VIS instrument after launch due to outgassing \citep{Schirmer-EP29}. Given that the VIS extended \ac{PSF} contains less power in its halo compared to NISP, as shown in \cref{fig:ExtendedPSF}, we hypothesise that a portion of the \ac{PSF} flux is uniformly extracted from the \ac{PSF} and scattered isotropically, contributing to the overall background in the instrument's cavity.

\begin{figure}[htbp!]
	\includegraphics[width=\linewidth]{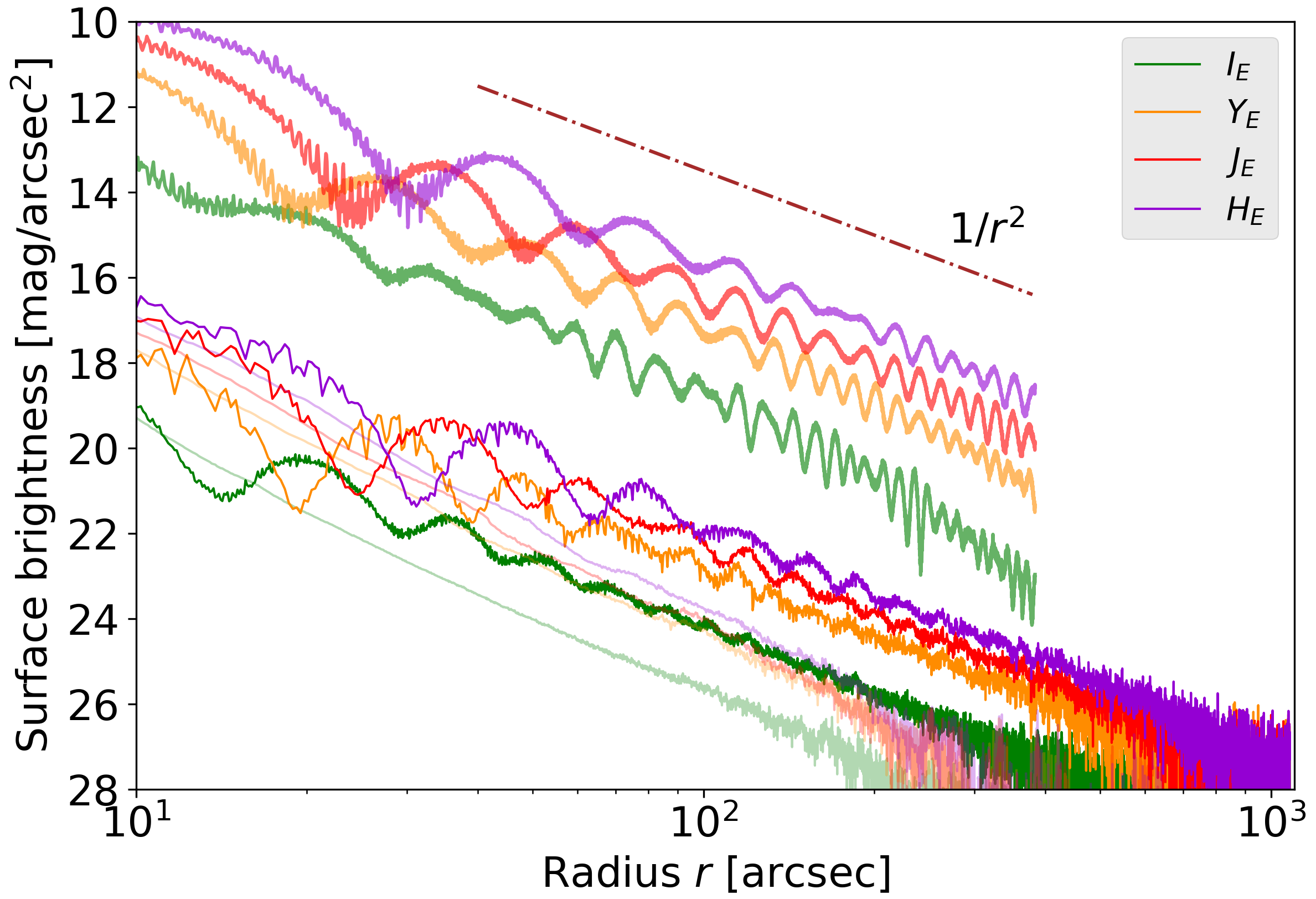}
    \caption{Diffraction spikes of the star HD\,1973 analysed through photometric extraction profiles at the bottom, juxtaposed with the star's halo profile in a corresponding lighter shade (as illustrated in \cref{fig:ExtendedPSF}). This comparison highlights the pronounced dominance of spike brightness over the halo, becoming evident beyond a radius of a few arcseconds. Additionally, a simulation of the four spikes extending to a \ang{;;400} radius (with matching colours, and the \HE-band arbitrarily set to $\mu_0=10$ for the purposes of this plot) corroborates the relative surface brightness of the spikes as observed across the four \Euclid bands. Unlike the halo that adheres more closely to a 1/$r^3$ function consistent with the Bessel envelope at high radius, the spike profiles, as anticipated, approximate a 1/$r^2$ function.     
    \label{fig:DiffractionSpikes}}
\end{figure}

\begin{table}
    \caption{Measured encircled energy of the \Euclid extended \ac{PSF} as a function of radius from \ang{;;0.1} to \ang{;;310} (there is no further signal detection beyond), derived from the \ac{ERO} data set. For context, the theoretical \ac{FWHM} are \IE=\ang{;;0.136}, \YE=\ang{;;0.179}, \JE=\ang{;;0.213}, and \HE=\ang{;;0.298}. }
    \label{tab:enclosedenergy}
    \centering
    \begin{tabular}{lcccc}
    \hline\hline
    \noalign{\vskip 1pt}
    Radius [arcsec] & \IE\ & \YE\ & \JE\ & \HE\ \\
    \hline
    \noalign{\vskip 1pt}
    0.1 & 0.36300 & 0.11880 & 0.11067 & 0.09818\\
    0.2 & 0.66571 & 0.39405 & 0.35903 & 0.31939\\
    0.3 & 0.78320 & 0.68071 & 0.61916 & 0.54811\\
    0.4 & 0.85492 & 0.82898 & 0.76611 & 0.68006\\
    0.5 & 0.87174 & 0.88110 & 0.84034 & 0.75903\\
    1.0 & 0.93556 & 0.91891 & 0.89640 & 0.88333\\
    1.5 & 0.95459 & 0.94315 & 0.93054 & 0.91843\\
    2.0 & 0.96443 & 0.95475 & 0.94425 & 0.93760\\
    4.0 & 0.98135 & 0.97364 & 0.96855 & 0.96615\\
    8.0 & 0.99161 & 0.98566 & 0.98242 & 0.98175\\
    16.0 & 0.99545 & 0.99258 & 0.99116 & 0.99091\\
    32.0 & 0.99717 & 0.99601 & 0.99551 & 0.99566\\
    64.0 & 0.99828 & 0.99783 & 0.99785 & 0.99786\\
    120.0 & 0.99911 & 0.99898 & 0.99913 & 0.99905\\
    240.0 & 0.99987 & 0.99983 & 0.99987 & 0.99982\\
    280.0 & 0.99999 & 0.99997 & 0.99997 & 0.99994\\
    310.0 & 1.00000 & 1.00000 & 1.00000 & 1.00000\\
    \hline
    \end{tabular}
\end{table}

\subsubsection{Energy in the diffraction spikes}

The \ac{PSF} halo profiles, established through {\tt AutoProf} employing median pixel values along the isophotes,\footnote{Except for the {\tt PSFex} sub-pixel stack for which the mean was used.} resulted in the initial analysis excluding the diffraction spike profiles from the \ac{PSF}. However, these spikes represent a significant aspect of the \ac{PSF} structure, and thus, the enclosed energy within the six spikes necessitates estimation in comparison to that derived from our ideal optical model. This step is crucial for a comprehensive understanding of the PSF's energy distribution and for ensuring the accuracy of our extended-\ac{PSF} model in reflecting the true performance of the optical system.

For our brightest star, HD\,1973, we extracted all six spikes across each band, subsequently rotating and median-stacking them to produce a single, uncontaminated spike per band. This process ensured each spike was averaged across its entire width (\ang{;;3.6} for all bands), culminating in the profiles depicted in \cref{fig:DiffractionSpikes}. The plot highlights the diffraction halo in lighter shades, clearly showing that the spike profiles exceed the halo's brightness at large radii. Furthermore, it can be seen that the overall surface brightness escalates with increasing wavelength. The data align closely with our straightforward simulation (\cref{apdx:OpticalModel}), as illustrated in \cref{fig:DiffractionSpikes}, particularly in terms of the relative brightness among the four bands, offering a vivid representation of the spike profiles' luminance in comparison to the halo.

\begin{figure*}[htbp!]
	\includegraphics[width=\linewidth]{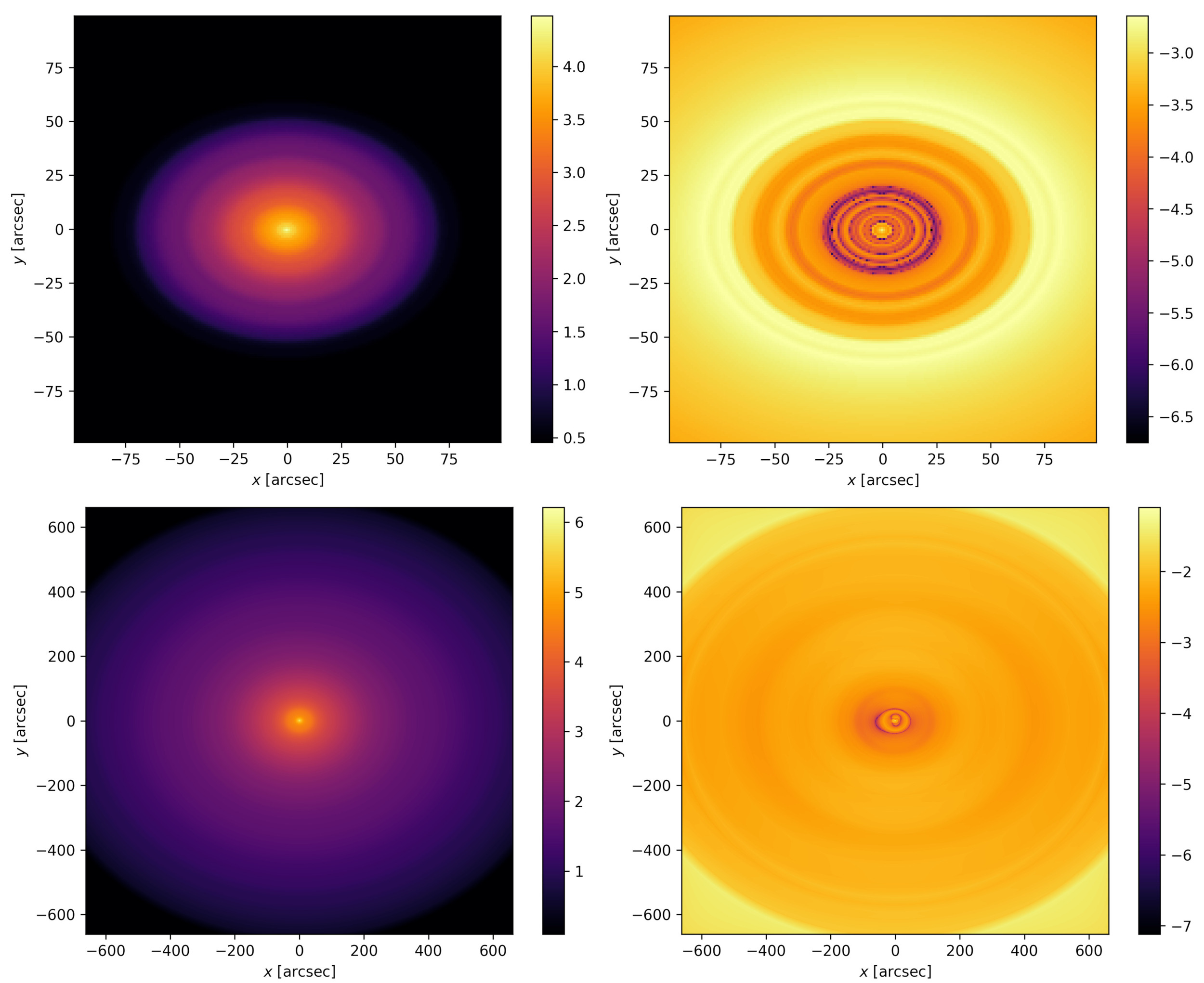}
    \caption{Negligible impact of the extended \ac{PSF} on physical parameters of very extended sources. \textit{Left panels}: On the top, an $\IE=14$ elliptical galaxy with a full extension less than that of the extended PSF ($r=\ang{;;300}$), and at the bottom, an $\IE=9$ galaxy (NGC\,1553) larger than the extended PSF -- colour intensity is in logarithmic scale on all four panels (-3.0 indicates a 10$^{-3}$ ratio, or 0.1\%). \textit{Right panels}: the relative variation of the flux redistribution after convolution with the extended PSF is at the sub-percent level and does not impact the derived physical properties such as their photometry or morphology.
    \label{fig:GalaxiesConvolutionPSF}}
\end{figure*}

We assessed the total flux present in the diffraction spikes of HD\,1973 relative to its overall brightness in the \Euclid bands. This evaluation was facilitated by the extended PSF, which allows for accurate magnitude estimation of any saturated star by aligning its profile outside the saturated core. Our findings are in line with the initial simulation estimates: beyond a \ang{;;10} radius, the combined total-light fraction in both the spikes and the halo amounts to approximately 1.5\% -- averaging 1\% for the halo and 0.5\% for the spikes across all four bands. However, there is a notable distinction in their spatial distribution; the spikes are distinctly localised, whereas the halo extends azimuthally, covering a substantially larger area. In fact, at a \ang{;;100} radius, the halo encompasses about two orders of magnitude more light than the spikes. Consequently, we opted not to incorporate the spike profiles into our extended \ac{PSF} model, given our aim for percent-level accuracy in the results.

\subsubsection{Consequence for the extended-emission science}

\Euclid boasts the most refined extended \ac{PSF} ever achieved by a wide-field, high-resolution imaging telescope. It sets a new benchmark far beyond past endeavours aimed at optimising telescopes for \ac{LSB} science, a leap of 8\,mag in minimising optical scattering of light in the optical while opening a whole new observational window in the \ac{NIR} with a comparable leap of 6\,mag in conjunction with the low background experienced at L2. See figure~8 of \cite{Liu2022} review of the best optical \ac{LSB} wide-field facilities around the world for an illustration, and our \cref{fig:ExtendedPSF} for a comparison.

Characterised by an exceptionally high energy concentration in the PSF's core, \Euclid outperforms its predecessors and maintains this dominance within a nearly pure diffraction-halo regime extending up to a radius of \ang{;;300}, marking the boundary of the extended PSF's influence. This unparalleled performance underscores \Euclid's revolutionary potential for astronomical observations in the realm of \ac{LSB} science from the optical to the \ac{NIR}.

Science focused on diffuse emission suffers through two main consequences of an extended \ac{PSF}. First, bright stars in the \ac{FOV} produce an accumulation of extended and overlapping halos that cause a modulation of the sky brightness, jeopardising the extraction of the \ac{LSB} astronomical signal that gets lost in that non-flatness noise. Second, bright extended objects contaminate their own extended emission, for example a big elliptical galaxy biasing its radial profile by pushing light from its core into its stellar halo. 

\subsubsection{Contamination by overlapping stellar halos}

The \ac{ERO} field Abell\,2764 -- featuring the $\HE=5$ star HD\,1973 -- presents a scenario that \Euclid will seldom face during its scientific mission. Despite its brightness, this star influences only about 0.022\,deg$^2$ (4\%) of the 0.5\,deg$^2$ field through its extended diffusion halo. The predominant background component is the zodiacal light that averages around 22.2\,mag\,arcsec$^{-2}$, and the influence of the extended halo drops below 1\% (equivalent to a 5\,mag difference) at a radius of \ang{;;200}, as illustrated in \cref{fig:AutoProfBlindingStar}.

To significantly affect the \ac{FOV} at the 0.01\% level at a radius of \ang{;;200} (which would be 1/100th of HD\,1973's impact, or 32\,mag\,arcsec$^{-2}$), one would need nearly 50 stars of magnitude 9 evenly scattered throughout the \ac{FOV}. However, at worst across the entire \ac{EWS} composed of 27\,571 pointings, one pointing (0.004\%) features at most 15 stars of this class (VIS or NISP), while on average across the \ac{EWS}, the density of such stars amounts to just 2.8 per \Euclid \ac{FOV}. 

Taking a more realistic perspective, considering the VIS saturation limit at magnitude 18.5 for long exposures and hypothesising the presence of one such star per square arcminute, the 1\% level of zodiacal-light background (27.2\,mag\,arcsec$^{-2}$, a problematic level for \ac{LSB} science) is attained within a radius of approximately \ang{;;5}, affecting merely 2\% of a square arcminute.
In even this pessimistic scenario, the overall impact on the detection of extended emission across scales of arcseconds and above, on average throughout the field, remains minimal. This analysis underscores \Euclid's robustness in handling the effects of bright stars on \ac{LSB} science, facilitated by its advanced design.

Across the 17 \ac{ERO} fields we encounter a range of stellar densities. Despite this variability, no field contains a sufficient number of stars to significantly impact the dominant zodiacal background. The noise in our images, across all scales, is primarily driven by photon statistics. Consequently, we deduce that our performance metric for assessing the \ac{LSB} capabilities of our data set should concentrate exclusively on the zodiacal background's brightness level. This approach remains applicable as long as the data processing is finely tuned for \ac{LSB} science, as exemplified by the \ac{ERO} pipeline. This optimisation ensures that the primary consideration in evaluating our \ac{LSB} performance is the zodiacal light contribution, rather than the collective influence of stellar contributions and data processing signatures.

\subsubsection{Self-contamination of a stellar halo}

A potential issue in \ac{LSB} science involves extended objects potentially contaminating their own signal \citep{Slater2009, Karabal2017}. This concern is particularly relevant when analysing large galaxies within the \ac{ERO} science framework. However, due to \Euclid's exceptionally clean extended PSF, no correction for energy redistribution is necessary. For instance, the brightest surface-brightness levels ($\mu_0$) were measured for NGC\,1549 and NGC\,1553 in the \ac{ERO} Dorado field, reaching approximately 14.3\,mag\,arcsec$^{-2}$ in the \IE band. When incorporating the extended \ac{PSF} (\cref{fig:ExtendedPSF}) at this brightness level, the self-contamination drops to around 30\,mag\,arcsec$^{-2}$ at a \ang{;;10} radius and to approximately 36\,mag\,arcsec$^{-2}$ at a \ang{;;100} radius.
The minor energy redistribution caused by the extended \ac{PSF} does not significantly impact \ac{ERO} science. This is because most galaxies under study exceed in size the 99.7\% encircled energy diameter of \ang{;2;} for both VIS and NISP, as detailed in \cref{tab:enclosedenergy}. This is demonstrated in \cref{fig:GalaxiesConvolutionPSF}, where the impact of the convolution on two extended galaxies (one smaller and one larger than the extended PSF), leads to residuals at the sub-percent level. Consequently, the physical properties of these galaxies remain accurate, affirming that \Euclid's extended \ac{PSF} allows for reliable \ac{LSB} measurements without the need for adjustments for self-contamination.

\subsubsection{Contamination by Galactic cirrus}

A notable consideration in assessing \Euclid's \ac{LSB} detection capabilities is that its actual limitation often stems from the interstellar medium foreground also known as Galactic cirrus \citep{Sandage1976}. This phenomenon is detectable across all Galactic latitudes at present with optical wide-field cameras \citep{Miville2016, Roman2020, Lim2023}, with brightness levels ranging between 26 and 28\,mag\,arcsec$^{-2}$. This factor is especially pertinent in the context of the \ac{ERO} Perseus cluster of galaxies, where \ac{LSB} performance is constrained by the non-uniformity of the background attributed to Galactic cirrus, as noted in studies like \cite{EROPerseusOverview, EROPerseusICL}. The pronounced visibility of such cirrus, even in observations targeting HD\,1973 (\cref{fig:VBPPhoenicis}, left), underscores \Euclid's exceptional ability to discern extremely low-contrast \ac{LSB} features, highlighting the challenge posed by the Galactic cirrus in accurately quantifying faint extragalactic astronomical signals. It is noteworthy that all 17 \ac{ERO} fields showcased in \cref{fig:EROfields} easily reveal a faint background of Galactic cirrus even at high galactic latitudes.

\begin{figure}
	\includegraphics[width=\linewidth]{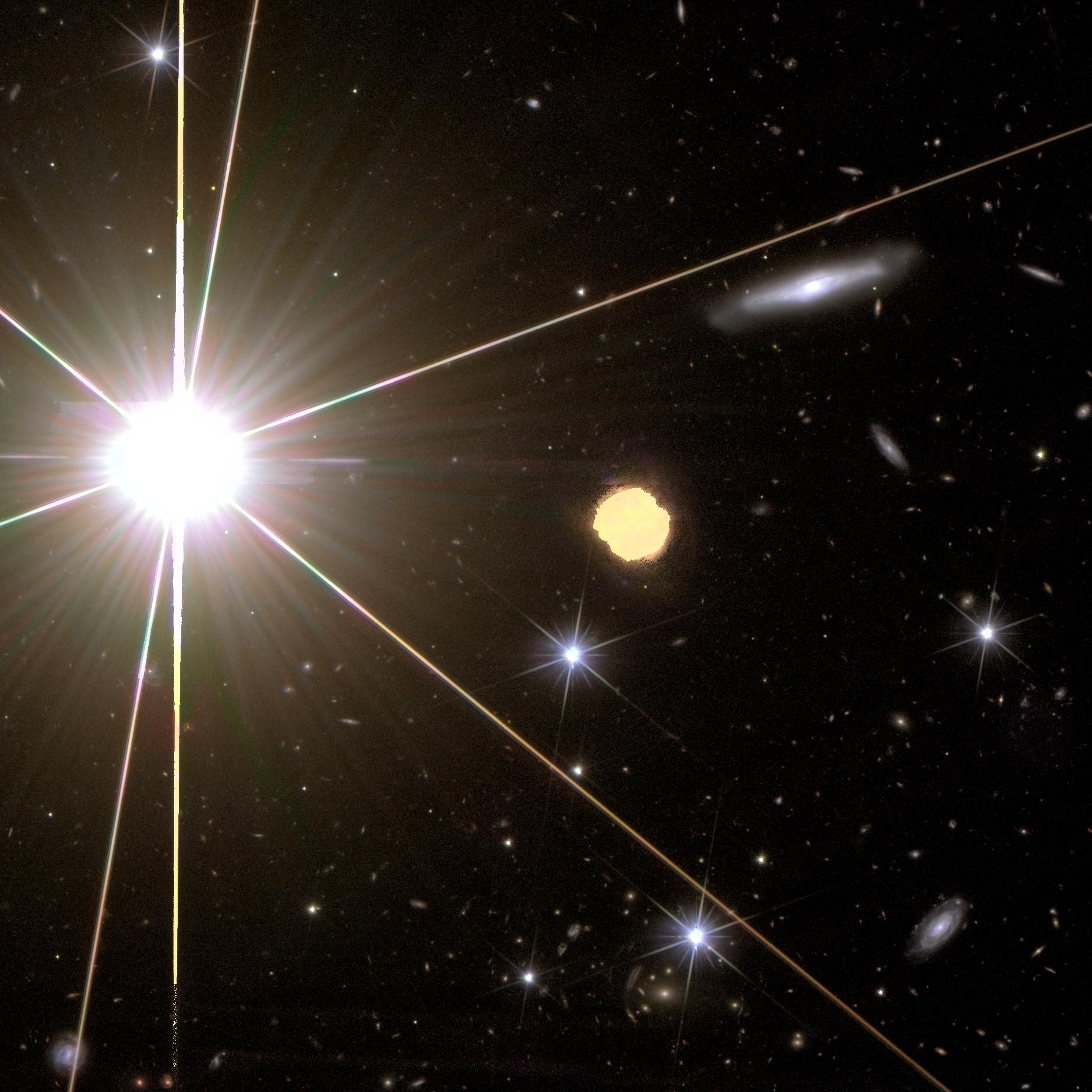}
    \caption{This paper quantifies \Euclid's capacity to measure faint extended emission in various ways. This capability is dramatically demonstrated through an extreme case where \ac{LSB} features on the largest spiral galaxy in the top right corner are clearly visible, despite its proximity -- just \ang{;2;} away -- from HD\,1973 ($\IE=6.3, \YE=5.5, \JE=5.3, \HE=4.9)$. The field size is \ang{;3;} by \ang{;3;}. Faint galaxies are easily perceived within tens of arcseconds from this M3II star that is nearly visible to the naked eye: $R$=6.6 Vega magnitude, \cite{Pickles2010}. This colour image was created by combining VIS and NISP data (without subtraction of the extended PSF model) using the \IE-band for the blue, \YE for the green, and \HE for the red channel. The colour image at the NISP resolution is cast onto the VIS channel through a LAB (luminance + colours) combination to showcase the angular resolution and the sensitivity to \ac{LSB} features. The oval structure at the centre is the result of a dichroic ghost in VIS caused by HD\,1973 which appears yellow because the colour image is dominated at that distance from the star by the NISP extended PSF (green+red = yellow).
    \label{fig:ExtremeLSB}}
\end{figure}

\begin{figure*}[htbp!]
\centering
	\includegraphics[width=0.99\textwidth]{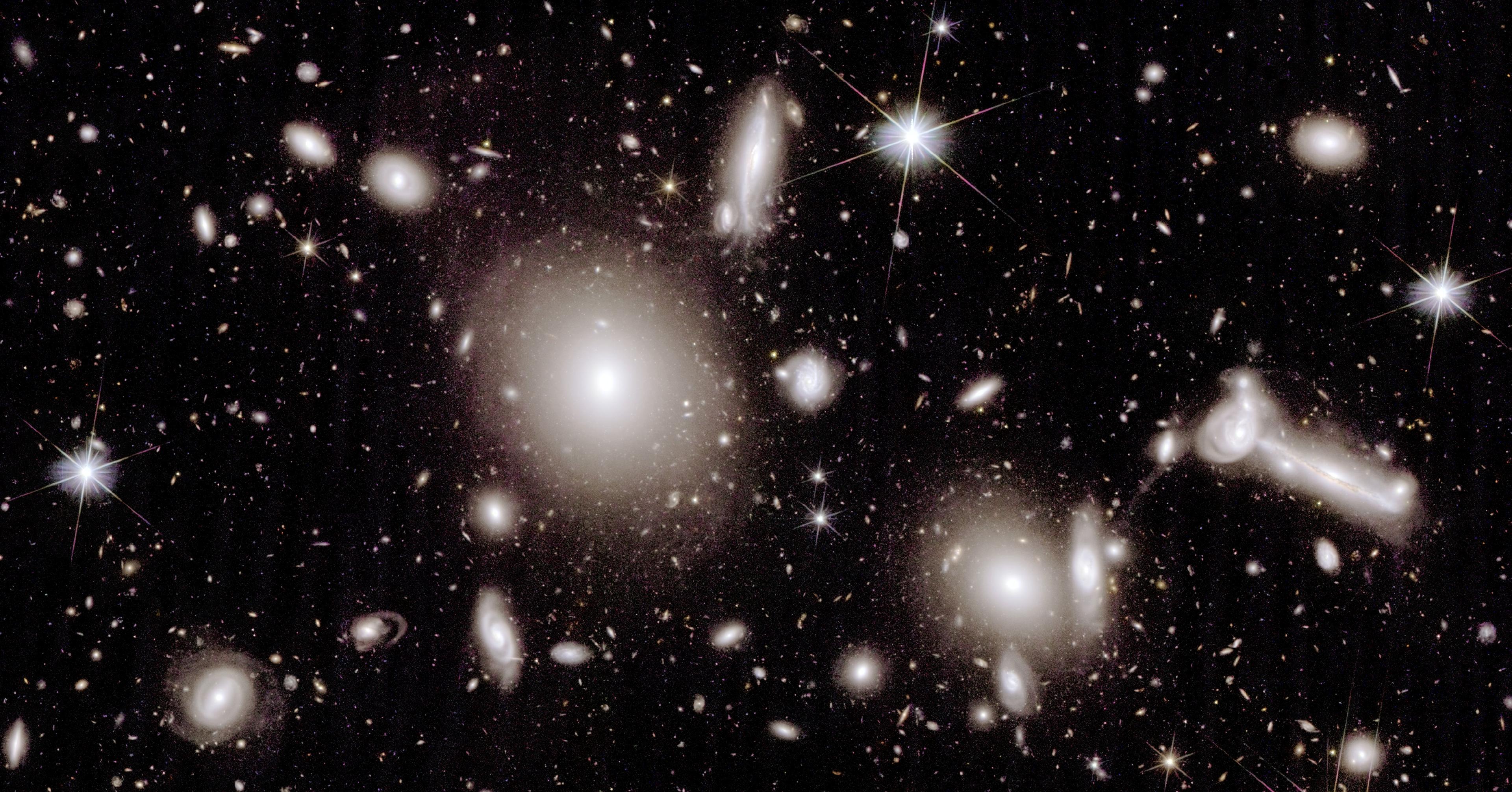}
    \caption{Centre of the Abell\,2764 cluster. Extended emission in the form of shells and stellar streams is traced in the \ac{NIR} (colour image with \IE-band in blue, \YE in green, and \HE in red) down to $27.0$\,mag\,arcsec$^{-2}$ in the \HE-band for a direct visual detection of physical features, such as a stellar stream, at 1\,$\sigma$ at the single square arcsecond scale.
    Structures within these faint features are well defined. The field size is \ang{;7;} by \ang{;3.5;}.
    \label{fig:Abell2764LSB}}
\end{figure*}

\subsection{LSB performance of the ERO data set}
\label{sec:LSBmetric}
In \cref{table:bigtable2} we present the limiting surface brightness for each \ac{ERO} field across the four \Euclid bands, defining the anticipated depth for \ac{LSB} science. This estimation is based on the assumption that the primary component of noise is pure photon statistics from the zodiacal light background, a justification rooted in the preceding discussions about the exceptional quality of the \Euclid extended \ac{PSF}. We determine the \ac{LSB} depth metric following the methodology outlined in \cite{Scaramella-EP1}, adapting the \ac{LSB} \textrm{asinh} metric introduced by \cite{Mihos2013} deriving from the \textrm{asinh} metric for compact sources by \cite{Lupton1999}, to more accurately mirror the real science capabilities in \ac{LSB}-optimised images. This strategy aims to predict \Euclid's future performance in conducting \ac{LSB} science with the understanding that the noise budget would primarily consist of photon-noise statistics from the zodiacal light. For a comprehensive explanation of this 1\,$\sigma$ \textrm{asinh} AB magnitude metric applied at the $\ang{;;10}\times\ang{;;10}$ scale -- used chiefly as a representative measure for various \ac{LSB} scientific analyses -- we refer to section 5.3 of \cite{Scaramella-EP1}.

Adopting this metric, the \ac{ERO} data set achieves depths of $\IE=29.9$, $\YE=28.2$, $\JE=28.4$, and $\HE=28.4$\,AB\,mag\,arcsec$^{-2}$, averaged across the 17 fields, aligning with predictions in \cite{Scaramella-EP1}. These depths illustrate the capability to detect a 100\,arcsec$^{2}$ extended feature at the 1\,$\sigma$ level based solely on photon statistics from the background, given the ultra-low impact of the extended PSF we have shown. The consistency in depth across our collection is attributable to the \ac{ERO} targets being located at ecliptic latitudes comparable to those surveyed by the \ac{EWS}, as depicted in \cref{fig:AllSky}. The photometric catalogues from the first public release of the \ac{ERO} data do not feature automated photometry for extended sources (scale above $\ang{;;10}\times\ang{;;10}$). This functionality is slated for inclusion in the subsequent public data release, expanding the scope of scientific exploration enabled by \Euclid.

As highlighted in the previous subsections, it is important to note that the \ac{LSB} performance achieved for each individual \ac{ERO} field is considerably influenced by the astronomical characteristics of the observed sky area, such as the presence of Galactic cirrus, stellar density, and -- in certain instances -- the density of large galaxies like in the \ac{ERO} Perseus field. These factors can substantially affect the ability to detect and analyse \ac{LSB} features. 
The \ac{ERO} Perseus cluster serves as a vivid illustration of how depth depends on the nature, scale, and analysis methods of specific astronomical entities: (i) the faintest dwarf galaxies in the new \ac{ERO} Perseus cluster catalogue \citep{EROPerseusDGs} present a typical effective radius of \ang{;;1} and reach down to an average effective surface brightness of $\langle  \mu_{\IE, \rm{e}} \rangle =26.3$\,mag\,arcsec$^{-2}$, and a surface brightness at the effective radius of $\mu_{\IE, \rm {e}}=28.7$\,mag\,arcsec$^{-2}$, at a total \ac{SNR} within the effective radius high enough to enable derivation of physical parameters (SNR$>12$, the performance being purely limited by background photon statistics at this scale); (ii) intra-cluster light reaches down to $\mu_{\IE}=29.4$\,mag\,arcsec$^{-2}$ at an \ac{SNR} of 1 by integrating the signal over very large areas \citep{EROPerseusICL}; and (iii) radial profiles of galaxies go down to $\mu_{\IE}=30.1$\,mag\,arcsec$^{-2}$ \citep{EROPerseusOverview} when integrating light at increasing radii, by combining over 360\,deg the signal of many 100\,arcsec$^{-2}$ areas, each at the \ac{SNR}$\sim$2 level. The full suite of \ac{ERO} papers related to \ac{LSB} science provides a deeper understanding of the challenges and successes encountered in capturing and interpreting \ac{LSB} phenomena within \Euclid's observations. For comprehensive insights into the actual scientific performance, we direct the reader to the first results of the \ac{ERO} programme \citep{EROLensData, ERONearbyGals, EROOrion, EROGalGCs, EROFornaxGCs, EROPerseusOverview, EROPerseusICL, EROPerseusDGs}. To encapsulate the main points of this section, \cref{fig:ExtremeLSB,fig:Abell2764LSB} demonstrate \Euclid's unique capacity to detect and measure \ac{LSB} features from the optical to the \ac{NIR}.

\section{\label{sec:summary}Summary}

We described the \ac{ERO} programme, aimed at demonstrating \Euclid's capabilities before its primary mission. It focuses on 17 unique astronomical targets and employs a range of advanced technical methods to process these observations. The scientific results obtained for the individual targets are presented in a set of accompanying papers. Besides the overview of the \ac{ERO} programme, the present paper focuses on the specific data processing that prepared the images for the individual \ac{ERO} projects, and an evaluation of \Euclid's unique \ac{LSB} performance. 

The \ac{ERO} pipeline, crucial for creating data products for initial science studies, emphasises preserving data quality. The FWHM of the PSF in the advanced products (stacks) for the four Euclid channels shows a diffraction-limited telescope with \ang{;;0.16} in the optical \IE-band, and \ang{;;0.49} in the \ac{NIR} bands \YE, \JE, and \HE (\cref{tab:summaryall}). The point-source and extended-source detection limits with the \Euclid survey nominal observing sequence match the pre-launch expectations of 25.3 and 23.2\,AB\,mag with a \ac{SNR} of 10 for galaxies, and 27.1 and 24.5\,AB\,mag at an \ac{SNR} of 5 for point sources for VIS and NISP, respectively.

The pipeline's main tasks involve removing instrumental signatures, calibrating astrometry and photometry, stacking images, and producing science-ready catalogues. Detrending involves several steps specific to each instrument to enhance image quality. For VIS, this includes using tools like {\tt deepCR} to remove cosmic rays and maintain accurate astrometry. Similar steps are taken for NISP, including additional corrections of, for example, charge persistence and row-correlated noise. The zodiacal-light background is the key for both instruments to produce a flat-field enabling \ac{LSB} science. Astrometric calibration starts by establishing an initial framework using {\tt Astrometry.net} with \textit{Gaia}-DR3, followed by refining global astrometry with {\tt SCAMP} to address geometric distortions. Calibration accuracy is high, with VIS achieving a median internal precision of 6\,mas RMS while NISP remains limited to 15\,mas RMS due to its coarser sampling of the PSF.

External catalogues were used to refine the photometric zero points by making adjustments based on empirical data and synthetic photometry. VIS calibration remains highly accurate, while NISP faces challenges due to model-dependent colour transformations. Detailed per-detector analysis in the VIS mosaic revealed statistically significant zero-point fluctuations, suggesting variability within and across detectors. The uncertainty of the VIS absolute flux calibration after tying it to \textit{Gaia} averages better than 1\%, with residual colour and brightness trends up to 10\%. For NISP, calibration accuracy is capped at about 10\% due to the model-dependent colour transformations and a limited early calibration data set.

 The resampling and stacking process involved merging exposures using {\tt SWarp}, producing two types of stacks: a background-removed stack for compact sources and a stack for studying extended emission, choosing interpolating functions like the Lanczos3 kernel for VIS and bilinear interpolation for NISP to preserve data quality. PSF models were created using {\tt PSFex}, however the accuracy of these models is limited for NISP data by factors like aliasing. The rich and detailed catalogues produced using {\tt SourceExtractor} include extensive data from both VIS and NISP, designed to support a broad range of scientific studies.

 This \ac{ERO} effort revealed that \Euclid boasts the best extended \ac{PSF} ever achieved by a wide-field, high-resolution imaging telescope, setting a new benchmark that surpasses previous efforts to optimise telescopes for \ac{LSB} science. It represents a leap of 8 magnitudes of surface brightness of the extended PSF halo in the optical range and opens in particular a new observational window into the \ac{NIR} \ac{LSB} Universe, facilitated by the low background experienced at L2. Such performance indicates that photometry of extended sources can be measured with high precision across the entire \ac{FOV} without needing to deconvolve the image. Also, the background of the \Euclid images is dominated by photon noise across the entire image, confirming pre-launch expectations, and reaching a surface brightness for the nominal observing sequence of $\IE=29.9$, $\YE=28.2$, $\JE=28.4$, $\HE=28.4$\,AB\,mag\,arcsec$^{-2}$ achieved in the \ac{ERO} data set for detecting a 100 arcsec$^2$ extended feature at the 1\,$\sigma$ level (\cref{tab:summaryall}).


This first \ac{ERO} data release includes science-ready source catalogues, with a total of 11 million objects in the VIS, and more than 5 million common detections in NISP, across the 17 \ac{ERO} fields. While this paper highlights this first release of the \ac{ERO} data (stacks and catalogues) to the global scientific community, future \ac{ERO} data releases will address existing challenges, such as photometry non-uniformity, improved stray-light correction, and introduce new features like automated \ac{LSB} photometry, and possibly include a curated spectroscopy collection. This initial \ac{ERO} release marks the beginning of a new era of scientific discoveries as \Euclid embarks on its main mission.\\

\vspace{1cm}

\begin{acknowledgements}
\AckERO  
\AckEC
This work presents results from the European Space Agency (ESA) space mission \textit{Gaia}. \textit{Gaia} data are being processed by the Gaia Data Processing and Analysis Consortium (DPAC). Funding for the DPAC is provided by national institutions, in particular the institutions participating in the \textit{Gaia} MultiLateral Agreement (MLA). 
The Pan-STARRS1 Surveys (PS1) and the PS1 public science archive have been made possible through contributions by the Institute for Astronomy, the University of Hawaii, the Pan-STARRS Project Office, the Max-Planck Society and its participating institutes, the Max Planck Institute for Astronomy, Heidelberg and the Max Planck Institute for Extraterrestrial Physics, Garching, The Johns Hopkins University, Durham University, the University of Edinburgh, the Queen's University Belfast, the Harvard-Smithsonian Center for Astrophysics, the Las Cumbres Observatory Global Telescope Network Incorporated, the National Central University of Taiwan, the Space Telescope Science Institute, the National Aeronautics and Space Administration under Grant No. NNX08AR22G issued through the Planetary Science Division of the NASA Science Mission Directorate, the National Science Foundation Grant No. AST–1238877, the University of Maryland, Eotvos Lorand University (ELTE), the Los Alamos National Laboratory, and the Gordon and Betty Moore Foundation.
This project used public archival data from the Dark Energy Survey (DES). Funding for the DES Projects has been provided by the U.S. Department of Energy, the U.S. National Science Foundation, the Ministry of Science and Education of Spain, the Science and Technology FacilitiesCouncil of the United Kingdom, the Higher Education Funding Council for England, the National Center for Supercomputing Applications at the University of Illinois at Urbana-Champaign, the Kavli Institute of Cosmological Physics at the University of Chicago, the Center for Cosmology and Astro-Particle Physics at the Ohio State University, the Mitchell Institute for Fundamental Physics and Astronomy at Texas A\&M University, Financiadora de Estudos e Projetos, Funda{\c c}{\~a}o Carlos Chagas Filho de Amparo {\`a} Pesquisa do Estado do Rio de Janeiro, Conselho Nacional de Desenvolvimento Cient{\'i}fico e Tecnol{\'o}gico and the Minist{\'e}rio da Ci{\^e}ncia, Tecnologia e Inova{\c c}{\~a}o, the Deutsche Forschungsgemeinschaft, and the Collaborating Institutions in the Dark Energy Survey.
CS acknowledges the support of the Natural Sciences and Engineering Research Council of Canada (NSERC). Cette recherche a été financée par le Conseil de recherches en sciences naturelles et en génie du Canada (CRSNG). CS also acknowledges support from the Canadian Institute for Theoretical Astrophysics (CITA) National Fellowship program.
\end{acknowledgements}

%
%
\clearpage
\bibliography{Euclid,EROplus.bib,EROpipelineRefs}

%

\begin{appendix}
  \onecolumn 

\section{\label{sec:obssummary}\Euclid \texorpdfstring{\ac{ERO}}{ERO} programme summary}

\begin{table*}[htbp!]  
\caption{Euclid ERO observations per proposal sorted by date. The project ID was assigned across all submitted proposals based on distance.}
\label{table:obsspecs}
\setlength{\tabcolsep}{5pt} 
\begin{tabular}{llllll}
\hline\hline
\noalign{\vskip 2pt}
\textbf{ERO ID} & \textbf{ERO project title} & \textbf{Overview paper} &\textbf{Field name} & \textbf{Date} & \textbf{Comments}  \\
\noalign{\vskip 2pt}
\hline
\noalign{\vskip 3pt}
ERO02 & A first glance at free- &\cite{EROOrion}& NGC\,1333   & 09-09-2023 & 1 ROS, bad guiding, few exposures  \\
      & floating baby Jupiters  && Taurus     &  09-09-2023 & 1 ROS, bad guiding, no data  \\ 
      & with \Euclid             && NGC\,1333   &  16-09-2023 & 1 ROS, bad guiding, few exposures  \\ 
      &                         && Taurus     &  16-09-2023 & 1 ROS, dither rotated  \\ 
      &                         && Horsehead  &  02-10-2023 &   1 ROS \\ 
      &                         && Messier 78 &  12-10-2023 &   1 ROS \\ 
      &                         && Barnard 30 &  12-10-2023 &   1 ROS \\ 
\noalign{\vskip 2pt}
\hline
\noalign{\vskip 3pt}
ERO03 & \Euclid view of Milky    &\cite{EROGalGCs}& NGC\,6254   &  09-09-2023 & 1 ROS, dither rotated  \\ 
      & Way globular clusters   && NGC\,6397   &  22-09-2023 &  1 ROS \\
\noalign{\vskip 2pt}
\hline
\noalign{\vskip 3pt}
ERO08 & A \Euclid Showcase       &\cite{ERONearbyGals}& IC\,10      &  03-09-2023 & 1 ROS, dither rotated  \\ 
      & of Nearby Galaxies      && IC\,10      &  06-09-2023 & 1 ROS, dither rotated  \\ 
      &                         && IC\,342     &  02-10-2023 &   1 ROS\\ 
      &                         && NGC\,6744   &  02-10-2023 &   1 ROS\\ 
      &                         && NGC\,6822   &  12-10-2023 &   1 ROS\\ 
      &                         && NGC\,2403   &  12-10-2023 &   1 ROS \\ 
      &                         && Holmberg II &  28-11-2023 & 1 ROS, hints of stray light  \\
\noalign{\vskip 2pt}   
\hline
\noalign{\vskip 3pt}
ERO09 & The Fornax galaxy       &\cite{EROFornaxGCs}& Fornax     &  23-08-2023 & 1 ROS, bad guiding, few exposures \\ 
      & cluster seen with       && Fornax     &  03-09-2023 & 1 ROS, bad guiding, few exposures \\ 
      & \Euclid                  && Fornax     &  06-09-2023 & 1 ROS, bad guiding, few exposures \\ 
      &                         && Dorado     &  28-11-2023 & 1 ROS \\
\noalign{\vskip 2pt}
\hline
\noalign{\vskip 3pt}
ERO10 & The Perseus cluster     &\cite{EROPerseusOverview}& Perseus    &  09-09-2023 & 2 ROS, dither rotated \\ 
      & of galaxies             && Perseus    &  16-09-2023 & 2 ROS, dither rotated \\
\noalign{\vskip 2pt}
\hline
\noalign{\vskip 3pt}
ERO11 & A glimpse into \Euclid's &\cite{EROLensData}& Abell 2390     &  28-11-2023 & 3 ROS\\
      & Universe through a      && Abell 2764     &  28-11-2023 & 3 ROS\\ 
      & giant magnifying lens   &&  &  &   \\
\noalign{\vskip 2pt}
\hline
\end{tabular}
\label{tab:tableXX}
\\
\end{table*}

\newpage

\section{\label{sec:datasummary}\Euclid \texorpdfstring{\ac{ERO}}{ERO} data summary}

\newcolumntype{R}{>{$}r<{$}}

\begin{table*}[htbp!] 
\caption{Main characteristics and properties of the 17 \ac{ERO} fields are consistent across the \YE, \JE, and \HE bands; therefore, we will report only on the \HE band. The right ascension (RA) and declination (Dec) coordinates represent the centres of the stacks. The average colour excess $E(B-V)$ from \cite{Planck_dust2014} is calculated across all valid pixels in each image, which cover approximately 0.6\,deg$^2$. For a summary of the data set's general properties, refer to \cref{tab:summaryall}.}
\label{table:bigtable1}
\begin{tabular}{lRRRRcrrrrrr}
  \hline\hline
  \noalign{\vskip 3pt}
ERO fields & $RA$\,\,\,\,\,  &  $Dec$\,\,\,  \hfil &  $$l$$\,\,\,\,\,\,\,\,\,  &  $$b$$\,\,\,\,\,\,\,  & \,\,$E(B-V)$\,\, & \multicolumn{2}{c}{\ac{FWHM}} & \multicolumn{2}{c}{Astrometry} & \multicolumn{2}{c}{Number of}\\
\noalign{\vskip 2pt}
per project & \textrm{[deg]}\,\,\,\, &\textrm{[deg]}\,\,\, & \textrm{[deg]}\,\,\,\, & \textrm{[deg]}\,\,\, & [\ac{FOV}] &  \multicolumn{2}{c}{[arcsec]}& \multicolumn{2}{c}{[mas]} & \multicolumn{2}{c}{objects}\\
\noalign{\vskip 2pt}
 &  & &  &  &  & \IE\,\, & \HE\,\, & \IE\,\, & \HE\,\, & \IE\,\,\,\, & \HE\,\,\,\,\\
\noalign{\vskip 2pt}
\hline
\noalign{\vskip 3pt}
Horsehead	& 		85.150 &	-2.613	& 	 207.053 & -16.931	& 		2.020	& 	   0.158 & 0.54	& 	   4.7 & 	14.2	& 	   157 264    &      250 942	\\
Messier\,78	& 		86.690 &	-0.015	& 	 205.414 & -14.356	& 		1.588	& 	   0.158 & 0.56	& 	   3.9 &	14.5	& 	   116 373    &      238 334	\\
Taurus		& 		64.983 &	+28.023	& 	 169.257 & -15.552	& 		0.468	& 	   0.165 & 0.53	& 	   3.8 &	13.7	& 	   123 726    &      223 725	\\
Barnard\,30	& 	 	82.880 &	+12.316	& 	 192.500 & -11.526	& 		1.001	& 	   0.158 & 0.54	& 	 	4.4  &	14.4	& 	   163 063    &      243 012	\\
\noalign{\vskip 2pt}
\hline
\noalign{\vskip 3pt}
NGC\,6254		& 		254.303 &	-4.100	& 	  15.145 & +23.062	& 		0.227	& 	   0.155 & 0.55	& 	   4.5 &	16.5	& 	   413 297    &      321 363	\\
NGC\,6397		& 		265.174 &	-53.658	& 	 338.178 & -11.951	& 		0.316	& 	   0.160 & 0.54	& 	   2.8 &	15.5	& 	   782 612    &      493 281	\\
\noalign{\vskip 2pt}
\hline
\noalign{\vskip 3pt}
NGC\,6822		& 		296.236 &	-14.788	& 	  25.352 & -18.389	& 		0.203	& 	   0.156 & 0.55	& 	   3.5 &	14.5	& 	   1 694 021   &      488 601	\\
Holmberg\,II	& 		124.790 &	+70.706	& 	 144.291 & +32.697	& 		0.035	& 	   0.157 & 0.60	& 	 	5.8  &	15.4	& 	   466 276    &      245 597	\\
IC\,10		& 		5.063 &	+59.288	& 	 118.952 &  -3.342	& 		0.822	& 	   0.157 & 0.55	& 	   3.0 & 	14.9	& 	   1 403 807   &      416 300	\\
IC\,342		& 		56.730 &	+68.084	& 	 138.188 & +10.577	& 		0.342	& 	   0.156 & 0.54	& 	   5.3 &	14.6	& 	   2 033 293   &      452 034	\\
NGC\,2403		& 		114.211 &	+65.586	& 	 150.587 & +29.184	& 		0.046	& 	   0.155 & 0.54	& 	   6.8 &	15.1	& 	   1 152 966   &      316 055	\\
NGC\,6744		& 		287.430 &	-63.842	& 	 332.240 & -26.139	& 		0.059	& 	   0.157 & 0.52	& 	   6.2 &	14.9	& 	   924 913    &      336 109	\\
\noalign{\vskip 2pt}
\hline
\noalign{\vskip 3pt}
Fornax	& 		54.017 &	-35.267	& 	 236.425 & -54.132	& 		0.018	& 	   0.162 & 0.63	& 	 	10.3 &	14.4	& 	   369 315    &      265 091	\\
Dorado		& 		64.014 &	-55.780	& 	 265.638 & -43.706	& 		0.018	& 	   0.159 & 0.65	& 	 	7.2  &	15.7	& 	   518 445    &      356 621	\\
\noalign{\vskip 2pt}
\hline
\noalign{\vskip 3pt}
Perseus		& 		49.638 &	+41.651	& 	 150.296 & -13.272	& 		0.156	& 	   0.156 & 0.56	& 	   8.2 &	15.4	& 	   546 563    &      335 340	\\
\noalign{\vskip 2pt}
\hline
\noalign{\vskip 3pt}
Abell\,2390	& 		328.397 &	+17.709	& 	  73.967 & -27.799	& 		0.106	& 	   0.157 & 0.56	& 	 	5.1  &	15.2	& 	   469 056    &      310 617	\\
Abell\,2764	& 		5.713 &	-49.249	& 	 315.028 & -67.201	& 		0.018	& 	   0.158 & 0.69	& 	 	7.9  &	15.8	& 	   542 729    &      321 407	\\
\noalign{\vskip 2pt}
\hline
\end{tabular}
\end{table*}

\begin{table*}[htbp!] 
\caption{Sky coverage and depth properties for the 17 \ac{ERO} fields, pertaining to both compact sources (galaxies and point sources) and extended emission, are detailed below. The limiting magnitude for extended emission (\ac{LSB} limit) is expressed in terms of the 1\,$\sigma$ \textrm{asinh} AB magnitude, utilising the $\ang{;;10}\times\ang{;;10}$ scale metric. For further information on the depth metrics, refer to \cref{sec:cats,sec:lsb}.}
\label{table:bigtable2}
\begin{tabular}{lcrrrrrrrrrrrr}
  \hline\hline
  \noalign{\vskip 3pt}
ERO fields & Area &   \multicolumn{4}{c}{Galaxies, 10\,$\sigma$ [AB mag]}&  \multicolumn{4}{c}{Point sources, 5\,$\sigma$ [AB mag]} & \multicolumn{4}{c}{\ac{LSB} limit [AB\,mag\,arcsec$^{-2}$]}\\
  \noalign{\vskip 2pt}
per project & [deg$^2$] & \IE\,\, & \YE\,\, & \JE\,\, & \HE\,\, & \IE\,\, & \YE\,\, & \JE\,\, & \HE\,\, & \IE\,\, & \YE\,\, & \JE\,\, & \HE\,\, \\
  \noalign{\vskip 2pt}
\hline
  \noalign{\vskip 3pt}
Horsehead	& 0.58 & 		24.83 & 22.63 & 22.80 & 22.77 & 27.05 & 24.32 & 24.46 & 24.34	& 29.53 & 28.04 & 28.25 & 28.22	\\
Messier\,78	& 0.60 &		24.90 & 22.61 & 22.71 & 22.82 & 27.10 & 24.34 & 24.47 & 24.41	& 29.66 & 28.08 & 28.24 & 28.26		\\
Taurus		& 0.61 &		24.93 & 22.77 & 23.01 & 23.05 & 27.09 & 24.34 & 24.49 & 24.45	& 29.58 & 28.06 & 28.24 & 28.28	\\
Barnard\,30	& 0.60 &		24.91 & 22.70 & 22.86 & 22.99 & 27.12 & 24.31 & 24.44 & 24.38	& 29.54 & 28.01 & 28.18 & 28.18	\\
  \noalign{\vskip 2pt}
\hline
  \noalign{\vskip 3pt}
NGC\,6254		& 0.60 &		25.11 & 22.97 & 23.22 & 23.26 & 27.23 & 24.37 & 24.51 & 24.46	& 29.66 & 28.05 & 28.22 & 28.26	\\
NGC\,6397		& 0.61 &		25.16 & 22.93 & 23.11 & 23.15 & 27.11 & 24.21 & 24.31 & 24.22	& 29.51 & 27.70 & 27.74 & 27.75	\\
  \noalign{\vskip 2pt}
\hline
  \noalign{\vskip 3pt}
NGC\,6822		& 0.60 &		25.19 & 22.81 & 23.08 & 23.11 & 26.79 & 24.20 & 24.34 & 24.27	& 29.53 & 27.83 & 27.98 & 27.99	\\
Holmberg\,II	& 0.60 &		25.54 & 23.20 & 23.44 & 23.47 & 27.45 & 24.69 & 24.85 & 24.78	& 29.97 & 28.43 & 28.60 & 28.61	\\
IC\,10		& 0.62 &		26.12 & 23.43 & 23.63 & 23.65 & 27.43 & 24.80 & 24.89 & 24.74	& 30.23 & 28.09 & 28.23 & 28.20	\\
IC\,342		& 0.59 &		25.41 & 22.87 & 23.23 & 23.31 & 26.70 & 24.45 & 24.60 & 24.52	& 29.81 & 28.04 & 28.18 & 28.18	\\
NGC\,2403		& 0.60 &		25.51 & 23.15 & 23.38 & 23.43 & 27.22 & 24.62 & 24.79 & 24.71	& 29.98 & 28.21 & 28.38 & 28.39	\\
NGC\,6744		& 0.60 &		25.29 & 22.97 & 23.21 & 23.31 & 27.09 & 24.50 & 24.64 & 24.58	& 29.81 & 28.05 & 28.20 & 28.21	\\
  \noalign{\vskip 2pt}
\hline
  \noalign{\vskip 3pt}
Fornax	& 0.57 &            25.07 & 22.97 & 23.38 & 23.37 & 26.89 & 24.56 & 24.87 & 24.81 & 29.66 & 28.29 & 28.56 & 28.56 \\
Dorado		& 0.60 &		25.52 & 23.15 & 23.37 & 23.37 & 27.40 & 24.68 & 24.84 & 24.76 & 30.05 & 28.41 & 28.58 & 28.60 \\
  \noalign{\vskip 2pt}
\hline
  \noalign{\vskip 3pt}
Perseus		& 0.70 &		26.12 & 23.77 & 24.04 & 24.08 & 28.03 & 25.20 & 25.38 & 25.32	& 30.57 & 28.77 & 28.95 & 28.94\\
  \noalign{\vskip 2pt}
\hline
  \noalign{\vskip 3pt}
Abell\,2390	& 0.75 &		25.88 & 23.60 & 23.85 & 23.87 & 27.83 & 25.08 & 25.23 & 25.17 & 30.42 & 28.70 & 28.86 & 28.88\\
Abell\,2764	& 0.75 &		26.06 & 23.71 & 23.94 & 23.96 & 27.83 & 25.19 & 25.34 & 25.27 & 30.56 & 28.78 & 28.96 & 28.98\\
  \noalign{\vskip 2pt}
\hline
\end{tabular}
\end{table*}

\begin{figure*}
\centering
	\includegraphics[width=0.955\linewidth]{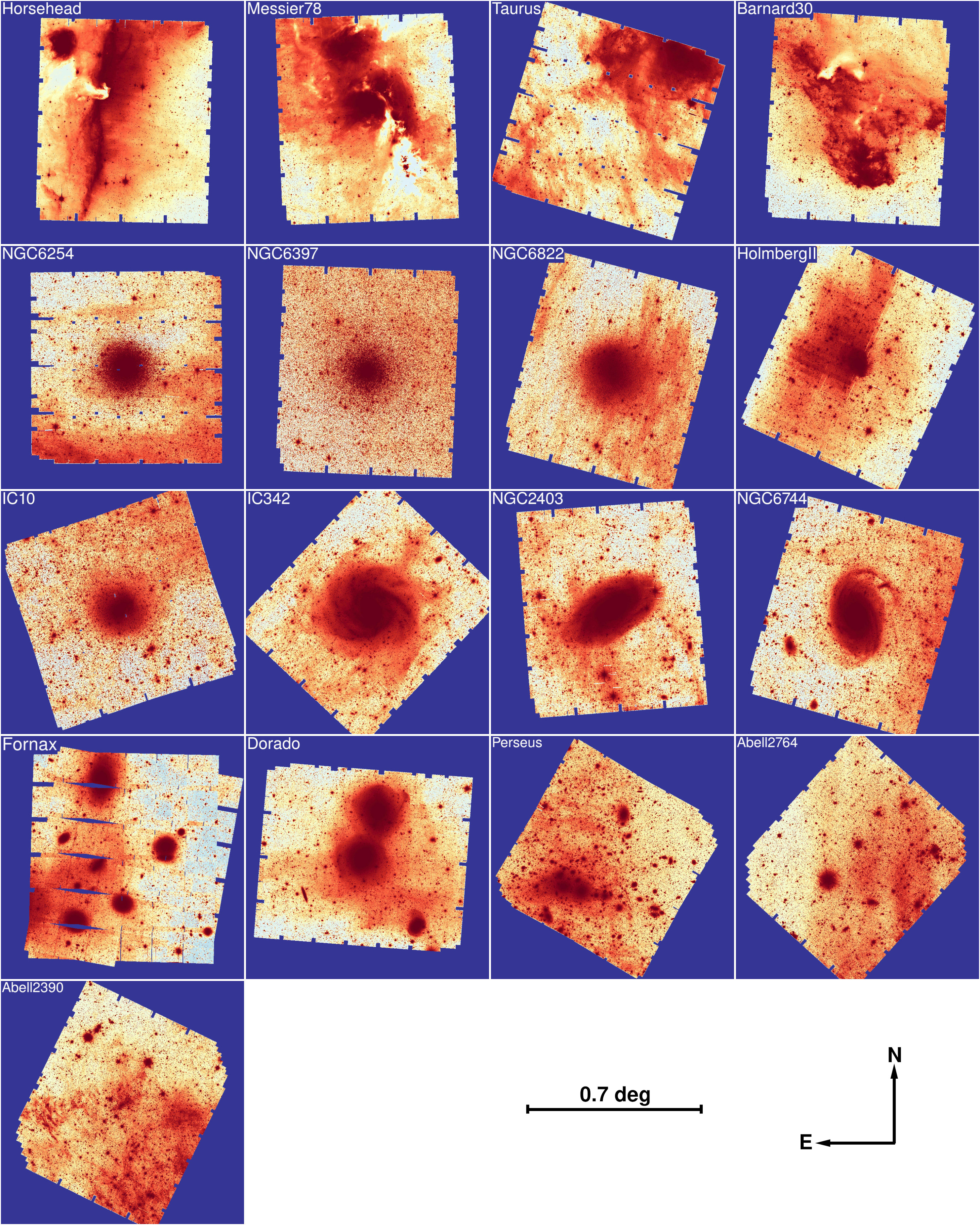}
    \caption{The 17 \ac{ERO} fields (approximately 0.6\,deg$^2$ each, \IE preview). The blue frame encompasses the entirety of the FITS image for each science stack, guaranteeing that no quality \Euclid data are omitted during projection into the equatorial coordinate system. Pixels identified as valid are those shared by both the VIS and NISP instruments. A dithering pattern impacted three fields -- Taurus, NGC\,6254, and IC\,10 -- resulting in incomplete sky coverage. The Fornax field is limited by having only two VIS exposures, leading to prominent gaps. Holmberg\,II, Fornax, and Dorado suffer from issues related to stray light, which will be addressed in the upcoming \ac{ERO} data release. Any observed variations in the background of the remaining images stem from the \ac{LSB} detection of Galactic nebulae (top row) or the presence of faint Galactic cirrus, illustrating the \Euclid's capability to easily capture these subtle astronomical features.
    \label{fig:EROfields}}
\end{figure*}

\begin{figure*}
    \centering
	\includegraphics[height=0.88\textheight]{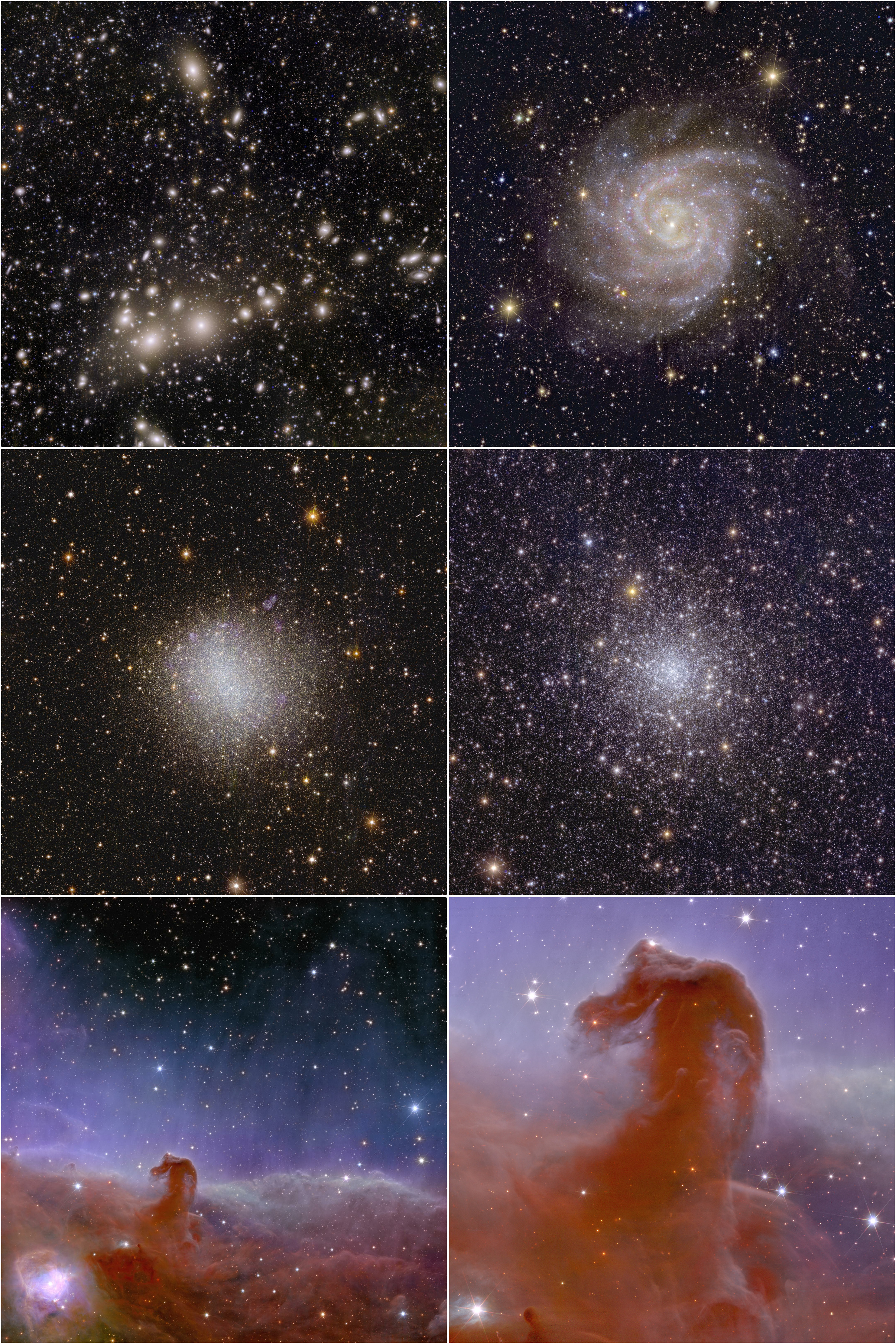}
    \caption{Released on 7 November 2023, the first set of five colour images unveiled \Euclid's capabilities to the global community. The images (cropped \ac{FOV}$=0.5$\,deg$^2$), starting from the top left, feature the Perseus cluster, IC\,342, NGC\,6822, NGC\,6397, and the Horsehead nebula, along with a cutout on the lower right ($\ang{;10;}\times\ang{;10;}$) that highlights the image resolution and depth achieved by \Euclid. The pipeline detailed in this paper produced each of the three channels that contributed to the initial RGB images. These images were subsequently refined using external tools. The chosen colour palette assigns assigns the \IE, \YE, and \HE bands to the blue, green, and red channels respectively, displaying the full sensitivity range of the observatory and offering a new perspective on these astronomical subjects. Credit: ESA/Euclid/Euclid Consortium/NASA, image processing by J.-C. Cuillandre (CEA Paris-Saclay), G. Anselmi.
    \label{fig:EROsplash2023}}
\end{figure*}

\newpage
\section{\label{sec:fluxcal}Selection of relevant stars along a given line of sight}
\label{apdx:DoradoFluxcal}

\begin{figure}[htb!]
\centerline{\includegraphics[clip=,width=0.8\textwidth]{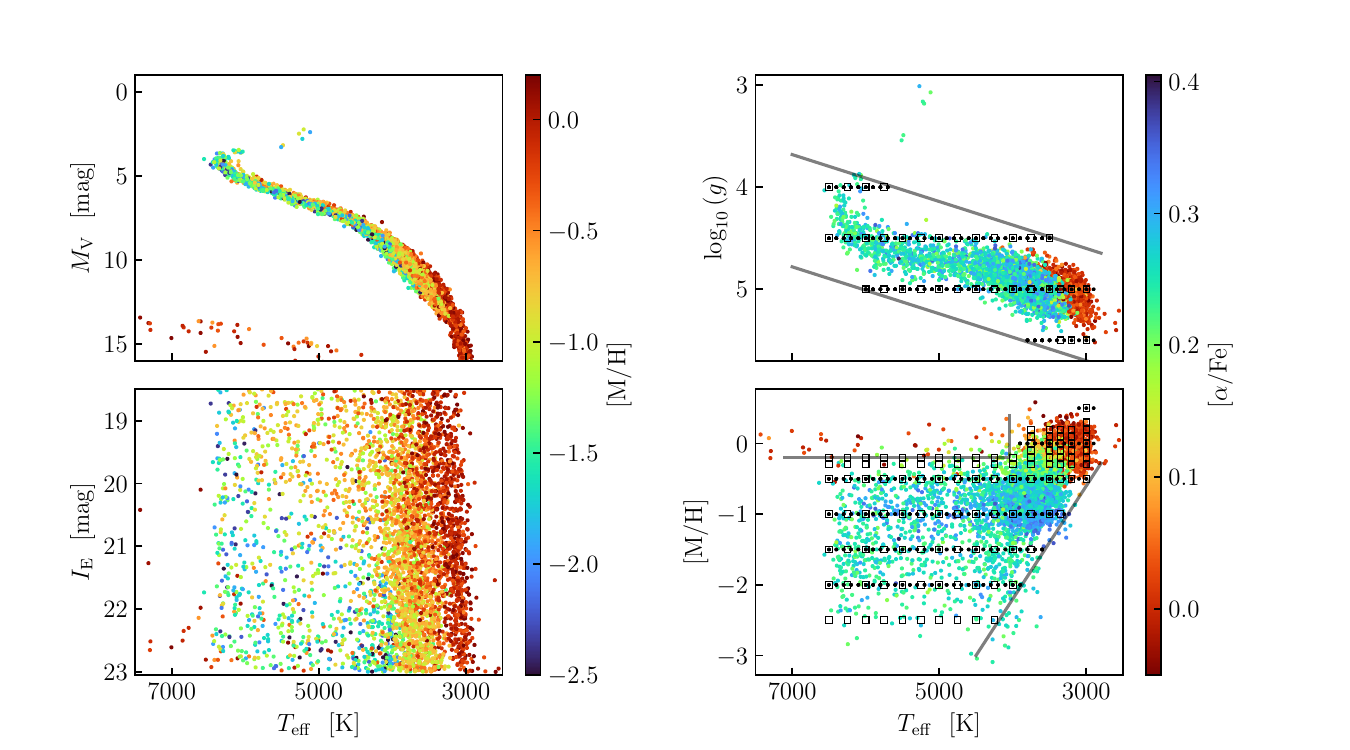}}
\caption[]{Relevant stars along the line of sight towards the Dorado field. This example showcases the selection process conducted prior to deriving colour transformations. The coloured dots depict the simulated stellar population along the line of sight, as provided by the Besan\c{c}on model of the Galaxy. In the two left panels, the colour coding represents metallicity [M/H]. In the two right panels, the colour coding corresponds to [$\alpha$/Fe], with black symbols marking the models available within the data-containing range of the simulation. Grey lines delineate the approximate boundaries we established for this selection process, ensuring that the subsequent colour transformations accurately reflect the characteristics of the stellar population along the observed line of sight.}
\label{fig:Dorado_ModSel}
\end{figure}

\begin{figure}[htb!]
\rule[0pt]{0.125\textwidth}{0pt} \includegraphics[clip=,width=0.71\textwidth]{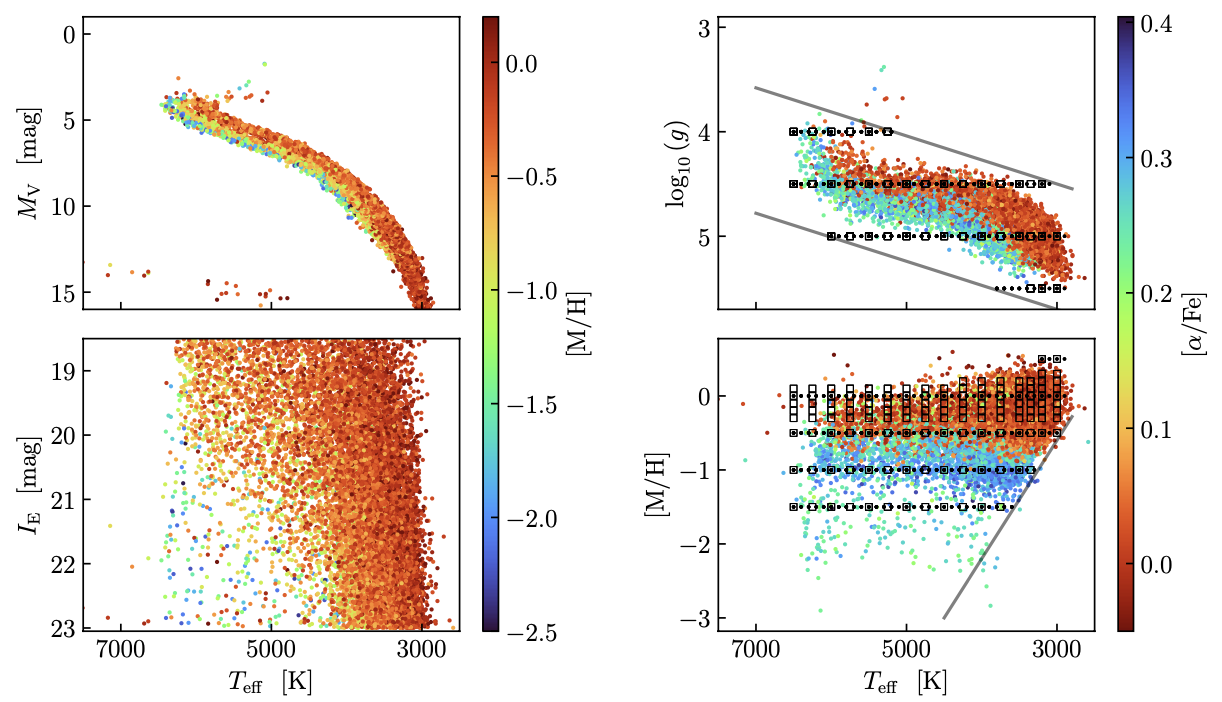}
\caption[]{Relevant stars along the line of sight towards the Perseus field. The layout is as in Fig.~\ref{fig:Dorado_ModSel}}
\label{fig:Perseus_ModSel}
\end{figure}

Colour transformations for comparing \Euclid photometry with existing literature are based on synthetic photometry of prevalent stars in the observation region, demonstrated through the Dorado and Perseus cases. Figure~\ref{fig:Dorado_ModSel} depicts outcomes from querying the Besançon model of the Galaxy \citep{Robin+2003,Czekaj2012,Lagarde+2021} for a 1 deg$^2$ area towards NGC\,1553, illustrating the process of deriving these transformations for accurate photometric comparisons.
Both the Besan\c{c}on model and the Planck 2013 dust map indicate an $A_V \simeq 0.04$ in the direction of NGC\,1553, suggesting that extinction effects are minimal and unlikely to be discernible in the figure. The right panels of the figure illustrate the models employed, with the {\tt SSED} models represented by black open squares and the {\tt TSED} models by black dots. It is important to note that [$\alpha$/Fe] is explicitly considered only in the selection process for {\tt TSED} models. Figure~\ref{fig:Perseus_ModSel} showcases the stellar populations expected along the line of sight towards Perseus, traversing extensive regions of the Galaxy's thin and thick discs.

\newpage
\section{Optical model of the telescope}
\label{apdx:OpticalModel}

To accurately correlate the flux observed at large radii with the energy concentrated in the core of the PSF, a model of the \Euclid \ac{PSF} was developed. This modelling encompassed the broadband \Euclid \ac{PSF} across five distinct wavelengths, each representing the \IE, \YE, \JE, and \HE bands. The approach involved calculating the modulus squared of the Fourier transform of the pupil function, adhering to the Fraunhofer far-field diffraction approximation \citep{Born1999}. The pupil function represents the complex amplitude of light within the pupil, assuming ideal optical conditions without any phase term. Consequently, the pupil function equates to the real transmission of the pupil, designated as 1 inside the pupil's boundary and 0 outside.
The \Euclid pupil was simulated on a substantial grid ($16\,384\times16\,384$) using an outer diameter of 1.21\,m for the primary mirror M1, a central obstruction of 0.395\,m by the secondary mirror M2, and the width of 12\,mm for the three spider arms. The spiders are connected tangentially to the structure supporting M2 \citep[see e.g.][]{racca2016}.

\subsection{Model of the encompassed energy of the \Euclid PSF}

To precisely determine the radial profile of the \ac{PSF} by examining the cumulative flux fraction as a function of radius, we implemented a sampling strategy for our model \ac{PSF} to achieve a resolution finer than \ang{;;0.01}. This necessitates oversampling by a factor of 5 relative to the Nyquist rate, resulting in a simulated field size of $\ang{;2.5;}\times\ang{;2.5;}$ on our $16\,384\times16\,384$ grid. Subsequently, we calculated the four \Euclid broadband \acp{PSF}, adjusting the pupil sampling to ensure a consistent pixel scale across all wavelengths. This enabled us to calculate the proportion of total flux within a circular aperture of incrementally increasing radius, aligning these values with measurements extracted from the \ac{ERO} data set.

For context, if we consider a perfect Airy disc for \Euclid, the \ac{FWHM} in radians is determined by the formula $1.025\times\lambda$/D (M1), which yields the following \ac{FWHM} values at the central wavelength for each of the four broadband filters: $\IE=\ang{;;0.136}$, $\YE=\ang{;;0.179}$, $\JE=\ang{;;0.213}$, and $\HE=\ang{;;0.298}$. This theoretical calculation aligns with the natural \ac{FWHM} observed in VIS non-resampled data at the finer \ang{;;0.1}\,pixel$^{-1}$ sampling rate. However, the coarser \ang{;;0.3}\,pixel$^{-1}$ sampling rate of NISP degrades the observed \ac{FWHM}, averaging approximately \ang{;;0.41} across its three bands.

\subsection{Energy in the diffraction spikes}

Both VIS and NISP images display six pronounced diffraction spikes around bright stars, a result of the structure created by the three supporting spiders. In the case of a nearly 5th magnitude star, such as that seen in one of our \ac{ERO} fields (\cref{fig:VBPPhoenicis}), each spike stretches from \ang{;12;} in VIS to as much as \ang{;20;} in NISP for the \HE-band. The extent of these spikes is influenced by the wavelength, with longer wavelengths exhibiting stronger effects and allowing for detection at greater distances. This phenomenon underscores two crucial points: first, the extended \ac{PSF} is likely to have minimal power at large radii, evidenced by our ability to trace the spikes over substantial distances; second, there is a need to specifically model and measure the energy contained within the diffraction spikes for each \Euclid band to affirm the methodology employed in determining the extended \ac{PSF}.

A preliminary estimate of the total flux present in the diffraction pattern can be derived geometrically by calculating the area ratio of the spiders to the transmissive area, resulting in a value of 1.89\%. However, due to the blinding effect of a 5th magnitude star on the \ac{ERO} images, it is not feasible to directly measure this total quantity within a radius of \ang{;;5} from our images. Instead, the proportion of energy situated beyond a specific radius, which is quantifiable based on the \ac{ERO} data, must initially be predicted through simulations.
To simulate the fraction of flux from the central object located within the six diffraction spikes, we created the largest feasible field for our simulation grid, realised when the \ac{PSF} is critically (Nyquist) sampled. In this scenario, the pupil diameter is halved in comparison to our grid size, rendering one pupil pixel equivalent to the diameter of M1 divided by our grid size: 148 microns on the primary mirror. This adjusted simulation extends to a radius of \ang{;7;}, with analyses of the \ac{ERO} data suggesting that a significant portion of the energy within the spikes is contained within this radius.
By creating identical pupil models both with and without the presence of spiders, we computed the corresponding \acp{PSF}, subtracted one from the other, and subsequently quantified the fraction of flux attributable to the spiders. In this way we ascertain the total fraction of flux from the central star located within all six diffraction spikes beyond a radius of \ang{;;10}, yielding the following percentages for each band: $\IE=0.22\%$, $\YE=0.41\%$, $\JE=0.60\%$, and $\HE=0.80\%$.

\begin{figure*}[htb!]
\includegraphics[width=1.0\textwidth]{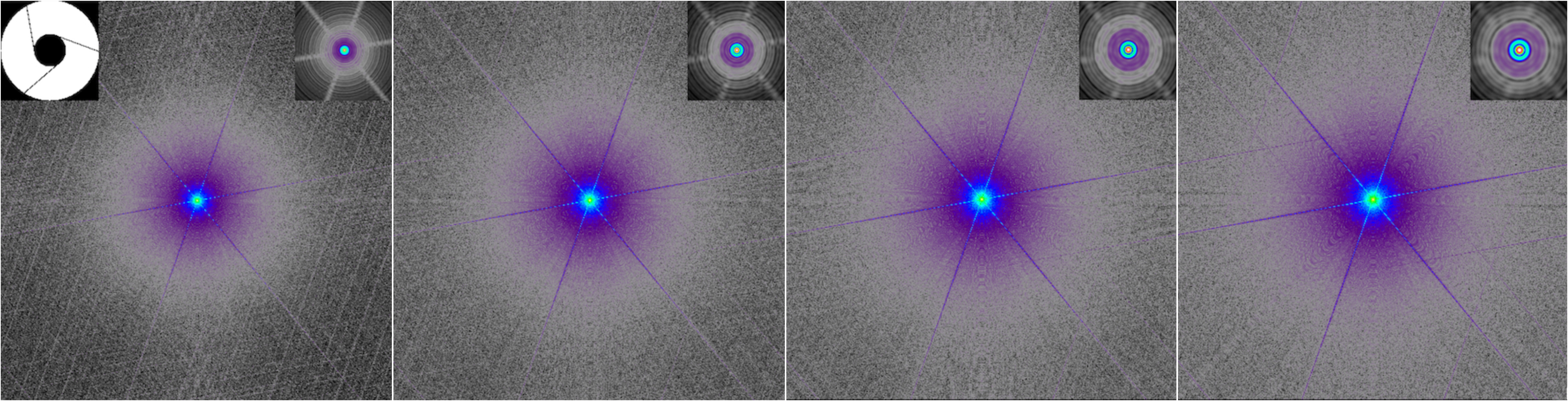}
\caption[]{\textit{Main panels from left to right}: The full \ang{;13;}$\times$\ang{;13;} field simulated PSF in the \IE, \YE, \JE, and \HE bands, in log scale, the core of the PSF in the inset (\ang{;;6}$\times$\ang{;;6}), and the pupil function in the top-left corner of the left panel. The pure diffraction halo grows larger towards the \ac{NIR}.}
\label{fig:PSFModelAllBands}
\end{figure*}

\newpage
\section{ERO catalogue parameters by {\tt SourceExtractor}}
\label{apdx:EROCatalogParameters}

\begin{table*}[htbp!]
\centering
\caption{Input parameters for the {\tt SourceExtractor} run producing the ERO science validation catalogues are outlined below. Some parameters generate multiple columns in the output catalogue, as exemplified by {\tt MAG\_APER}, which covers 10 different apertures as described in \cref{sec:cats}. This is indicated by the jump in column numbers in this table. In total, the ERO catalogues feature 364 columns. The parameter descriptions are from {\tt SourceExtractor}.}
\label{table:erocatalogparams}
\small
\begin{tabular*}{1.0\linewidth}{@{\extracolsep{\fill}}llll}
 \hline\hline
  \noalign{\vskip 3pt}
Column & Parameter & Description\hspace{7cm} &  Unit \\
  \noalign{\vskip 2pt}
\hline
 \noalign{\vskip 3pt}
   1 & {\tt NUMBER                 }& Running object number                                      & \\
   2 & {\tt X\_IMAGE                }& Object position along x                                    & [pixel]\\
   3 & {\tt Y\_IMAGE                }& Object position along y                                    & [pixel]\\
   4 & {\tt ID\_PARENT              }& Parent ID (before deblending)                              & \\
   5 & {\tt EXT\_NUMBER             }& FITS extension number                                      & \\
   6 & {\tt FLUX\_ISO               }& Isophotal flux                                             & [ADU]\\
   7 & {\tt FLUXERR\_ISO            }& RMS error for isophotal flux                               & [ADU]\\
   8 & {\tt MAG\_ISO                }& Isophotal magnitude                                        & [mag]\\
   9 & {\tt MAGERR\_ISO             }& RMS error for isophotal magnitude                          & [mag]\\
  10 & {\tt FLUX\_ISOCOR            }& Corrected isophotal flux                                   & [ADU]\\
  11 & {\tt FLUXERR\_ISOCOR         }& RMS error for corrected isophotal flux                     & [ADU]\\
  12 & {\tt MAG\_ISOCOR             }& Corrected isophotal magnitude                              & [mag]\\
  13 & {\tt MAGERR\_ISOCOR          }& RMS error for corrected isophotal magnitude                & [mag]\\
  14 & {\tt FLUX\_APER              }& Flux vector within fixed circular aperture(s)              & [ADU]\\
  24 & {\tt FLUXERR\_APER           }& RMS error vector for aperture flux(es)                     & [ADU]\\
  34 & {\tt MAG\_APER               }& Fixed aperture magnitude vector                            & [mag]\\
  44 & {\tt MAGERR\_APER            }& RMS error vector for fixed aperture mag.                   & [mag]\\
  54 & {\tt FLUX\_AUTO              }& Flux within a Kron-like elliptical aperture                & [ADU]\\
  55 & {\tt FLUXERR\_AUTO           }& RMS error for AUTO flux                                    & [ADU]\\
  56 & {\tt MAG\_AUTO               }& Kron-like elliptical aperture magnitude                    & [mag]\\
  57 & {\tt MAGERR\_AUTO            }& RMS error for AUTO magnitude                               & [mag]\\
  58 & {\tt KRON\_RADIUS            }& Kron apertures in units of A or B                          & \\
  59 & {\tt FLUX\_PETRO             }& Flux within a Petrosian-like elliptical aperture           & [ADU]\\
  60 & {\tt FLUXERR\_PETRO          }& RMS error for Petrosian flux                               & [ADU]\\
  61 & {\tt MAG\_PETRO              }& Petrosian-like elliptical aperture magnitude               & [mag]\\
  62 & {\tt MAGERR\_PETRO           }& RMS error for Petrosian magnitude                          & [mag]\\
  63 & {\tt PETRO\_RADIUS           }& Petrosian apertures in units of A or B                     & \\
  64 & {\tt FLUX\_GROWTH            }& Cumulated growth-curve                                     & [ADU]\\
  65 & {\tt FLUX\_GROWTHSTEP        }& Step for growth-curves                                     & [pixel]\\
  66 & {\tt MAG\_GROWTH             }& Cumulated magnitude growth-curve                           & [mag]\\
  67 & {\tt MAG\_GROWTHSTEP         }& Step for growth-curves                                     & [pixel]\\
  68 & {\tt FLUX\_RADIUS            }& Fraction-of-light radii                                    & [pixel]\\
  69 & {\tt BACKGROUND             }& Background at centroid position                            & [ADU]\\
  70 & {\tt MU\_MAX                 }& Peak surface brightness above background                   & [mag arcsec$^{-2}$]\\
  71 & {\tt X\_FOCAL                }& Barycenter position along focal-plane x axis               & \\
  72 & {\tt Y\_FOCAL                }& Barycenter position along focal-plane y axis               & \\
  73 & {\tt X\_WORLD                }& Barycenter position along world x axis                     & [deg]\\
  74 & {\tt Y\_WORLD                }& Barycenter position along world y axis                     & [deg]\\
  75 & {\tt ALPHA\_SKY              }& Right ascension of barycenter (native)                     & [deg]\\
  76 & {\tt DELTA\_SKY              }& Declination of barycenter (native)                         & [deg]\\
  77 & {\tt ALPHA\_J2000            }& Right ascension of barycenter (J2000)                      & [deg]\\
  78 & {\tt DELTA\_J2000            }& Declination of barycenter (J2000)                          & [deg]\\
  79 & {\tt ALPHA\_B1950            }& Right ascension of barycenter (B1950)                      & [deg]\\
  80 & {\tt DELTA\_B1950            }& Declination of barycenter (B1950)                          & [deg]\\
  81 & {\tt ERRX2\_IMAGE            }& Variance of position along x                               & [pixel$^{2}$]\\
  82 & {\tt ERRY2\_IMAGE            }& Variance of position along y                               & [pixel$^{2}$]\\
  83 & {\tt ERRXY\_IMAGE            }& Covariance of position between x and y                     & [pixel$^{2}$]\\
  84 & {\tt ERRA\_IMAGE             }& RMS position error along major axis                        & [pixel]\\
  85 & {\tt ERRB\_IMAGE             }& RMS position error along minor axis                        & [pixel]\\
  86 & {\tt ERRTHETA\_IMAGE         }& Error ellipse position angle (CCW/x)                       & [deg]\\
  87 & {\tt ERRCXX\_IMAGE           }& Cxx error ellipse parameter                                & [pixel$^{-2}$]\\
  88 & {\tt ERRCYY\_IMAGE           }& Cyy error ellipse parameter                                & [pixel$^{-2}$]\\
  89 & {\tt ERRCXY\_IMAGE           }& Cxy error ellipse parameter                                & [pixel$^{-2}$]\\
  90 & {\tt XPEAK\_IMAGE            }& x-coordinate of the brightest pixel                        & [pixel]\\
  91 & {\tt YPEAK\_IMAGE            }& y-coordinate of the brightest pixel                        & [pixel]\\
  92 & {\tt XPEAK\_FOCAL            }& Focal-plane x coordinate of the brightest pixel            & \\
  93 & {\tt YPEAK\_FOCAL            }& Focal-plane y coordinate of the brightest pixel            & \\
  94 & {\tt XPEAK\_WORLD            }& World-x coordinate of the brightest pixel                  & [deg]\\
  95 & {\tt YPEAK\_WORLD            }& World-y coordinate of the brightest pixel                  & [deg]\\
\noalign{\vskip 2pt}
\hline
\end{tabular*}
\end{table*}

\newpage

\begin{table*}[htbp!]
\centering
\small
\begin{tabular*}{1.0\linewidth}{@{\extracolsep{\fill}}llll}
\hline\hline
\noalign{\vskip 3pt}
Column & Parameter & Description\hspace{7cm} &  Unit \\
\noalign{\vskip 2pt}
\hline
\noalign{\vskip 3pt}
   96 & {\tt ALPHAPEAK\_SKY          }& Right ascension of brightest pix (native)                  & [deg]\\
  97 & {\tt DELTAPEAK\_SKY          }& Declination of brightest pix (native)                      & [deg]\\
  98 & {\tt ALPHAPEAK\_J2000        }& Right ascension of brightest pix (J2000)                   & [deg]\\
  99 & {\tt DELTAPEAK\_J2000        }& Declination of brightest pix (J2000)                       & [deg]\\
  100 & {\tt ALPHAPEAK\_B1950        }& Right ascension of brightest pix (B1950)                   & [deg]\\
 101 & {\tt DELTAPEAK\_B1950        }& Declination of brightest pix (B1950)                       & [deg]\\
 102 & {\tt XMIN\_IMAGE             }& Minimum x-coordinate among detected pixels                 & [pixel]\\
 103 & {\tt YMIN\_IMAGE             }& Minimum y-coordinate among detected pixels                 & [pixel]\\
 104 & {\tt XMAX\_IMAGE             }& Maximum x-coordinate among detected pixels                 & [pixel]\\
 105 & {\tt YMAX\_IMAGE             }& Maximum y-coordinate among detected pixels                 & [pixel]\\
 106 & {\tt XWIN\_IMAGE             }& Windowed position estimate along x                         & [pixel]\\
 107 & {\tt YWIN\_IMAGE             }& Windowed position estimate along y                         & [pixel]\\
 108 & {\tt ERRX2WIN\_IMAGE         }& Variance of windowed position along x                           & [pixel$^{2}$]\\
 109 & {\tt ERRY2WIN\_IMAGE         }& Variance of windowed position along y                           & [pixel$^{2}$]\\
 110 & {\tt ERRXYWIN\_IMAGE         }& Covariance of windowed position between x and y                 & [pixel$^{2}$]\\
 111 & {\tt ERRAWIN\_IMAGE          }& RMS windowed position error along major axis                    & [pixel]\\
 112 & {\tt ERRBWIN\_IMAGE          }& RMS windowed position error along minor axis                    & [pixel]\\
 113 & {\tt ERRTHETAWIN\_IMAGE      }& Windowed error ellipse position angle (CCW/x)                   & [deg]\\
 114 & {\tt ERRCXXWIN\_IMAGE        }& Cxx windowed error ellipse parameter                       & [pixel$^{-2}$]\\
 115 & {\tt ERRCYYWIN\_IMAGE        }& Cyy windowed error ellipse parameter                       & [pixel$^{-2}$]\\
 116 & {\tt ERRCXYWIN\_IMAGE        }& Cxy windowed error ellipse parameter                       & [pixel$^{-2}$]\\
 117 & {\tt X2\_IMAGE               }& Variance along x                                           & [pixel$^{2}$]\\
 118 & {\tt Y2\_IMAGE               }& Variance along y                                           & [pixel$^{2}$]\\
 119 & {\tt XY\_IMAGE               }& Covariance between x and y                                 & [pixel$^{2}$]\\
 120 & {\tt A\_IMAGE                }& Profile RMS along major axis                               & [pixel]\\
 121 & {\tt B\_IMAGE                }& Profile RMS along minor axis                               & [pixel]\\
 122 & {\tt THETA\_IMAGE            }& Position angle (CCW/x)                                     & [deg]\\
 123 & {\tt ELONGATION             }& {\tt A\_IMAGE/B\_IMAGE}                                            & \\
 124 & {\tt ELLIPTICITY            }& 1 - {\tt B\_IMAGE/A\_IMAGE}                                        & \\
 125 & {\tt CXX\_IMAGE              }& Cxx object ellipse parameter                               & [pixel$^{-2}$]\\
 126 & {\tt CYY\_IMAGE              }& Cyy object ellipse parameter                               & [pixel$^{-2}$]\\
 127 & {\tt CXY\_IMAGE              }& Cxy object ellipse parameter                               & [pixel$^{-2}$]\\
 128 & {\tt ISOAREAF\_IMAGE         }& Isophotal area (filtered) above Detection threshold        & [pixel$^{2}$]\\
 129 & {\tt ISOAREA\_IMAGE          }& Isophotal area above analysis threshold                    & [pixel$^{2}$]\\
 130 & {\tt X2WIN\_IMAGE            }& Windowed variance along x                                  & [pixel$^{2}$]\\
 131 & {\tt Y2WIN\_IMAGE            }& Windowed variance along y                                  & [pixel$^{2}$]\\
 132 & {\tt XYWIN\_IMAGE            }& Windowed covariance between x and y                        & [pixel$^{2}$]\\
 133 & {\tt CXXWIN\_IMAGE           }& Windowed Cxx object ellipse parameter                      & [pixel$^{-2}$]\\
 134 & {\tt CYYWIN\_IMAGE           }& Windowed Cyy object ellipse parameter                      & [pixel$^{-2}$]\\
 135 & {\tt CXYWIN\_IMAGE           }& Windowed Cxy object ellipse parameter                      & [pixel$^{-2}$]\\
 136 & {\tt AWIN\_IMAGE             }& Windowed profile RMS along major axis                      & [pixel]\\
 137 & {\tt BWIN\_IMAGE             }& Windowed profile RMS along minor axis                      & [pixel]\\
 138 & {\tt THETAWIN\_IMAGE         }& Windowed position angle (CCW/x)                            & [deg]\\
 139 & {\tt CLASS\_STAR             }& S/G classifier output                                      & \\
 140 & {\tt FWHM\_IMAGE             }& FWHM assuming a gaussian core                              & [pixel]\\
 141 & {\tt XPSF\_IMAGE             }& X coordinate from PSF-fitting                              & [pixel]\\
 142 & {\tt YPSF\_IMAGE             }& Y coordinate from PSF-fitting                              & [pixel]\\
 143 & {\tt ALPHAPSF\_J2000         }& Right ascension of the fitted PSF (J2000)                  & [deg]\\
 144 & {\tt DELTAPSF\_J2000         }& Declination of the fitted PSF (J2000)                      & [deg]\\
 145 & {\tt FLUX\_PSF               }& Flux from PSF-fitting                                      & [ADU]\\
 146 & {\tt FLUXERR\_PSF            }& RMS flux error for PSF-fitting                             & [ADU]\\
 147 & {\tt MAG\_PSF                }& Magnitude from PSF-fitting                                 & [mag]\\
 148 & {\tt MAGERR\_PSF             }& RMS magnitude error from PSF-fitting                       & [mag]\\
 149 & {\tt FLUX\_POINTSOURCE       }& Point source flux from fitting                             & [ADU]\\
 150 & {\tt FLUXERR\_POINTSOURCE    }& RMS error on fitted point source total flux                & [ADU]\\
 151 & {\tt MAG\_POINTSOURCE        }& Point source total magnitude from fitting                  & [mag]\\
 152 & {\tt MAGERR\_POINTSOURCE     }& RMS error on fitted point source total magnitude           & [mag]\\
 153 & {\tt FLUX\_DISK              }& Disk total flux from fitting                               & [ADU]\\
 154 & {\tt FLUXERR\_DISK           }& RMS error on fitted disk total flux                        & [ADU]\\
 155 & {\tt MAG\_DISK               }& Disk total magnitude from fitting                          & [mag]\\
 156 & {\tt MAGERR\_DISK            }& RMS error on fitted disk total magnitude                   & [mag]\\
 157 & {\tt MU\_MAX\_DISK            }& Peak disk surface brightness above background              & [mag arcsec$^{-2}$]\\
 158 & {\tt MU\_EFF\_DISK            }& Effective disk surface brightness above background         & [mag arcsec$^{-2}$]\\
 159 & {\tt MU\_MEAN\_DISK           }& Mean effective disk surface brightness above background    & [mag arcsec$^{-2}$]\\
 160 & {\tt FLUX\_SPHEROID          }& Spheroid total flux from fitting                           & [ADU]\\
 161 & {\tt FLUXERR\_SPHEROID       }& RMS error on fitted spheroid total flux                    & [ADU]\\
 \noalign{\vskip 2pt}
\hline
\end{tabular*}
\end{table*}

\newpage

\begin{table*}[htbp!] 
\centering
\small
\begin{tabular*}{1.0\linewidth}{@{\extracolsep{\fill}}llll}
\hline\hline
\noalign{\vskip 3pt}
Column & Parameter & Description\hspace{7cm} &  Unit \\
\noalign{\vskip 2pt}
\hline
\noalign{\vskip 3pt}
 162 & {\tt MAG\_SPHEROID           }& Spheroid total magnitude from fitting                      & [mag]\\
 163 & {\tt MAGERR\_SPHEROID        }& RMS error on fitted spheroid total magnitude               & [mag]\\
 164 & {\tt MU\_MAX\_SPHEROID        }& Peak spheroid surface brightness above background          & [mag arcsec$^{-2}$]\\
 165 & {\tt MU\_EFF\_SPHEROID        }& Effective spheroid surface brightness above background     & [mag arcsec$^{-2}$]\\
 166 & {\tt MU\_MEAN\_SPHEROID       }& Mean effective spheroid surface brightness above background & [mag arcsec$^{-2}$]\\
 167 & {\tt DISK\_INCLINATION       }& Disk inclination from fitting                              & [deg]\\
 168 & {\tt DISK\_INCLINATIONERR    }& RMS error on disk inclination from fitting                 & [deg]\\
 169 & {\tt DISK\_THETA\_IMAGE       }& Disk position angle (CCW/x) from fitting                   & [deg]\\
 170 & {\tt DISK\_THETAERR\_IMAGE    }& RMS error on fitted disk position angle                    & [deg]\\
 171 & {\tt SPHEROID\_SERSICN       }& Spheroid Sersic index from fitting                         & \\
 172 & {\tt SPHEROID\_SERSICNERR    }& RMS error on fitted spheroid Sersic index                  & \\
 173 & {\tt SPHEROID\_THETA\_IMAGE   }& Spheroid position angle (CCW/x) from fitting               & [deg]\\
 174 & {\tt SPHEROID\_THETAERR\_IMAG }& RMS error on spheroid position angle                       & [deg]\\
 175 & {\tt SPHEROID\_REFF\_IMAGE    }& Spheroid effective radius from fitting                     & [pixel]\\
 176 & {\tt SPHEROID\_REFFERR\_IMAGE }& RMS error on fitted spheroid effective radius              & [pixel]\\
 177 & {\tt FLUX\_MODEL             }& Flux from model-fitting                                    & [ADU]\\
 178 & {\tt FLUXERR\_MODEL          }& RMS error on model-fitting flux                            & [ADU]\\
 179 & {\tt MAG\_MODEL              }& Magnitude from model-fitting                               & [mag]\\
 180 & {\tt MAGERR\_MODEL           }& RMS error on model-fitting magnitude                       & [mag]\\
 181 & {\tt MU\_MAX\_MODEL           }& Peak model surface brightness above background             & [mag arcsec$^{-2}$]\\
 182 & {\tt FLUX\_MAX\_MODEL         }& Peak model flux above background                           & [ADU]\\
 183 & {\tt FLUX\_EFF\_MODEL         }& Effective model flux above background                      & [ADU]\\
 184 & {\tt FLUX\_MEAN\_MODEL        }& Mean effective model flux above background                 & [ADU]\\
 185 & {\tt MU\_EFF\_MODEL           }& Effective model surface brightness above background        & [mag arcsec$^{-2}$]\\
 186 & {\tt MU\_MEAN\_MODEL          }& Mean effective model surface brightness above background   & [mag arcsec$^{-2}$]\\
 187 & {\tt XMODEL\_IMAGE           }& X coordinate from model-fitting                            & [pixel]\\
 188 & {\tt YMODEL\_IMAGE           }& Y coordinate from model-fitting                            & [pixel]\\
 189 & {\tt CXXMODEL\_IMAGE         }& Cxx ellipse parameter from model-fitting                   & [pixel$^{-2}$]\\
 190 & {\tt CYYMODEL\_IMAGE         }& Cyy ellipse parameter from model-fittinh                   & [pixel$^{-2}$]\\
 191 & {\tt CXYMODEL\_IMAGE         }& Cxy ellipse parameter from model-fitting                   & [pixel$^{-2}$]\\
 192 & {\tt SPREAD\_MODEL           }& Spread parameter from model-fitting                        & \\
 193 & {\tt SPREADERR\_MODEL        }& Spread parameter error from model-fitting                  & \\
 194 & {\tt NOISEAREA\_MODEL        }& Equivalent noise area of the fitted model                  & [pixel$^{2}$]\\
 195 & {\tt NITER\_MODEL            }& Number of iterations for model-fitting                     & \\
 196 & {\tt VECTOR\_MODEL           }& Model-fitting coefficients                                 & \\
 208 & {\tt VECTOR\_MODELERR        }& Model-fitting coefficient uncertainties                    & \\
 220 & {\tt MATRIX\_MODELERR       } & Model-fitting covariance matrix                            & \\
 364 & {\tt CHI2\_MODEL             }& Reduced Chi2 of the fit                                    & \\
\noalign{\vskip 2pt}
\hline
\end{tabular*}
\end{table*}

\end{appendix}

\end{document}